# The Distribution of the Elements in the Galactic Disk II. Azimuthal and Radial Variation in Abundances from Cepheids

R.E. Luck[1], S.M. Andrievsky[2,3], V.V. Kovtyukh[2], W. Gieren[4], and D. Graczyk[4]

[1]Department of Astronomy, Case Western Reserve University
10900 Euclid Avenue, Cleveland, OH 44106-7215
luck@fafnir.astr.cwru.edu

[2]Department of Astronomy and Astronomical Observatory, Odessa National University
Isaac Newton Institute of Chile, Odessa Branch
Shevchenko Park, 65014, Odessa, Ukraine

[3]GEPI, Observatoire de Paris-Meudon, CNRS, Université Paris Diderot, 92125 Meudon Cedex, France
scan@deneb1.odessa.ua, val@deneb1.odessa.ua

[4]Departamento de Astronomia, Universidad de Concepción,
Casilla 160-C, Concepcion, Chile
wgieren@astro-udec.cl, darek@astro-udec.cl

## Abstract

This paper reports on the spectroscopic investigation of 101 Cepheids in the Carina region. These Cepheids extend previous samples by about 35% in number and increase the amount of the galactic disk coverage especially in the direction of $l \approx 270°$. The new Cepheids do not add much information to the radial gradient, but provide a substantial increase in azimuthal coverage. We find no azimuthal dependence in abundance over an 80° angle from the galactic center in an annulus of 1 kpc depth centered on the Sun. A simple linear fit to the Cepheid data yields a gradient $d[Fe/H]/dR_G = -0.055 \pm 0.003$ dex/kpc which is somewhat shallower than found from our previous, smaller Cepheid sample.

## 1. Introduction

In previous papers from this series (Andrievsky et al. 2002a, 2002b, 2002c, 2004, Luck et al. 2003, 2006 (collectively Papers I-VI) and Kovtyuhk, Wallerstein, & Andrievsky (2005, (KWA)) we described the characteristic features of the metallicity distribution across galactic disk, as derived from Cepheid variable stars. Since our earlier work, the gradient has been investigated by variety of technique's, but the derived value of the gradient has not significantly changed from our 2006 value (Maciel & Costa 2010 and references therein). In Andrievsky et al. (2004) we noted an apparent step-like character in the iron (and also some other elements) abundance distribution at galactocentric distances of about 10 − 11 kpc. More importantly, in the 2006 work, we found an ostensible inhomogeneity, a "metallicity island", in the direction of l = 130° at a distance of about 3 kpc from the Sun.

Any structure within the metallicity distribution, such as the alleged "metallicity island", if it is proved to be real, may have important consequences for scenarios of the formation and evolution of our Galaxy. In order to further explore the distribution of elements in the galactic disk we have acquired data on a large sample of Cepheids in the southern sky. These Cepheids stretch along the galactic plane from Puppis to Scorpius and are concentrated in the Carina / Centaurus region. These stars have galactocentric distances of about 8 kpc, but spread out along a significant arc relative to the galactic center. They will add little information about the radial gradient, but will significantly add to the galactic surface area investigated.

We present here the results of the abundance determinations for 101 galactic Cepheids, the majority of which have not been investigated spectroscopically before (at least for detailed elemental abundance determination). We obtained spectra for four catalog Cepheids which are not included in this paper. The bright Cepheid X Sgr and the much fainter Cepheid HQ Car were observed but both were found to have double lines making an analysis difficult if not impossible. FO Car is listed in the GCVS as an 10.4 day Cepheid, but the spectrum appears to be more like that of an SRd variable – very cool at the time of observation. The last star eliminated is AF Cru. This star is listed as a Cepheid by SIMBAD, but is actually a 10 day spectroscopic binary field G dwarf (Berdnikov & Turner 1997). The list of Cepheids considered can be found in Table 1 along with some basic data.

## 2. Spectroscopic material

High signal-to-noise spectra were obtained in the period 25 March – 1 April 2010 using the 2.2m MPG telescope and FEROS spectrograph at ESO La Silla. The spectra cover a continuous wavelength range from 400 to 785 nm with a resolving power of about 48000. Typical maximum S/N values (per pixel) for the spectra are in excess of 150. Each night we observed a broad lined B star with a S/N exceeding that of the program stars to enable cancellation of telluric lines where necessary. Table 1 also contains details concerning our program Cepheid observations.

We used IRAF[1] to perform CCD processing, scattered light subtraction, and echelle order extraction. For these spectra two extractions were done, one uses a zero-order (i.e., the mean) normalization of the flat field which removes the blaze from the extracted spectra. The second uses a high-order polynomial to normalize the flat-field which leaves the blaze function in the extracted spectrum. The latter spectrum reflects more accurately the true counts along the orders. A Windows based graphical package (ASP) developed by REL was used to process the blaze removed spectra. This included Beer's law removal of telluric lines, smoothing with a fast Fourier transform procedure, continuum normalization, and wavelength calibration using template spectra. Echelle orders show significant S/N variations from edge to maximum due to blaze efficiency. To maximize the S/N in these spectra we have co-added the order overlap region using as weights the counts from the second data extraction. The co-added spectra were then inspected and the continua sometimes modified by minor amounts in the overlap regions. Equivalent widths from the co-added spectra were then measured using the Windows compatible DECH20 package (Galazutdinov 1992). The line-list is the same as used in our previous studies and derives from Kovtyukh & Andrievsky (1999). The equivalent widths can be obtained from the authors by request.

---

[1] IRAF is distributed by the National Optical Astronomy Observatories, which are operated by the Association of Universities for Research in Astronomy, Inc., under cooperative agreement with the National Science Foundation.

It would be useful to compare equivalent widths from a variety of sources for our program stars but Cepheids do not lend themselves to such efforts. The Cepheids would have to be at the same phase from different authors/measurers and should be from different sources (spectrographs). With respect to the stars in the current paper such data does not exist. As a basic check, we have measured equivalent widths for HR 3177 (an non-variable G1 Ib star) from both the FEROS spectrograph and the Sandiford spectrograph on the McDonald Observatory 2.2m Struve telescope. There is no scale difference between the two datasets and the average fractional difference is of order 7 percent.

## 3. Methods

### 3.1. Atmospheric Models

Atmospheric models were interpolated for each Cepheid using the Kurucz ATLAS9 model atmosphere grid (Kurucz 1992). The ATLAS9 models are 1D LTE models with ODF line-blanketing and a standard treatment of convection (using the default overshoot scale length). The models used adopt a microturbulent velocity of 3 km s$^{-1}$. At some phases Cepheids can have microturbulent velocities significantly deviating from this model value; however, our previous test calculations (see Luck et al. 2000) showed that changes of several kilometers per second in the microturbulent velocity used to compute the model atmosphere has little effect on the structure of the model. Thus, a mismatch between the derived microturbulent velocity for a specific star and the microturbulence used in the model computation at a 2 – 3 km s$^{-1}$ level has an insignificant impact on the resulting elemental abundances.

### 3.2. Atmosphere parameters: temperatures, gravities, microturbulent velocities

The effective temperature for each Cepheid was determined using effective temperature relations which originate in the work of Kovtyukh & Gorlova (2000) and Kovtyukh (2007). These relations combine the effective temperature with a set of spectral line depth ratios. The internal accuracy of the effective temperature determined in this way is rather high in the temperature range 5000 K to 6500 K: typically 150 K or less (standard deviation or a standard error ($\sigma/\sqrt{N}$) of 10 to 20 K). Note that this method uses multiple measures (ratios) each obtained from a single observation. In Table 2 we give the standard deviation of the mean temperature and number of ratios used at each phase (observation). Another important advantage of this method (or any spectroscopic method) is that it produces reddening-free $T_{eff}$ estimates.

The method used for gravity and microturbulent velocity determination in supergiant stars such as Cepheids is described in detail by Kovtyukh & Andrievsky (1999). This method determines the microturbulent velocity using Fe II lines: the dominant ionization species of iron and hence less susceptible to any NLTE effects which might be in play in supergiant atmospheres. The gravity value is found by enforcing the ionization balance condition; i.e., the mean iron abundance from Fe II lines equals the iron abundance which results from the Fe I – EW relation extrapolated to zero equivalent width.

The final results of the determinations of $T_{eff}$, $log\ g$ and $V_t$ are given in Table 2. The uncertainty in the microturbulent velocity and the gravity is more difficult to assess than the statistical error in the effective temperature. For the microturbulence, a variation of ±0.5 km s$^{-1}$ from the adopted velocity causes a significant slope in the relation between Fe II line abundance and equivalent width. We therefore adopt ±0.5 km s$^{-1}$ as the uncertainty in the microturbulence. For $log\ g$ we adopt ±0.1 dex as the formal uncertainty based on the numerical result that a change in gravity at that level will result in a difference of 0.05 dex between the total iron abundance as computed from the Fe I and Fe II lines. Since we have forced an ionization balance we do not allow a spread larger than 0.05 dex in the total abundance of iron as derived from the two ions and thus our uncertainty estimate. Iron abundance details are also given in Table 2 along with the stellar parameters.

### 3.3. Distances

For the determination of the Cepheid galactocentric distances the following standard formula was used:

$$R_G = [R_{G,\odot}^2 + (d\cos b)^2 - 2R_{G,\odot} d\cos b \cos l]^{1/2}$$

where $R_{G,\odot}$ is the galactocentric distance of the Sun (7.9 kpc, McNamara et al. 2000), $d$ is the heliocentric distance of the Cepheid, and $l$ and $b$ are the galactic longitude and latitude respectively.

To determine the heliocentric distance we need the absolute V magnitude, the mean apparent V magnitude, and the reddening. To estimate the absolute magnitude we used "absolute magnitude – pulsational period" relation of Fouqué et al. (2007) for Johnson V magnitudes. The formal uncertainty of this relation is about ±0.2 mag which leads to a distance uncertainty of 9%. Periods are from Berdnikov (2006 – private communication to S. M. Andrievsky)) or the GCVS. For first overtone pulsators (DCEPS), we assume $P_0 = P_1 / 0.71$ where $P_1$ is the observed period. The mean visual (V) magnitudes are from Fernie et al. (1995) or as noted in Table 1. Reddenings are from Fouqué et al. (2007) or Fernie et al. (1995) with the systematic correction of Fouqué et al. applied (i.e., E(B-V) = 0.952 E(B-V)$_{Fernie}$). These reddening are on the E(B-V) system of Laney & Caldwell (2007). We follow Fouqué et al. and adopt $R_V=A_V/E(B-V) = 3.23$. These distances show minor differences with respect to distances found in Papers I-VI leading us to recompute all distances for the stars of Papers I-VI and KWA using this procedure. Galactocentric distances for program stars are listed in Table 1 along with the adopted E(B-V) and <V>. Table 3 contains the galactocentric distance information for the stars from Papers I-VI and KWA.

Adopting a distance modulus combined uncertainty of ±0.3 mag the distances to our Cepheids are imprecise at the 13% level. Propagating this uncertainty into the galactocentric distances we find possible distance errors that range from 10's of parsecs for Cepheids at the solar galactocentric radius to more than a kiloparsec for stars well outside the solar circle. We indicate in Figure 1 two representative distance uncertainties.

## 4. Results

### 4.1. Elemental abundances

The elemental abundances in our program stars are in Table 4 − mean [x/H] ratios for each star including the stars of Papers I-VI and KWA. Iron abundance details for our new southern Cepheids are given in Table 2 along with the stellar parameters. Note that the oscillator strengths used in this (and all preceding Cepheid analyses of this series) are based on an inverted solar analysis which used the solar abundances of Grevesse et al. 1996.

The mean abundances of Table 4 have been averaged using a weighting equal to the number of spectra for each star. For example, BF Oph is included in the new southern Cepheid dataset and is also found in Luck et al. (2006). Each analysis has a single spectrum so they are added with equal weight. Note that the Table 4 abundances for the stars in prior programs can be different from the previously published values. This is due to the averaging over the gradient papers. Additionally, a number of the Cepheids have been considered in multi-phase analyses (Luck & Andrievsky 2004, Kovtyukh et al. 2005, Andrievsky, Luck, & Kovtyukh 2005, and Luck et al. 2008 – sources 9 through 12 in Table 4) and the average from these studies is used in the formation of Table 4 in preference to the abundances found in Papers I-VI.

#### 4.1.1. Comparison with other Abundance Studies

Abundance reliability in Cepheids as a function of phase has been addressed in a recent series of papers (Luck & Andrievsky 2004, Kovtyukh et al. 2005, Andrievsky, Luck, & Kovtyukh 2005, and Luck et al. 2008). Our finding is that accurate parameters and [Fe/H] ratios can be derived at any phase in a Cepheid independent of the period of the Cepheid. Based on this we would not expect any untoward effects in the abundances due to phase (stellar) parameter problems.

A number of the new southern Cepheids have previous analyses. The largest number of analyses are from our previous work – 12 stars, and Romaniello et al. (2008) – 18 stars. The mean difference between Papers I-VI and this work for [Fe/H] is +0.02 ($\sigma$ = 0.09) while for Romaniello et al. the mean [Fe/H] difference is +0.03 ($\sigma$ = 0.11). The stars in common and the [Fe/H] data can be found in Table 5. Another limited source of comparison is Yong et al. (2006). There are two stars in common, NT Pup and CE Pup, for which we obtain [Fe/H] values higher by 0.25 and 0.2 dex respectively. HQ Car was also analyzed by Yong et al. but we were not able to obtain abundances from our spectrum due to line doubling and emission. The agreement with our previous analyses and the work of Romaniello et al. is excellent; however, the abundance results of Yong et al. appear to be somewhat low relative to our values. The latter conclusion was also reached by Lemasle et al. (2008) in regard to their abundance determinations.

### 4.2. Spatial Abundance Distributions

Using the distance and abundance information from Tables 2 – 4 we have constructed radial (1D) distributions of elemental abundances. The radial gradients are found in Table 6 for all elements with more than 30 determinations.

#### 4.2.1. [Fe/H] Distribution

In Figure 1 we show the radial gradient for iron discriminating between the older data and the data from the southern Cepheids of this study. As can be seen the stars of this study are concentrated at a distance of about 8 kpc from the galactic center. The radial gradient is thus dominated by the stars from the older studies. Fitting a simple linear model to the data we find that [Fe/H] = −0.055 (±0.003) * $R_G$ + 0.475 (±0.027). This is somewhat shallower than our previous determinations: slope = −0.068, but in excellent agreement with the values of Romaniello et al. (2008) and Pedicelli et al. (2009).

If we divide the whole range into three parts (zone I: 4.0 − 6.6 kpc, zone II 6.6 − 10.6 kpc, and zone III 10.6 – 14.6 kpc), then the statistics for each zone and the total sample for iron (the most reliable abundance) are the following:

Zone I:     [Fe/H] = −0.100 (±0.041) * $R_G$ + 0.794 (±0.245)
            <[Fe/H]> = +0.201 (s.d. = 0.116, n = 18)

Zone II:    [Fe/H] = −0.032 (0.006) * $R_G$ + 0.282 (± 0.049)
            <[Fe/H]> = +0.025 (s.d. = 0.092, n = 217)

Zone III :  [Fe/H] = −0.060 (± 0.014) * $R_G$ + 0.516 (± 0.172)
            <[Fe/H]> = −0.226 (s.d. = 0.141, n = 35)

The slope in Zone II is heavily influenced by the high metallicity stars at galactocentric radius ($R_G$) ~ 9 to 10 kpc. This is the "metallicity island" noted by Luck et al. (2006). Confirmation (or direct contradiction) or the reality of this anomaly has yet to be provided, but we do note that it depends on the abundances of fewer than 10 stars. With the use of our internally consistent abundances the outer zone gradient is much the same as the overall gradient. In agreement with other studies (Romaniello et al. 2008, Pedicelli et al. 2009), as well as our previous work, we find the gradient in the inner zone shows a steepening towards the galactic center (see Fig 1).

Maciel & Costa (2010) emphasize the flattening of the gradient outside of galactocentric radius 10 kpc. Our data contains no evidence for such behavior. The behavior of the gradient in Zone II and particularly in the region of $R_G$ of 9 – 10 kpc is crucial in theoretical studies of the gradient such as the work of Cruz et al. (2011). Their contention is that the change in slope at about 9 kpc is due to dynamical effects at the co-rotation radius. While there may be dynamical effects which can cause a break in an abundance gradient, the evidence from Cepheids on such a discontinuity is scant, and depends in all elements on the presence of the "metallicity island." If the island disappears, then so does the discontinuity in the gradient.

The five stars at the greatest distance from the galactic center in this work show rather different abundances: [Fe/H] = −0.26, −0.04, −0.51, −0.34, and −0.74 for CU Mon, CE Pup, EE Mon, ER Aur, and EF Tau respectively (see Figure 1). The two most discrepant objects from the overall trend are CE Pup and EF Tau. Their distances and abundances are uncertain at the ±1 kpc and ±0.1 dex levels respectively . A distance change of 5 kpc or an abundance change of 0.3 dex (or some combination of simultaneous changes in distance and abundance) would put them into better accord with the general trend, but this does not seem very realistic given the expected uncertainties. The range in $R_G$ for these stars is about 2 kpc (all Cepheids exterior to Rg = 14.0 kpc) but the physical separation spans about 15 kpc (see Figure 3). A typical range in abundance for a span of 2 kpc in galactocentric radius (in Zone II especially) is approximately 0.15 dex. Could these stars be indicating that the outer regions of the Galaxy show much more significant variations in abundance than do the regions closer to the solar circle? This would not be especially surprising given the overall lower gas densities and star formation rates which could be significantly influenced by local events. This also means that gradient values can be influenced in the outer region quite strongly by candidate selection. Probing to larger distances especially in the direction of l = 135° is needed to test the consistency of abundance in the outer Galaxy.

The primary aim of the new abundance work presented here is to extend the azimuthal range of our Cepheid studies at the distance of the Sun from the galactic center. In Figure 2 we show two cuts of the data both being an annulus of 1 kpc width, the first centered at a galactocentric radius of 6.9 kpc, and the second at 7.9 kpc. The latter is our assumed solar distance from the galactic center (McNamara et al. 2000). The inner sample spans a total angle of about 60 degrees (as seen from the galactic center) while the outer annulus spans a slightly larger total angle.

What is obvious from Figure 2 is that there is no believable evidence in the data for any variation in abundance as a function of azimuthal angle. A least squares fit to the data in the bottom panel of Figure 2 yields an angular gradient -4.56E-4 dex/degree with an uncertainty of ±5.6E-4. To put this into more familiar terms, note that the arc spans about 12 kpc. The abundance difference is ≈0.038 over the span which yields a gradient of d[Fe/H]/dA of -0.003 dex kpc$^{-1}$ (A being the distance along the arc). Considering the confidence interval on the solution, the data is entirely consistent with no azimuthal variation. The result of Luck et al. (2006) says variations may exist at larger radius, but this result says that there are no systematic variations in the 80 degree region centered on the solar direction at the solar galactocentric distance and somewhat inwards.

In Figure 3 we show the spatial distribution of our total sample relative to the galactic center. As expected, the density of observed stars decreases with galactocentric distance and distance from the Sun. The reason is twofold: one, stars far from the solar region require the investment of significant observing time; and two, as one approaches the edge of the Galaxy ($R_G$ ≈ 16 kpc or d ≈ 8 kpc from the Sun towards *l* = 180°) the stellar density decreases. The galactic stellar scale length is of order 3 kpc for the Milky Way (Zheng et al. 2001, López-Corredoira et al. 2002) which implies a stellar disk of about 12 kpc based on external spirals (de Grijs, Kregel & Wesson 2001). López-Corredoira et al. however find no indication of a cut-off in the galactic stellar disk at radii $R_G < 15$ kpc. Additionally, there is evidence of spiral structure from molecular cloud and H I mapping at $R_G$ from 15 to 24 kpc (McClure-Griffiths et al. 2004, Nakagawa et al. 2005, Strasser et al. 2007, Dickey et al. 2009). Thus, it is possible that the Cepheids thus far sampled do not approach the full radius of the Milky Way. The addition of the Carina Cepheids fills out a previously sparsely occupied zone making the region within 3 kpc of the Sun well sampled.

With the addition of the Carina stars our Cepheid sample has grown to 270 individual stars. While Cepheids are primarily located near the galactic plane, our sample is now large enough to investigate the Z (distance from the plane) distribution in the stars themselves. Figure 4 shows the Z distance as a function of galactocentric distance. A number of things are readily apparent: 1) there are a number of Cepheids at large distance from the galactic plane, 2) there is a tendency for those that deviate the most to lie at larger galactocentric radii, and 3) the mean Z height of the Cepheids is −30 to −40 pc which reflects the solar position above the galactic plane. One wonders if the Cepheids lying away from the plane exert an undue influence on the abundance gradient. If we eliminate all stars with $|Z| > 300$ pc the value of the gradient changes very little: the slope in [Fe/H] decreases from −0.055 to −0.053. We conclude that they do not influence the gradient, but that they do warrant further consideration especially in terms of kinematics. The question is, if they are truly young massive stars, how did they get to distances approaching a kiloparsec from the plane?

As a last demonstration concerning the behavior of iron in the local region of the galaxy we show in Figure 5 a contour map of position (X,Y relative to the Sun) versus [Fe/H]. What is most obvious about the behavior of iron is the simple linear behavior independent of azimuth. The "metallicity island" in the Cygnus region is present as expected (this is essentially the same data as Luck et al. (2006) in that region). The reality of that region needs further exploration to increase the sample size and to check abundance consistency. Otherwise, what is seen is a monotonic gradient with an increase in slope starting at about $R_G$ about 6.6 kpc and proceeding inwards. The outer slope appears overall to be well behaved, but it is possible, if not probable, that the outer disk shows a larger dispersion in abundance than does the region at 12 kpc and inwards.

### 4.2.2. Light Element Gradients

While the primary result of these studies is the [Fe/H] gradient, we have information on other elements. The light elements carbon, nitrogen, and oxygen are especially important as they are abundant species and the product of primary energy generation nucleosynthesis. Additionally, they are the species that are most often investigated in gradient studies involving B stars or H II regions. The difficulty with these species in Cepheids is that they are affected by nucleosynthesis and subsequent mixing events during the first dredge-up which alter the surface composition of carbon and nitrogen. While standard evolution leaves the surface oxygen content unaltered, abundance studies starting with Luck (1978) have shown that oxygen in intermediate-mass stars is subsolar (Luck & Lambert 1981, 1985, 1992, Luck 1994, Luck et al. 2006). While the current best solar oxygen value ($\log \varepsilon_O \approx 8.75$ − Asplund et al. 2009) alleviates this discrepancy somewhat, the problem still persists. Note that the newest solar results for oxygen discussed in Asplund et al. (2009) reflect a near convergence of 3D models and modern 1D models for the oxygen content. Another element that is affected by nucleosynthesis and mixing is Na. In this case the agent is Ne-Na cycle which brings sodium enriched material to the stellar surface (Sasselov 1986; Luck 1994; Denissenkov 1993a, 1993b, 1994). Before considering gradients in these species derived from Cepheids, we must try to ascertain if evolutionary processes have left differential effects in these stars that would vitiate any gradient result. To examine this question we look at our abundance data for these species as a function of period.

In Figure 6 we show [Fe/H], [C/Fe], [O/Fe], and [Na/Fe] as a function of log (Period). The Cepheids span a period range of about 2 to 50 days and thus have a significant mass range. We show [Fe/H] to emphasize that there is no meaningful dependence on period in a species not affected by the first dredge-up (if there were, it could mean there is a problem in gravity determinations). The remaining species are normalized to iron to remove the overall metallicity from affecting the result. There is scant evidence in Figure 6 for there being period dependent mixing affecting these stars.

Perhaps the most striking aspect of Figure 6 is that a few of these Cepheids are extremely carbon deficient; and thus, may be related to the weak G-band stars. The difficulty is that weak G-band stars are typically G/K giants of $1 - 1.5$ $M_\odot$ and show strong Li features (Sneden et al 1978, Lambert & Sawyer 1984, Gratton 1991). These Cepheids are more massive and show no Li, but do show evidence for deep-mixing in their extreme carbon deficiencies. This phenomenon needs further explication as do the weak G-band stars.

To return to the oxygen problem briefly, we note that in Figure 6 the [O/Fe] ratios are generally subsolar. In fact, the mean [O/Fe] value for the total sample is −0.09. Because these oxygen abundances are based on solar gf values, the lower oxygen abundance of Asplund et al. (2009) will not resolve the problem. This is because the lower solar oxygen abundance in an inverted solar analysis will yield a correspondingly high gf value; and hence, the same differential abundance. What is needed to resolve this difficulty is a systematic re-examination of Cepheid oxygen abundances using the same gf values employed to determine the solar abundance.

In Figures 7 and 8 (and Table 6) we present gradient information for carbon and oxygen. The carbon gradient (Figure 7) is well defined presenting a gradient d[C/H]/dR = −0.077 dex/kpc. This is somewhat steeper than found for iron. There is a gradient determination for carbon from B stars by Daflon & Cunha (2004). They find d[C/H]dR = −0.037 dex/kpc which is significantly lower than the value determined here. Perhaps evolution in Cepheids is altering the carbon abundance in a non-homogeneous manner. A common feature of gradient determinations is a flattening of the gradient at larger galactocentric radii (Maciel & Costa 2010). It may be that [C/H] shows such a behavior. Most gradients show the flattening beginning at about 10 kpc, but the [C/H] data indicates the flattening begins at about 13 kpc. We caution against over-interpretation of this result as it depends on a very small number of stars. In lower panel of Figure 7 we show [C/Fe] against galactocentric distance. The slope for this data is d[C/Fe] = −0.023 dex/kpc reflecting the difference in the gradients in the two species. If one looks at the confidence interval on the gradient, one notes that d[C/Fe]/dR = 0 is not excluded.

Studies of H II regions, B stars, and planetary nebula most often quote gradients in terms of d[O/H]/dR$_G$. Maciel & Costa (2010) in their review of gradients note that H II regions yield gradients in the range of −0.044 dex/kpc (Esteban et al. 2005) to −0.060 dex/kpc (Rudolph et al. 2006). The more common determination for the oxygen gradient from H II regions is about −0.04 dex/kpc. Our information for [O/H] is presented in Figure 8 (and Table 6). It is obvious from Figure 8 that the scatter in oxygen is much larger than for most other species. The bulk of the determinations here are from the blended O I lines near 615 nm. What is needed to improve the data quality is an analysis of the [O I] line at 630 nm and/or a non-LTE study of the O I triplet at 775 nm. The gradient d[O/H]/dR found here is −0.037 dex/kpc which is in good accord with the lower, more prevalent, determinations from H II regions. Since Cepheids are young stars freshly minted from H II regions and GMCs, it is not surprising they give the same gradient.

B stars are the direct predecessors of Cepheids and one expects that gradients determined from them should be very similar to those found in Cepheids. Daflon & Cunha (2004) have examined a number of OB associations and derive gradients for a mix of elements. For oxygen they derive d[O/H]/dR = −0.031 dex/kpc. This is in excellent agreement with our value.

Gradients from studies of planetary nebulae suffer from a lack of homogeneity in age. In fact, the ages of planetary nebulae range from about 1 Gyr to 8 Gyr (Maciel et al. 2009). This means that gradients derived from planetary nebula are not strictly comparable to those determined from a young object such as a Cepheid. Nonetheless, the gradient in $d[O/H]/dR_G$ for planetary nebulae is about −0.04 to −0.09 dex/kpc (Maciel & Costa 2010). They also note that a number of studies indicate very flat to non-existent gradients. The latter result brings to mind the effects of galactic "churning" (Sellwood & Binney 2002, Schönrich & Binney 2008, Casagrande et al. 2011) which erases gradients over time due to orbit migration.

### 4.2.3. Alpha and Heavy Element Gradients

Figures 9 shows an array of radial gradients. Si is a typical α species while Y, Nd, and Eu are s/r process elements. These figures are very similar to our previous work. What is different is that now a gradient in [Y/H] is certain and there may well be one in [Eu/H]. Only [Nd/H] definitely does not show a gradient but one must wonder if the large scatter is hiding the actual slope. Of interest is the Si gradient of Daflon & Cunha (2004) derived from O/B stars. They find d[Si/H]/dR = −0.040 dex/kpc which compares very well with our value of −0.049 dex/kpc. Daflon & Cunha also derive gradients for Mg, Al and S of −0.052, −0.048 and −0.040 dex/kpc respectively while we find for the same elements gradients of −0.048, −0.053, and −0.076. The only discrepant value here is for sulfur. Neutral sulfur lines in moderate temperature Cepheids are weak and give rather uncertain abundances while the ionized sulfur lines in the spectra of B stars can be quite strong and hence, also uncertain.

## 5. Summary

With the newly analyzed galactic Cepheids (101 new stars) we have enlarged our homogeneous sample of stars used for galactic abundance gradient investigation to 270 objects. The primary results obtained in present study in conjunction with those reported in our previous work are:

- The abundance distribution in galactic disk shows what is on the whole a linear structure. The galactic abundance gradient in the present epoch is approximately $d[Fe/H]/dR_G = -0.06$ dex / kpc.
- There is little or no evidence in our data (especially [Fe/H]) for a flattening of the gradient beyond 10 kpc from the galactic center. However, that region shows an increased dispersion in abundance which given inadequate sampling could lead to an under or over estimate of the gradient. We do not claim that we have definitive evidence one way or the other, only that our results are more consistent with a large variation in abundance.
- We have found no evidence for azimuthal variations in the [Fe/H] ratios at the solar galactocentric radius in an arc of 80° relative to the galactic center.

What is next in the study of the distribution of the elements in the Galaxy? As alluded to several times previously, the reality of the "metallicity island" rests upon a small number of stars. Obviously needed is a re-examination of that region utilizing better data and more stars.

Acknowledgements: We thank J.R.D. Lépine for comments on this work. WG and DG gratefully acknowledge financial support for this work from the Chilean Center for Astrophysics FONDAP 15010003, and from the BASAL Centro de Astrofisica y Tecnologias Afines CATA under PFB-06/2007. We are grateful to the ESO staff at La Silla whose expert support is much appreciated.

| Name | Type | P (days) | $M_V$ | <V> | E(B-V) | D (kpc) | Rg (kpc) | UT Date | Phase | Exp (s) | S/N |
|---:|---|---:|---:|---:|---:|---:|---:|---|---:|---:|---:|
| T Ant | DCEP: | 5.898 | -3.34 | 9.34 | 0.300 | 2.195 | 8.384 | 2010-03-29 | 0.173 | 990 | 279 |
| l Car | DCEP | 35.557 | -5.43 | 3.72 | 0.147 | 0.544 | 7.794 | 2010-03-25 | 0.094 | 15 | 429 |
| U Car | DCEP | 38.801 | -5.53 | 6.29 | 0.265 | 1.557 | 7.537 | 2010-03-25 | 0.968 | 60 | 187 |
| V Car | DCEP | 6.697 | -3.49 | 7.36 | 0.169 | 1.150 | 7.877 | 2010-03-31 | 0.889 | 165 | 246 |
| SX Car | DCEP | 4.860 | -3.11 | 9.09 | 0.318 | 1.718 | 7.586 | 2010-03-31 | 0.943 | 900 | 288 |
| UW Car | DCEP | 5.346 | -3.22 | 9.43 | 0.435 | 1.774 | 7.617 | 2010-03-25 | 0.035 | 885 | 231 |
| UX Car | DCEP | 3.682 | -2.79 | 8.31 | 0.112 | 1.404 | 7.663 | 2010-04-01 | 0.902 | 450 | 213 |
| UY Car | DCEP | 5.544 | -3.27 | 8.97 | 0.180 | 2.141 | 7.548 | 2010-03-30 | 0.946 | 720 | 230 |
| UZ Car | DCEP | 5.205 | -3.19 | 9.32 | 0.178 | 2.445 | 7.544 | 2010-03-31 | 0.044 | 975 | 264 |
| VY Car | DCEP | 18.937 | -4.70 | 7.44 | 0.237 | 1.882 | 7.582 | 2010-03-31 | 0.169 | 270 | 292 |
| WW Car | DCEP | 4.677 | -3.07 | 9.74 | 0.379 | 2.078 | 7.515 | 2010-03-27 | 0.852 | 1200 | 218 |
| WZ Car | DCEP | 23.013 | -4.92 | 9.25 | 0.370 | 3.934 | 7.573 | 2010-03-26 | 0.009 | 915 | 231 |
| XX Car | DCEP | 15.716 | -4.48 | 9.32 | 0.347 | 3.436 | 7.382 | 2010-03-25 | 0.692 | 810 | 215 |
| XY Car | DCEP | 12.435 | -4.21 | 9.30 | 0.411 | 2.721 | 7.356 | 2010-03-25 | 0.242 | 780 | 199 |
| XZ Car | DCEP | 16.650 | -4.55 | 8.60 | 0.365 | 2.475 | 7.414 | 2010-04-01 | 0.072 | 560 | 327 |
| YZ Car | DCEP | 18.166 | -4.65 | 8.71 | 0.381 | 2.668 | 7.629 | 2010-03-31 | 0.860 | 555 | 230 |
| AQ Car | DCEP | 9.769 | -3.93 | 8.85 | 0.165 | 2.810 | 7.631 | 2010-03-31 | 0.998 | 630 | 273 |
| CN Car | DCEP | 4.933 | -3.13 | 10.70 | 0.399 | 3.225 | 7.802 | 2010-03-31 | 0.066 | 2840 | 276 |
| CY Car | DCEP | 4.266 | -2.96 | 9.78 | 0.370 | 2.040 | 7.471 | 2010-03-30 | 0.095 | 1720 | 228 |
| DY Car | DCEP | 4.675 | -3.07 | 11.31 | 0.372 | 4.326 | 7.686 | 2010-04-01 | 0.956 | 4020 | 199 |
| ER Car | DCEP | 7.719 | -3.65 | 6.82 | 0.096 | 1.079 | 7.597 | 2010-03-26 | 0.872 | 105 | 258 |
| FI Car | DCEP | 13.458 | -4.30 | 11.61 | 0.691 | 5.435 | 8.107 | 2010-03-28 | 0.215 | 5535 | 196 |
| FR Car | DCEP | 10.717 | -4.03 | 9.66 | 0.334 | 3.335 | 7.387 | 2010-03-25 | 0.011 | 1110 | 227 |
| GH Car | DCEPS | 5.726 | -3.70 | 9.18 | 0.394 | 2.096 | 7.414 | 2010-03-25 | 0.965 | 705 | 232 |
| GX Car | DCEP | 7.197 | -3.57 | 9.36 | 0.386 | 2.177 | 7.762 | 2010-03-30 | 0.996 | 990 | 255 |
| HW Car | DCEP | 9.199 | -3.86 | 9.16 | 0.184 | 3.056 | 7.556 | 2010-03-26 | 0.947 | 810 | 255 |
| IO Car | CEP | 13.597 | -4.31 | 11.10 | 0.502 | 5.730 | 8.078 | 2010-03-27 | 0.072 | 3300 | 212 |
| IT Car | DCEP | 7.533 | -3.62 | 8.10 | 0.184 | 1.680 | 7.451 | 2010-03-25 | 0.950 | 315 | 218 |
| V397 Car | DCEPS | 2.063 | -2.52 | 8.32 | 0.266 | 0.990 | 7.674 | 2010-03-25 | 0.903 | 330 | 214 |
| V Cen | DCEP | 5.494 | -3.26 | 6.84 | 0.292 | 0.676 | 7.426 | 2010-03-28 | 0.781 | 60 | 246 |
| V Cen | DCEP | 5.494 | -3.26 | 6.84 | 0.292 | 0.676 | 7.426 | 2010-03-29 | 0.952 | 180 | 304 |
| QY Cen | DCEP | 17.752 | -4.62 | 11.76 | 1.447 | 2.196 | 6.638 | 2010-03-27 | 0.004 | 6150 | 282 |
| XX Cen | DCEP | 10.953 | -4.06 | 7.82 | 0.266 | 1.598 | 6.997 | 2010-03-30 | 0.116 | 350 | 219 |
| AY Cen | DCEP | 5.310 | -3.22 | 8.83 | 0.295 | 1.655 | 7.424 | 2010-03-30 | 0.999 | 800 | 248 |
| AZ Cen | DCEPS | 3.212 | -3.03 | 8.64 | 0.168 | 1.678 | 7.413 | 2010-03-27 | 0.947 | 510 | 239 |
| BB Cen | DCEPS | 3.998 | -3.28 | 10.07 | 0.377 | 2.680 | 7.126 | 2010-03-31 | 0.183 | 1750 | 229 |
| KK Cen | DCEP | 12.180 | -4.18 | 11.48 | 0.611 | 5.466 | 7.542 | 2010-03-25 | 0.012 | 4620 | 197 |
| KN Cen | DCEP | 34.024 | -5.38 | 9.87 | 0.797 | 3.424 | 6.404 | 2010-03-25 | 0.488 | 1230 | 236 |
| MZ Cen | CEP | 10.353 | -3.99 | 11.53 | 0.869 | 3.494 | 6.531 | 2010-03-26 | 0.207 | 5520 | 242 |
| V339 Cen | DCEP | 9.466 | -3.89 | 8.75 | 0.413 | 1.827 | 6.774 | 2010-03-25 | 0.142 | 555 | 288 |
| V378 Cen | DCEPS | 6.460 | -3.84 | 8.46 | 0.376 | 1.651 | 7.054 | 2010-03-30 | 0.131 | 660 | 235 |
| V381 Cen | CEP | 5.079 | -3.17 | 7.65 | 0.195 | 1.090 | 7.236 | 2010-04-01 | 0.896 | 230 | 266 |
| V419 Cen | DCEPS | 5.507 | -3.66 | 8.19 | 0.168 | 1.822 | 7.411 | 2010-04-01 | 0.905 | 400 | 275 |

**Table 1**
Program Stars

| Name | Type | P (days) | $M_V$ | <V> | E(B-V) | D (kpc) | Rg (kpc) | UT Date | Phase | Exp (s) | S/N |
|---|---|---|---|---|---|---|---|---|---|---|---|
| V496 Cen | DCEP | 4.424 | -3.00 | 9.97 | 0.541 | 1.757 | 7.059 | 2010-03-29 | 0.025 | 1800 | 288 |
| V659 Cen | DCEP | 5.623 | -3.28 | 6.60 | 0.128 | 0.783 | 7.447 | 2010-04-01 | 0.856 | 105 | 274 |
| V737 Cen | DCEP | 7.066 | -3.55 | 6.72 | 0.206 | 0.833 | 7.335 | 2010-04-01 | 0.967 | 110 | 297 |
| AV Cir | DCEP | 3.065 | -2.58 | 7.44 | 0.378 | 0.574 | 7.516 | 2010-03-30 | 0.207 | 235 | 218 |
| AX Cir | DCEP | 5.273 | -3.21 | 5.88 | 0.146 | 0.529 | 7.530 | 2010-03-28 | 0.745 | 60 | 258 |
| AX Cir | DCEP | 5.273 | -3.21 | 5.88 | 0.146 | 0.529 | 7.530 | 2010-03-30 | 0.119 | 90 | 297 |
| BP Cir | CEP | 2.398 | -2.29 | 7.56 | 0.224 | 0.670 | 7.431 | 2010-04-01 | 0.033 | 215 | 283 |
| R Cru | DCEP | 5.826 | -3.32 | 6.77 | 0.183 | 0.794 | 7.539 | 2010-03-28 | 0.100 | 120 | 319 |
| S Cru | DCEP | 4.690 | -3.07 | 6.60 | 0.166 | 0.672 | 7.553 | 2010-03-29 | 0.983 | 140 | 347 |
| T Cru | DCEP | 6.733 | -3.49 | 6.57 | 0.184 | 0.782 | 7.546 | 2010-03-30 | 0.958 | 125 | 268 |
| X Cru | DCEP | 6.220 | -3.40 | 8.40 | 0.272 | 1.531 | 7.201 | 2010-03-30 | 0.125 | 520 | 196 |
| VW Cru | DCEP | 5.265 | -3.21 | 9.62 | 0.643 | 1.415 | 7.275 | 2010-03-29 | 0.175 | 1260 | 272 |
| AD Cru | DCEP | 6.398 | -3.43 | 11.05 | 0.647 | 3.011 | 6.987 | 2010-03-27 | 0.030 | 3240 | 237 |
| AG Cru | DCEP | 3.837 | -2.84 | 8.23 | 0.212 | 1.191 | 7.346 | 2010-03-29 | 0.039 | 360 | 288 |
| BG Cru | DCEPS | 3.343 | -3.08 | 5.49 | 0.132 | 0.424 | 7.694 | 2010-03-30 | 0.150 | 135 | 496 |
| GH Lup | CEP | 9.277 | -3.87 | 7.64 | 0.335 | 1.213 | 6.944 | 2010-03-28 | 0.142 | 240 | 313 |
| R Mus | DCEP | 7.510 | -3.62 | 6.30 | 0.114 | 0.812 | 7.502 | 2010-03-29 | 0.184 | 110 | 338 |
| S Mus | DCEP | 9.660 | -3.91 | 6.12 | 0.212 | 0.740 | 7.564 | 2010-03-29 | 0.106 | 135 | 412 |
| RT Mus | DCEP | 3.086 | -2.59 | 9.02 | 0.344 | 1.257 | 7.426 | 2010-03-31 | 0.204 | 800 | 266 |
| TZ Mus | DCEP | 4.945 | -3.13 | 11.70 | 0.664 | 3.456 | 7.063 | 2010-03-28 | 0.024 | 5730 | 223 |
| UU Mus | DCEP | 11.636 | -4.13 | 9.78 | 0.399 | 3.344 | 7.054 | 2010-03-25 | 0.041 | 1230 | 257 |
| S Nor | DCEP | 9.754 | -3.92 | 6.39 | 0.179 | 0.887 | 7.169 | 2010-03-30 | 0.124 | 125 | 281 |
| U Nor | DCEP | 12.644 | -4.23 | 9.24 | 0.862 | 1.367 | 6.815 | 2010-04-01 | 0.476 | 900 | 270 |
| SY Nor | DCEP | 12.646 | -4.23 | 9.51 | 0.756 | 1.818 | 6.442 | 2010-03-26 | 0.164 | 1140 | 279 |
| TW Nor | DCEP | 10.786 | -4.04 | 11.70 | 1.157 | 2.521 | 5.843 | 2010-04-01 | 0.094 | 5580 | 271 |
| GU Nor | DCEP | 3.453 | -2.72 | 10.41 | 0.651 | 1.603 | 6.553 | 2010-03-27 | 0.145 | 2105 | 270 |
| V340 Nor | CEP | 11.288 | -4.09 | 8.38 | 0.321 | 1.934 | 6.306 | 2010-03-25 | 0.430 | 405 | 265 |
| BF Oph | DCEP | 4.068 | -2.91 | 7.34 | 0.235 | 0.789 | 7.121 | 2010-04-01 | 0.033 | 180 | 250 |
| AP Pup | DCEP | 5.084 | -3.17 | 7.37 | 0.198 | 0.954 | 8.189 | 2010-03-31 | 0.052 | 165 | 281 |
| AT Pup | DCEP | 6.665 | -3.48 | 7.96 | 0.191 | 1.460 | 8.412 | 2010-03-29 | 0.658 | 285 | 286 |
| CE Pup | DCEP | 49.530 | -5.81 | 11.80 | 0.740 | 11.085 | 14.739 | 2010-03-31 | 0.804 | 6300 | 248 |
| MY Pup | DCEPS | 5.694 | -3.70 | 5.68 | 0.061 | 0.684 | 8.028 | 2010-03-25 | 0.114 | 45 | 272 |
| NT Pup | DCEP | 15.565 | -4.47 | 12.14 | 0.670 | 7.739 | 12.416 | 2010-03-26 | 0.219 | 8310 | 223 |
| RV Sco | DCEP | 15.560 | -4.47 | 7.04 | 0.349 | 1.191 | 6.734 | 2010-03-29 | 0.914 | 230 | 299 |
| V482 Sco | DCEP | 4.528 | -3.03 | 7.97 | 0.336 | 0.960 | 6.945 | 2010-03-29 | 0.113 | 280 | 299 |
| V636 Sco | DCEP | 6.797 | -3.50 | 6.65 | 0.207 | 0.791 | 7.148 | 2010-04-01 | 0.969 | 125 | 247 |
| V950 Sco | DCEPS | 3.380 | -3.09 | 7.30 | 0.254 | 0.821 | 7.099 | 2010-03-30 | 0.126 | 180 | 274 |
| R TrA | DCEP | 3.389 | -2.69 | 6.66 | 0.142 | 0.601 | 7.475 | 2010-03-28 | 0.167 | 105 | 297 |
| S TrA | DCEP | 6.323 | -3.42 | 6.40 | 0.084 | 0.811 | 7.283 | 2010-03-30 | 0.176 | 140 | 307 |
| LR TrA | DCEPS | 2.428 | -2.71 | 7.81 | 0.268 | 0.851 | 7.291 | 2010-03-28 | 0.086 | 270 | 274 |
| T Vel | DCEP | 4.640 | -3.06 | 8.02 | 0.289 | 1.072 | 8.054 | 2010-03-25 | 0.827 | 300 | 238 |
| V Vel | DCEP | 4.371 | -2.99 | 7.59 | 0.186 | 0.990 | 7.849 | 2010-04-01 | 0.043 | 200 | 242 |
| RY Vel | DCEP | 28.131 | -5.16 | 8.40 | 0.547 | 2.276 | 7.731 | 2010-03-25 | 0.412 | 345 | 246 |

**Table 1**
Program Stars

| Name | Type | P (days) | $M_V$ | <V> | E(B-V) | D (kpc) | Rg (kpc) | UT Date | Phase | Exp (s) | S/N |
|---|---|---|---|---|---|---|---|---|---|---|---|
| RZ Vel | DCEP | 20.399 | -4.78 | 7.08 | 0.299 | 1.510 | 8.225 | 2010-03-25 | 0.808 | 135 | 205 |
| ST Vel | DCEP | 5.858 | -3.33 | 9.70 | 0.479 | 1.985 | 8.183 | 2010-03-29 | 0.978 | 1365 | 317 |
| SV Vel | DCEP | 14.097 | -4.35 | 8.52 | 0.373 | 2.159 | 7.594 | 2010-03-26 | 0.939 | 495 | 271 |
| SW Vel | DCEP | 23.440 | -4.94 | 8.12 | 0.344 | 2.458 | 8.427 | 2010-03-30 | 0.607 | 345 | 338 |
| SX Vel | DCEP | 9.550 | -3.90 | 8.25 | 0.263 | 1.821 | 8.245 | 2010-03-25 | 0.852 | 375 | 253 |
| XX Vel | DCEP | 6.985 | -3.54 | 10.65 | 0.545 | 3.063 | 7.708 | 2010-03-28 | 0.011 | 2700 | 269 |
| AE Vel | DCEP | 7.134 | -3.56 | 10.26 | 0.635 | 2.261 | 7.983 | 2010-04-01 | 0.732 | 1500 | 150 |
| AH Vel | DCEPS | 4.227 | -3.35 | 5.70 | 0.070 | 0.580 | 7.996 | 2010-03-30 | 0.383 | 180 | 519 |
| BG Vel | DCEP | 6.924 | -3.53 | 7.64 | 0.426 | 0.905 | 7.922 | 2010-04-01 | 0.849 | 210 | 244 |
| CS Vel | DCEP | 5.905 | -3.34 | 11.68 | 0.737 | 3.374 | 8.198 | 2010-03-30 | 0.678 | 7500 | 199 |
| CX Vel | DCEP | 6.256 | -3.41 | 11.37 | 0.723 | 3.087 | 8.360 | 2010-03-29 | 0.003 | 4335 | 258 |
| DK Vel | DCEP | 2.482 | -2.33 | 10.61 | 0.287 | 2.536 | 8.126 | 2010-03-25 | 0.009 | 2295 | 194 |
| DR Vel | DCEP | 11.199 | -4.08 | 9.52 | 0.656 | 1.982 | 8.036 | 2010-03-26 | 0.165 | 1170 | 304 |
| EX Vel | DCEP | 13.234 | -4.28 | 11.56 | 0.775 | 4.651 | 8.872 | 2010-04-01 | 0.954 | 5145 | 250 |
| FG Vel | DCEP | 6.453 | -3.44 | 11.81 | 0.810 | 3.374 | 8.286 | 2010-03-27 | 0.756 | 6465 | 187 |
| FN Vel | CEP | 5.324 | -3.22 | 10.29 | 0.588 | 2.102 | 7.850 | 2010-03-28 | 0.221 | 2010 | 251 |

Notes:

| | |
|---|---|
| Type: | Variability type from the General Catalog of Variable Stars (Samus , N. N. et al. 2007-2009) |
| P: | Period in days from Berdnikov (2006– private communication to S. Andrievsky) or GCVS. |
| $M_V$: | Absolute V magnitude computed assuming listed period is the fundamental except for DCEPS where $P_0$ = P/0.71. Period-Luminosity relation and R (=Av/E(B-V) = 3.23) from Fouqué et al. (2007). |
| <V>: | Intensity mean apparent V magnitude from Fernie (1995). Exception is NT Pup which is from Yong et al. (2006) |
| E(B-V): | From Fouqué et al. (2007) or Fernie (1995) as modified by Fouqué et al. Reddening for CE Pup, HQ Car, and NT Pup from Yong et al. (2006). FI Car reddening from van Leeuven et al. (2007) and T Ant reddening from Turner & Berdnikov (2003). |
| d and Rg: | Distance in kiloparsecs from the Sun and the galactic center respectively. The Sun is assumed to be 7.9 kpc from the galactic center (McNamara et al. 2000). |
| UT Date: | UT date of observation. |
| Phase: | Phase at time of observation. |
| Exp: | Integration time in seconds. |
| S/N: | Signal to noise (per pixel) near 500 nm at order center. |

**Table 2**
Stellar Parameter and Iron Abundance Details

| Star | T (Kelvin) | σ | N | log g | $V_t$ (km/s) | Fe I ([Fe/H]) | σ | N | Fe II ([Fe/H]) | σ | N |
|---|---|---|---|---|---|---|---|---|---|---|---|
| T Ant | 6265 | 130 | 54 | 2.1 | 4.4 | -0.24 | 0.09 | 153 | -0.24 | 0.08 | 26 |
| l Car | 5298 | 102 | 53 | 1.1 | 4.5 | 0.05 | 0.10 | 275 | 0.04 | 0.08 | 22 |
| U Car | 4934 | 118 | 25 | 0.8 | 5.8 | 0.01 | 0.07 | 53 | 0.03 | 0.09 | 2 |
| V Car | 5906 | 61 | 32 | 2.1 | 5.2 | 0.01 | 0.07 | 127 | 0.02 | 0.04 | 14 |
| SX Car | 6513 | 45 | 32 | 2.1 | 4.5 | -0.09 | 0.07 | 121 | -0.08 | 0.06 | 21 |
| UW Car | 6618 | 76 | 39 | 2.1 | 4.3 | -0.06 | 0.09 | 150 | -0.05 | 0.05 | 22 |
| UX Car | 6442 | 54 | 39 | 2.6 | 4.7 | 0.02 | 0.08 | 267 | 0.04 | 0.05 | 37 |
| UY Car | 6376 | 73 | 68 | 2.4 | 5.7 | 0.03 | 0.08 | 193 | 0.05 | 0.05 | 27 |
| UZ Car | 6079 | 91 | 40 | 2.0 | 4.3 | 0.07 | 0.07 | 210 | 0.08 | 0.05 | 29 |
| VY Car | 4895 | 60 | 12 | 1.6 | 5.2 | 0.26 | 0.09 | 47 | 0.26 | 0.04 | 3 |
| WW Car | 5858 | 122 | 51 | 2.1 | 5.3 | -0.07 | 0.08 | 206 | -0.05 | 0.08 | 30 |
| WZ Car | 5770 | 100 | 42 | 1.8 | 6.8 | 0.03 | 0.10 | 166 | 0.06 | 0.10 | 23 |
| XX Car | 5942 | 95 | 50 | 1.6 | 4.5 | 0.11 | 0.08 | 245 | 0.13 | 0.05 | 30 |
| XY Car | 5738 | 87 | 64 | 1.7 | 4.6 | 0.04 | 0.09 | 230 | 0.06 | 0.08 | 32 |
| XZ Car | 6170 | 37 | 36 | 2.1 | 5.7 | 0.14 | 0.07 | 128 | 0.16 | 0.04 | 14 |
| YZ Car | 5655 | 69 | 49 | 1.6 | 6.5 | 0.02 | 0.08 | 155 | 0.00 | 0.05 | 12 |
| AQ Car | 5815 | 45 | 61 | 1.9 | 4.9 | 0.06 | 0.10 | 197 | 0.05 | 0.04 | 21 |
| CN Car | 6331 | 49 | 51 | 2.2 | 4.3 | 0.06 | 0.08 | 169 | 0.07 | 0.04 | 23 |
| CY Car | 6042 | 50 | 49 | 2.2 | 4.0 | 0.10 | 0.08 | 192 | 0.12 | 0.06 | 25 |
| DY Car | 6533 | 71 | 40 | 2.3 | 5.3 | -0.07 | 0.09 | 211 | -0.06 | 0.09 | 43 |
| ER Car | 5874 | 82 | 52 | 2.1 | 6.0 | 0.03 | 0.09 | 199 | 0.05 | 0.05 | 19 |
| FI Car | 5215 | 68 | 55 | 1.2 | 3.8 | 0.06 | 0.08 | 162 | 0.04 | 0.12 | 23 |
| FR Car | 5905 | 79 | 60 | 2.0 | 5.0 | 0.02 | 0.10 | 263 | 0.04 | 0.06 | 27 |
| GH Car | 6336 | 73 | 59 | 2.0 | 5.5 | -0.01 | 0.08 | 246 | -0.01 | 0.07 | 37 |
| GX Car | 6297 | 60 | 40 | 2.2 | 5.3 | 0.01 | 0.09 | 245 | 0.03 | 0.07 | 34 |
| HW Car | 5671 | 42 | 71 | 2.0 | 5.2 | 0.04 | 0.11 | 298 | 0.04 | 0.06 | 41 |
| IO Car | 6145 | 63 | 60 | 1.8 | 6.0 | -0.05 | 0.08 | 183 | -0.04 | 0.07 | 30 |
| IT Car | 5794 | 50 | 69 | 1.8 | 4.7 | 0.06 | 0.07 | 241 | 0.07 | 0.06 | 28 |
| V397 Car | 6036 | 102 | 55 | 2.3 | 4.8 | 0.03 | 0.09 | 293 | 0.04 | 0.08 | 39 |
| V Cen #1 | 5616 | 89 | 41 | 1.8 | 4.8 | 0.02 | 0.07 | 125 | 0.01 | 0.07 | 16 |
| V Cen #2 | 6427 | 90 | 45 | 2.3 | 4.6 | -0.07 | 0.08 | 201 | -0.08 | 0.06 | 41 |
| QY Cen | 6205 | 74 | 48 | 2.0 | 6.3 | 0.16 | 0.09 | 145 | 0.14 | 0.05 | 15 |
| XX Cen | 5953 | 69 | 49 | 2.0 | 5.0 | 0.16 | 0.09 | 158 | 0.17 | 0.05 | 18 |
| AY Cen | 5933 | 50 | 54 | 2.1 | 4.5 | 0.01 | 0.07 | 307 | 0.03 | 0.05 | 36 |
| AZ Cen | 6425 | 66 | 52 | 2.3 | 4.0 | -0.05 | 0.08 | 305 | -0.03 | 0.10 | 48 |
| BB Cen | 6202 | 63 | 53 | 2.3 | 4.6 | 0.13 | 0.10 | 330 | 0.15 | 0.07 | 39 |
| KK Cen | 5894 | 72 | 72 | 1.9 | 5.5 | 0.12 | 0.09 | 304 | 0.12 | 0.07 | 32 |
| KN Cen | 4870 | 137 | 12 | 1.7 | 4.5 | 0.35 | 0.09 | 117 | 0.34 | 0.04 | 3 |
| MZ Cen | 5816 | 89 | 64 | 1.9 | 5.2 | 0.20 | 0.11 | 247 | 0.22 | 0.09 | 33 |
| V339 Cen | 5923 | 53 | 65 | 2.0 | 5.3 | 0.04 | 0.10 | 338 | 0.04 | 0.03 | 28 |
| V378 Cen | 6184 | 49 | 57 | 2.0 | 4.7 | -0.02 | 0.10 | 315 | -0.02 | 0.09 | 47 |
| V381 Cen | 6224 | 97 | 56 | 2.3 | 5.1 | 0.02 | 0.07 | 244 | 0.02 | 0.05 | 33 |
| V419 Cen | 6276 | 74 | 48 | 2.1 | 5.6 | 0.07 | 0.09 | 171 | 0.09 | 0.07 | 27 |

**Table 2**
Stellar Parameter and Iron Abundance Details

| Star | T (Kelvin) | σ | N | log g | $V_t$ (km/s) | Fe I ([Fe/H]) | σ | N | Fe II ([Fe/H]) | σ | N |
|---|---|---|---|---|---|---|---|---|---|---|---|
| V496 Cen | 6194 | 52 | 45 | 2.2 | 3.8 | 0.00 | 0.09 | 261 | 0.02 | 0.06 | 31 |
| V659 Cen | 6067 | 62 | 46 | 2.2 | 5.4 | 0.07 | 0.10 | 155 | 0.09 | 0.05 | 17 |
| V737 Cen | 5865 | 44 | 56 | 2.0 | 4.5 | 0.13 | 0.08 | 176 | 0.14 | 0.03 | 14 |
| AV Cir | 6169 | 56 | 54 | 2.1 | 3.3 | 0.10 | 0.08 | 212 | 0.11 | 0.05 | 34 |
| AX Cir #1 | 5648 | 128 | 58 | 1.9 | 4.5 | -0.07 | 0.08 | 245 | -0.07 | 0.06 | 28 |
| AX Cir #2 | 5761 | 84 | 73 | 1.8 | 3.7 | -0.05 | 0.09 | 351 | -0.06 | 0.07 | 46 |
| BP Cir | 6533 | 73 | 56 | 2.4 | 3.7 | -0.06 | 0.10 | 252 | -0.08 | 0.08 | 38 |
| R Cru | 6203 | 35 | 57 | 2.0 | 4.0 | 0.08 | 0.08 | 245 | 0.08 | 0.04 | 26 |
| S Cru | 6464 | 67 | 42 | 2.1 | 4.1 | -0.12 | 0.08 | 229 | -0.11 | 0.07 | 37 |
| T Cru | 5940 | 54 | 42 | 2.2 | 5.1 | 0.09 | 0.08 | 143 | 0.11 | 0.03 | 12 |
| X Cru | 5948 | 78 | 46 | 2.0 | 4.2 | 0.14 | 0.08 | 149 | 0.14 | 0.05 | 15 |
| VW Cru | 5880 | 48 | 49 | 2.1 | 4.4 | 0.10 | 0.09 | 294 | 0.11 | 0.07 | 36 |
| AD Cru | 6000 | 52 | 60 | 2.0 | 4.0 | 0.06 | 0.08 | 248 | 0.07 | 0.07 | 35 |
| AG Cru | 6628 | 74 | 31 | 2.2 | 4.7 | -0.13 | 0.09 | 239 | -0.12 | 0.08 | 45 |
| BG Cru | 6309 | 69 | 48 | 2.3 | 4.2 | 0.06 | 0.10 | 198 | 0.07 | 0.08 | 36 |
| GH Lup | 5460 | 89 | 61 | 1.6 | 4.6 | 0.08 | 0.09 | 279 | 0.09 | 0.08 | 27 |
| R Mus | 5985 | 54 | 55 | 2.0 | 4.0 | 0.10 | 0.08 | 153 | 0.12 | 0.05 | 21 |
| S Mus | 5752 | 58 | 67 | 1.8 | 4.4 | -0.02 | 0.09 | 300 | -0.02 | 0.08 | 33 |
| RT Mus | 6236 | 61 | 46 | 2.4 | 4.0 | 0.02 | 0.12 | 343 | 0.02 | 0.09 | 46 |
| TZ Mus | 6348 | 61 | 27 | 2.1 | 4.6 | -0.01 | 0.09 | 105 | -0.01 | 0.09 | 20 |
| UU Mus | 6175 | 57 | 42 | 1.8 | 5.4 | 0.05 | 0.07 | 154 | 0.05 | 0.05 | 18 |
| GU Nor | 6036 | 79 | 65 | 2.2 | 3.6 | 0.15 | 0.08 | 366 | 0.16 | 0.07 | 46 |
| S Nor | 5859 | 63 | 69 | 2.0 | 5.0 | 0.07 | 0.09 | 278 | 0.07 | 0.05 | 29 |
| U Nor | 5426 | 75 | 45 | 1.4 | 3.6 | 0.15 | 0.08 | 254 | 0.14 | 0.07 | 25 |
| SY Nor | 5641 | 140 | 43 | 1.8 | 4.7 | 0.31 | 0.11 | 161 | 0.33 | 0.06 | 16 |
| TY Nor | 5979 | 103 | 42 | 2.0 | 5.6 | 0.28 | 0.09 | 111 | 0.27 | 0.06 | 13 |
| V340 Nor | 5733 | 107 | 43 | 2.0 | 5.3 | 0.08 | 0.08 | 117 | 0.09 | 0.03 | 12 |
| BF Oph | 6246 | 54 | 50 | 2.3 | 4.6 | 0.05 | 0.09 | 261 | 0.05 | 0.07 | 43 |
| AP Pup | 6233 | 51 | 49 | 2.1 | 4.2 | 0.06 | 0.08 | 162 | 0.05 | 0.05 | 24 |
| AT Pup | 6455 | 84 | 36 | 2.1 | 5.0 | -0.14 | 0.08 | 211 | -0.13 | 0.06 | 36 |
| CE Pup | 5846 | 92 | 30 | 1.3 | 5.8 | -0.04 | 0.09 | 115 | -0.04 | 0.07 | 14 |
| MY Pup | 6317 | 52 | 44 | 2.2 | 4.8 | -0.14 | 0.10 | 254 | -0.12 | 0.07 | 37 |
| NT Pup | 5561 | 123 | 50 | 1.2 | 4.0 | -0.15 | 0.12 | 276 | -0.15 | 0.08 | 32 |
| RV Sco | 6172 | 74 | 56 | 2.3 | 5.4 | 0.03 | 0.08 | 269 | 0.03 | 0.06 | 38 |
| V482 Sco | 6129 | 46 | 56 | 2.2 | 4.0 | 0.07 | 0.10 | 284 | 0.08 | 0.08 | 35 |
| V636 Sco | 5348 | 45 | 59 | 1.6 | 4.3 | 0.07 | 0.07 | 255 | 0.07 | 0.06 | 27 |
| V950 Sco | 6322 | 45 | 47 | 2.2 | 4.2 | 0.11 | 0.10 | 220 | 0.09 | 0.06 | 27 |
| R TrA | 6121 | 72 | 49 | 2.2 | 3.8 | 0.06 | 0.08 | 157 | 0.07 | 0.05 | 22 |
| S TrA | 5976 | 67 | 63 | 2.1 | 4.2 | 0.12 | 0.09 | 258 | 0.14 | 0.05 | 27 |
| LR TrA | 5943 | 117 | 57 | 2.2 | 4.4 | 0.25 | 0.10 | 214 | 0.25 | 0.05 | 23 |
| T Vel | 5692 | 149 | 61 | 2.0 | 4.8 | -0.02 | 0.09 | 272 | -0.01 | 0.05 | 29 |
| V Vel | 6364 | 98 | 46 | 2.1 | 5.0 | -0.23 | 0.10 | 222 | -0.24 | 0.10 | 42 |
| RY Vel | 5466 | 93 | 59 | 1.3 | 4.6 | 0.10 | 0.10 | 282 | 0.11 | 0.06 | 25 |

**Table 2**
Stellar Parameter and Iron Abundance Details

| Star | T (Kelvin) | σ | N | log g | $V_t$ (km/s) | Fe I ([Fe/H]) | σ | N | Fe II ([Fe/H]) | σ | N |
|---|---|---|---|---|---|---|---|---|---|---|---|
| RZ Vel | 5234 | 95 | 55 | 1.4 | 5.8 | 0.07 | 0.08 | 136 | 0.08 | 0.06 | 11 |
| ST Vel | 6255 | 69 | 46 | 2.1 | 4.3 | 0.00 | 0.09 | 256 | 0.01 | 0.06 | 35 |
| SV Vel | 6064 | 51 | 65 | 2.0 | 6.0 | 0.08 | 0.11 | 261 | 0.07 | 0.08 | 33 |
| SW Vel | 6652 | 95 | 14 | 1.8 | 7.0 | -0.16 | 0.09 | 146 | -0.15 | 0.07 | 25 |
| SX Vel | 6280 | 49 | 59 | 2.0 | 4.5 | 0.00 | 0.10 | 230 | 0.01 | 0.06 | 28 |
| XX Vel | 6521 | 80 | 46 | 2.2 | 4.9 | -0.05 | 0.10 | 222 | -0.05 | 0.07 | 39 |
| AE Vel | 5460 | 87 | 69 | 1.7 | 4.5 | 0.05 | 0.09 | 288 | 0.05 | 0.10 | 36 |
| AH Vel | 6037 | 45 | 56 | 2.2 | 4.3 | 0.10 | 0.11 | 183 | 0.11 | 0.04 | 23 |
| BG Vel | 5727 | 114 | 43 | 1.9 | 5.3 | -0.01 | 0.08 | 144 | 0.01 | 0.07 | 18 |
| CS Vel | 5430 | 50 | 29 | 1.6 | 4.2 | 0.08 | 0.08 | 100 | 0.08 | 0.04 | 9 |
| CX Vel | 6251 | 87 | 54 | 2.2 | 5.3 | 0.06 | 0.10 | 234 | 0.06 | 0.08 | 34 |
| DK Vel | 6448 | 63 | 52 | 2.2 | 3.7 | -0.02 | 0.09 | 203 | -0.02 | 0.06 | 32 |
| DR Vel | 5482 | 70 | 64 | 1.5 | 3.8 | 0.08 | 0.09 | 326 | 0.07 | 0.07 | 32 |
| EX Vel | 5902 | 105 | 71 | 2.0 | 5.0 | 0.05 | 0.11 | 306 | 0.04 | 0.08 | 39 |
| FG Vel | 5411 | 110 | 62 | 1.5 | 4.4 | -0.05 | 0.08 | 233 | -0.05 | 0.06 | 23 |
| FN Vel | 5809 | 63 | 72 | 1.9 | 3.7 | 0.06 | 0.11 | 269 | 0.05 | 0.09 | 32 |

Notes:

T: Effective temperature determined using the line ratio method of Kovtyukh (2007). Sigma (σ) the standard deviation of determination. N is the number of ratios utilized.

log g: The base 10 logarithm of the surface gravity (cm s$^{-2}$)

$V_t$: Mictroturbulent velocity in km s$^{-1}$.

Fe I: The [Fe/H] value as determined from Fe I lines. The two columns that follow are the standard deviation of the abundance and N is the number of individual lines utilized.

Fe II: The [Fe/H] value as determined from Fe II lines. The two columns that follow are the standard deviation of the abundance and N is the number of individual lines utilized.

**Table 3**
Distance Information for Cepheids in Papers I-VI

| Star | Type | P (days) | $M_V$ | <V> | E(B-V) | D (kpc) | Rg (kpc) |
|---|---|---|---|---|---|---|---|
| Eta Aql | DCEP | 7.177 | -3.57 | 3.90 | 0.130 | 0.256 | 7.713 |
| SZ Aql | DCEP | 17.139 | -4.58 | 8.60 | 0.537 | 1.945 | 6.421 |
| TT Aql | DCEP | 13.755 | -4.32 | 7.14 | 0.438 | 1.023 | 7.099 |
| FF Aql | DCEPS | 4.471 | -3.42 | 5.37 | 0.196 | 0.427 | 7.629 |
| FM Aql | DCEP | 6.114 | -3.38 | 8.27 | 0.589 | 0.891 | 7.290 |
| FN Aql | DCEPS | 9.482 | -4.29 | 8.38 | 0.483 | 1.668 | 6.678 |
| V496 Aql | DCEPS | 6.807 | -3.90 | 7.75 | 0.397 | 1.187 | 6.884 |
| V600 Aql | DCEP | 7.239 | -3.58 | 10.04 | 0.798 | 1.612 | 6.831 |
| V733 Aql | DCEP | 6.179 | -3.39 | 9.97 | 0.106 | 4.019 | 6.187 |
| V1162 Aql | DCEPS | 5.376 | -3.63 | 7.80 | 0.195 | 1.444 | 6.742 |
| V1359 Aql | DCEPS: | 3.732 | -2.81 | 9.06 | 0.661 | 0.884 | 7.257 |
| V340 Ara | DCEP | 20.809 | -4.81 | 10.16 | 0.546 | 4.374 | 4.344 |
| Y Aur | DCEP | 3.859 | -2.85 | 9.61 | 0.375 | 1.771 | 9.628 |
| RT Aur | DCEP | 3.728 | -2.81 | 5.45 | 0.059 | 0.409 | 8.304 |
| RX Aur | DCEP | 11.624 | -4.13 | 7.66 | 0.263 | 1.537 | 9.397 |
| SY Aur | DCEP | 10.145 | -3.97 | 9.07 | 0.432 | 2.136 | 9.975 |
| YZ Aur | DCEP | 18.193 | -4.65 | 10.33 | 0.538 | 4.454 | 12.283 |
| AN Aur | DCEP | 10.291 | -3.99 | 10.46 | 0.565 | 3.339 | 11.157 |
| AO Aur | DCEP | 6.763 | -3.50 | 10.86 | 0.431 | 3.918 | 11.813 |
| BK Aur | DCEP | 8.002 | -3.69 | 9.43 | 0.425 | 2.238 | 10.010 |
| CY Aur | DCEP | 13.848 | -4.33 | 11.85 | 0.768 | 5.498 | 13.208 |
| ER Aur | DCEP | 15.691 | -4.48 | 11.52 | 0.494 | 7.589 | 15.362 |
| V335 Aur | DCEP | 3.413 | -2.70 | 12.46 | 0.626 | 4.247 | 12.111 |
| RW Cam | DCEP | 16.414 | -4.53 | 8.69 | 0.633 | 1.719 | 9.354 |
| RX Cam | DCEP | 7.912 | -3.68 | 7.68 | 0.532 | 0.849 | 8.614 |
| TV Cam | DCEP | 5.295 | -3.21 | 11.66 | 0.613 | 3.788 | 11.196 |
| AB Cam | DCEP | 5.788 | -3.32 | 11.85 | 0.656 | 4.069 | 11.437 |
| AD Cam | DCEP | 11.261 | -4.09 | 12.56 | 0.864 | 5.924 | 13.048 |
| BD Cas | CWB | 3.651 | -2.78 | 11.00 | 1.006 | 1.277 | 8.574 |
| RW Cas | DCEP | 14.795 | -4.41 | 9.12 | 0.380 | 2.882 | 9.962 |
| RY Cas | DCEP | 12.137 | -4.18 | 9.93 | 0.618 | 2.642 | 9.336 |
| SU Cas | DCEPS | 1.949 | -2.45 | 5.97 | 0.259 | 0.329 | 8.127 |
| SW Cas | DCEP | 5.441 | -3.25 | 9.71 | 0.467 | 1.942 | 8.747 |
| SY Cas | DCEP | 4.071 | -2.91 | 9.87 | 0.442 | 1.861 | 8.928 |
| SZ Cas | DCEPS | 13.621 | -4.71 | 9.85 | 0.794 | 2.511 | 9.832 |
| TU Cas | CEP(B) | 2.139 | -2.16 | 7.73 | 0.109 | 0.809 | 8.312 |
| XY Cas | DCEP | 4.502 | -3.02 | 9.94 | 0.533 | 1.768 | 8.979 |
| CE Cas A | DCEP | 5.141 | -3.18 | 10.92 | 0.556 | 2.891 | 9.549 |
| CE Cas B | DCEP | 4.479 | -3.02 | 11.06 | 0.527 | 2.989 | 9.615 |
| CF Cas | DCEP | 4.875 | -3.12 | 11.14 | 0.553 | 3.115 | 9.702 |
| CH Cas | DCEP | 15.086 | -4.43 | 10.97 | 0.894 | 3.187 | 9.601 |
| CY Cas | DCEP | 14.377 | -4.38 | 11.64 | 0.963 | 3.810 | 10.062 |
| DD Cas | DCEP | 9.812 | -3.93 | 9.88 | 0.450 | 2.956 | 9.602 |

**Table 3**
Distance Information for Cepheids in Papers I-VI

| Star | Type | P (days) | $M_v$ | <V> | E(B-V) | D (kpc) | Rg (kpc) |
|---|---|---|---|---|---|---|---|
| DF Cas | DCEP | 3.832 | -2.84 | 10.85 | 0.570 | 2.337 | 9.718 |
| DL Cas | DCEP | 8.001 | -3.69 | 8.97 | 0.488 | 1.649 | 8.846 |
| FM Cas | DCEP | 5.809 | -3.32 | 9.13 | 0.325 | 1.904 | 8.941 |
| V379 Cas | DCEPS | 4.306 | -3.37 | 9.05 | 0.600 | 1.251 | 8.592 |
| V636 Cas | DCEPS | 8.376 | -4.15 | 7.20 | 0.666 | 0.689 | 8.337 |
| V Cen | DCEP | 5.494 | -3.26 | 6.84 | 0.292 | 0.676 | 7.426 |
| Delta Cep | DCEP | 5.366 | -3.23 | 3.95 | 0.075 | 0.244 | 7.968 |
| CP Cep | DCEP | 17.859 | -4.63 | 10.59 | 0.649 | 4.208 | 9.600 |
| CR Cep | DCEP | 6.233 | -3.40 | 9.66 | 0.709 | 1.425 | 8.442 |
| IR Cep | DCEP | 2.114 | -2.15 | 7.78 | 0.413 | 0.524 | 8.037 |
| V351 Cep | CWB | 2.806 | -2.47 | 9.41 | 0.436 | 1.245 | 8.313 |
| RW CMa | DCEP | 5.730 | -4.81 | 10.16 | 0.546 | 4.374 | 4.344 |
| RY CMa | DCEP | 4.678 | -3.07 | 8.11 | 0.239 | 1.206 | 8.781 |
| RZ CMa | DCEP | 4.255 | -2.96 | 9.70 | 0.443 | 1.758 | 9.106 |
| TW CMa | DCEP | 6.995 | -3.54 | 9.56 | 0.329 | 2.554 | 9.764 |
| VZ CMa | DCEPS | 3.126 | -3.00 | 9.38 | 0.461 | 1.509 | 8.753 |
| BG Cru | DCEPS | 3.343 | -3.08 | 5.49 | 0.132 | 0.424 | 7.694 |
| X Cyg | DCEP | 16.386 | -4.53 | 6.39 | 0.228 | 1.087 | 7.726 |
| SU Cyg | DCEP | 3.846 | -2.84 | 6.86 | 0.080 | 0.773 | 7.603 |
| SZ Cyg | DCEP | 15.110 | -4.43 | 9.43 | 0.571 | 2.536 | 8.059 |
| TX Cyg | DCEP | 14.710 | -4.40 | 9.51 | 1.130 | 1.129 | 7.869 |
| VX Cyg | DCEP | 20.133 | -4.77 | 10.07 | 0.753 | 3.025 | 8.064 |
| VY Cyg | DCEP | 7.857 | -3.67 | 9.59 | 0.606 | 1.827 | 7.885 |
| VZ Cyg | DCEP | 4.864 | -3.11 | 8.96 | 0.266 | 1.750 | 8.132 |
| BZ Cyg | DCEP | 10.142 | -3.97 | 10.21 | 0.888 | 1.831 | 7.946 |
| CD Cyg | DCEP | 17.074 | -4.58 | 8.95 | 0.493 | 2.432 | 7.474 |
| DT Cyg | DCEPS | 2.499 | -2.74 | 5.77 | 0.042 | 0.474 | 7.805 |
| MW Cyg | DCEP | 5.955 | -3.35 | 9.49 | 0.635 | 1.437 | 7.553 |
| V386 Cyg | DCEP | 5.258 | -3.21 | 9.64 | 0.841 | 1.059 | 7.888 |
| V402 Cyg | DCEP | 4.365 | -2.99 | 9.87 | 0.391 | 2.088 | 7.600 |
| V532 Cyg | DCEPS | 3.284 | -3.06 | 9.09 | 0.494 | 1.286 | 7.981 |
| V924 Cyg | DCEPS | 5.571 | -3.67 | 10.71 | 0.261 | 5.100 | 7.529 |
| V1154 Cyg | CEP | 4.925 | -3.13 | 9.19 | 0.319 | 1.811 | 7.704 |
| V1334 Cyg | DCEPS | 3.333 | -3.07 | 5.87 | 0.025 | 0.593 | 7.856 |
| V1726 Cyg | DCEPS | 4.237 | -3.35 | 9.01 | 0.339 | 1.792 | 8.177 |
| TX Del | CWB: | 6.166 | -3.39 | 9.03 | 0.222 | 2.191 | 6.820 |
| Beta Dor | DCEP | 9.843 | -3.93 | 3.73 | 0.052 | 0.316 | 7.896 |
| Zeta Gem | DCEP | 10.150 | -3.97 | 3.92 | 0.014 | 0.370 | 8.249 |
| W Gem | DCEP | 7.914 | -3.68 | 6.95 | 0.255 | 0.915 | 8.776 |
| RZ Gem | DCEP | 5.529 | -3.26 | 10.01 | 0.563 | 1.952 | 9.838 |
| AA Gem | DCEP | 11.303 | -4.10 | 9.72 | 0.309 | 3.662 | 11.550 |
| AD Gem | DCEP | 3.788 | -2.82 | 9.86 | 0.173 | 2.657 | 10.481 |
| BB Gem | DCEP | 2.308 | -2.25 | 11.43 | 0.430 | 2.863 | 10.641 |

**Table 3**
Distance Information for Cepheids in Papers I-VI

| Star | Type | P (days) | $M_V$ | <V> | E(B-V) | D (kpc) | Rg (kpc) |
|---|---|---|---|---|---|---|---|
| DX Gem | DCEPS | 3.137 | -3.00 | 10.75 | 0.430 | 2.965 | 10.755 |
| BB Her | DCEP | 7.508 | -3.62 | 10.09 | 0.392 | 3.074 | 6.054 |
| V Lac | DCEP | 4.983 | -3.14 | 8.94 | 0.335 | 1.583 | 8.485 |
| X Lac | DCEPS | 5.445 | -3.64 | 8.41 | 0.336 | 1.560 | 8.477 |
| Y Lac | DCEP | 4.324 | -2.98 | 9.15 | 0.207 | 1.955 | 8.419 |
| Z Lac | DCEP | 10.886 | -4.05 | 8.42 | 0.370 | 1.796 | 8.564 |
| RR Lac | DCEP | 6.416 | -3.44 | 8.85 | 0.319 | 1.782 | 8.554 |
| BG Lac | DCEP | 5.332 | -3.22 | 8.88 | 0.300 | 1.687 | 8.158 |
| V473 Lyr | DCEPS: | 1.491 | -1.74 | 6.18 | 0.025 | 0.370 | 7.726 |
| T Mon | DCEP | 27.019 | -5.11 | 6.12 | 0.181 | 1.348 | 9.150 |
| SV Mon | DCEP | 15.233 | -4.44 | 8.22 | 0.234 | 2.405 | 10.143 |
| TW Mon | CEP | 7.097 | -3.55 | 12.58 | 0.663 | 6.278 | 13.608 |
| TX Mon | DCEP | 8.702 | -3.79 | 10.96 | 0.485 | 4.335 | 11.742 |
| TZ Mon | DCEP | 7.428 | -3.61 | 10.76 | 0.420 | 4.003 | 11.439 |
| UY Mon | DCEPS | 2.398 | -2.69 | 9.39 | 0.064 | 2.372 | 10.085 |
| WW Mon | DCEP | 4.662 | -3.07 | 12.51 | 0.605 | 5.291 | 12.943 |
| XX Mon | DCEP | 5.456 | -3.25 | 11.90 | 0.567 | 4.600 | 11.946 |
| AA Mon | DCEP | 3.938 | -2.87 | 12.71 | 0.792 | 4.014 | 11.364 |
| AC Mon | DCEP | 8.015 | -3.70 | 10.07 | 0.484 | 2.755 | 10.121 |
| CU Mon | DCEP | 4.708 | -3.08 | 13.61 | 0.751 | 7.104 | 14.451 |
| CV Mon | DCEP | 5.379 | -3.23 | 10.30 | 0.722 | 1.737 | 9.461 |
| EE Mon | DCEP | 4.809 | -3.10 | 12.94 | 0.465 | 8.098 | 15.018 |
| EK Mon | DCEP | 3.958 | -2.88 | 11.05 | 0.556 | 2.663 | 10.191 |
| FG Mon | DCEP | 4.497 | -3.02 | 13.31 | 0.651 | 7.015 | 13.944 |
| FI Mon | DCEP | 3.288 | -2.66 | 12.92 | 0.513 | 6.098 | 13.106 |
| V465 Mon | DCEP | 2.714 | -2.44 | 10.38 | 0.244 | 2.544 | 10.088 |
| V495 Mon | DCEP | 4.097 | -2.92 | 12.43 | 0.609 | 4.730 | 12.105 |
| V504 Mon | CEP | 2.774 | -2.46 | 11.81 | 0.538 | 3.219 | 10.676 |
| V508 Mon | DCEP | 4.134 | -2.93 | 10.52 | 0.307 | 3.091 | 10.710 |
| V510 Mon | DCEP | 7.457 | -3.61 | 12.68 | 0.802 | 5.505 | 12.956 |
| V526 Mon | DCEPS | 2.675 | -2.82 | 8.60 | 0.089 | 1.682 | 9.325 |
| S Nor | DCEP | 9.754 | -3.92 | 6.39 | 0.179 | 0.887 | 7.169 |
| V340 Nor | DCEP | 11.288 | -4.09 | 8.38 | 0.321 | 1.934 | 6.306 |
| Y Oph | DCEPS | 17.126 | -4.98 | 6.17 | 0.645 | 0.649 | 7.305 |
| BF Oph | DCEP | 4.068 | -2.91 | 7.34 | 0.235 | 0.789 | 7.121 |
| RS Ori | DCEP | 7.567 | -3.63 | 8.41 | 0.352 | 1.516 | 9.363 |
| CS Ori | DCEP | 3.889 | -2.85 | 11.38 | 0.383 | 3.980 | 11.738 |
| GQ Ori | DCEP | 8.616 | -3.78 | 8.97 | 0.249 | 2.444 | 10.226 |
| VX Per | DCEP | 10.888 | -4.05 | 9.31 | 0.475 | 2.322 | 9.627 |
| SV Per | DCEP | 11.129 | -4.08 | 9.02 | 0.408 | 2.269 | 10.088 |
| UX Per | DCEP | 4.566 | -3.04 | 11.66 | 0.512 | 4.075 | 11.104 |
| AS Per | DCEP | 4.973 | -3.14 | 9.72 | 0.674 | 1.372 | 9.154 |
| AW Per | DCEP | 6.464 | -3.45 | 7.49 | 0.489 | 0.744 | 8.622 |

**Table 3**
Distance Information for Cepheids in Papers I-VI

| Star | Type | P (days) | $M_V$ | <V> | E(B-V) | D (kpc) | Rg (kpc) |
|---|---|---|---|---|---|---|---|
| BM Per | DCEP | 22.959 | -4.92 | 10.39 | 0.871 | 3.154 | 10.851 |
| HQ Per | CEP | 8.639 | -3.78 | 11.60 | 0.564 | 5.146 | 12.901 |
| MM Per | DCEP | 4.117 | -2.92 | 10.80 | 0.490 | 2.678 | 10.304 |
| V440 Per | DCEPS | 7.573 | -4.03 | 6.28 | 0.260 | 0.784 | 8.478 |
| X Pup | DCEP | 25.965 | -5.06 | 8.46 | 0.402 | 2.785 | 9.730 |
| RS Pup | DCEP | 41.388 | -5.60 | 6.95 | 0.457 | 1.641 | 8.540 |
| VW Pup | DCEP | 4.285 | -2.97 | 11.37 | 0.489 | 3.551 | 10.340 |
| VX Pup | CEP(B) | 3.012 | -2.56 | 8.33 | 0.129 | 1.240 | 8.638 |
| VZ Pup | DCEP | 23.174 | -4.93 | 9.62 | 0.459 | 4.109 | 10.404 |
| WW Pup | DCEP | 5.517 | -3.26 | 10.55 | 0.379 | 3.298 | 10.068 |
| AD Pup | DCEP | 13.596 | -4.31 | 9.86 | 0.314 | 4.283 | 10.611 |
| AQ Pup | DCEP | 30.077 | -5.23 | 8.79 | 0.518 | 2.953 | 9.487 |
| BC Pup | DCEP | 3.544 | -2.75 | 13.84 | 0.800 | 6.320 | 12.440 |
| BN Pup | DCEP | 13.673 | -4.32 | 9.88 | 0.416 | 3.724 | 9.920 |
| HW Pup | DCEP | 13.457 | -4.30 | 12.05 | 0.688 | 6.684 | 12.332 |
| MY Pup | DCEPS | 5.694 | -3.70 | 5.68 | 0.061 | 0.684 | 8.028 |
| V335 Pup | DCEPS | 4.861 | -3.51 | 8.72 | 0.154 | 2.219 | 9.193 |
| RV Sco | DCEP | 6.061 | -3.37 | 7.04 | 0.349 | 0.719 | 7.196 |
| RY Sco | DCEP | 20.320 | -4.78 | 8.00 | 0.718 | 1.237 | 6.668 |
| KQ Sco | DCEP | 28.694 | -5.18 | 9.81 | 0.869 | 2.728 | 5.409 |
| V500 Sco | DCEP | 9.317 | -3.87 | 8.73 | 0.593 | 1.370 | 6.530 |
| Z Sct | DCEP | 12.902 | -4.25 | 9.60 | 0.492 | 2.832 | 5.522 |
| SS Sct | DCEP | 3.671 | -2.79 | 8.21 | 0.325 | 0.977 | 7.029 |
| UZ Sct | DCEP | 14.748 | -4.40 | 11.31 | 1.020 | 3.043 | 5.125 |
| EW Sct | CEP(B) | 5.823 | -3.32 | 7.98 | 1.074 | 0.369 | 7.568 |
| V367 Sct | CEP(B) | 6.293 | -3.41 | 11.60 | 1.231 | 1.610 | 6.431 |
| BQ Ser | CEP(B) | 4.271 | -2.96 | 9.50 | 0.815 | 0.926 | 7.166 |
| S Sge | DCEP | 8.382 | -3.75 | 5.62 | 0.100 | 0.645 | 7.552 |
| U Sgr | DCEP | 6.745 | -3.50 | 6.70 | 0.403 | 0.599 | 7.321 |
| W Sgr | DCEP | 7.595 | -3.63 | 4.67 | 0.108 | 0.389 | 7.512 |
| Y Sgr | DCEP | 5.773 | -3.31 | 5.74 | 0.191 | 0.488 | 7.425 |
| VY Sgr | DCEP | 13.558 | -4.31 | 11.51 | 1.221 | 2.369 | 5.584 |
| WZ Sgr | DCEP | 21.850 | -4.86 | 8.03 | 0.431 | 1.995 | 5.964 |
| XX Sgr | CEP | 6.424 | -3.44 | 8.85 | 0.521 | 1.323 | 6.632 |
| AP Sgr | DCEP | 5.058 | -3.16 | 6.96 | 0.178 | 0.809 | 7.101 |
| AV Sgr | DCEP | 15.411 | -4.46 | 11.39 | 1.206 | 2.456 | 5.475 |
| BB Sgr | DCEP | 6.637 | -3.48 | 6.95 | 0.281 | 0.800 | 7.138 |
| V350 Sgr | DCEP | 5.154 | -3.18 | 7.48 | 0.299 | 0.871 | 7.065 |
| YZ Sgr | DCEP | 9.554 | -3.90 | 7.36 | 0.281 | 1.175 | 6.799 |
| ST Tau | DCEP | 4.034 | -2.90 | 8.22 | 0.368 | 0.966 | 8.834 |
| SZ Tau | DCEPS | 3.149 | -3.01 | 6.53 | 0.295 | 0.521 | 8.394 |
| AE Tau | CEP | 3.896 | -2.86 | 11.68 | 0.575 | 3.433 | 11.326 |
| EF Tau | DCEP | 3.448 | -2.71 | 13.10 | 0.360 | 8.499 | 16.323 |

**Table 3**
Distance Information for Cepheids in Papers I-VI

| Star | Type | P (days) | $M_V$ | <V> | E(B-V) | D (kpc) | Rg (kpc) |
|---|---|---|---|---|---|---|---|
| EU Tau | DCEPS | 2.102 | -2.54 | 8.09 | 0.164 | 1.048 | 8.933 |
| T Vel | DCEP | 4.640 | -3.06 | 8.02 | 0.289 | 1.072 | 8.054 |
| RY Vel | DCEP | 28.131 | -5.16 | 8.40 | 0.547 | 2.276 | 7.731 |
| SW Vel | DCEP | 23.440 | -4.94 | 8.12 | 0.344 | 2.458 | 8.427 |
| SX Vel | DCEP | 9.550 | -3.90 | 8.25 | 0.263 | 1.821 | 8.245 |
| S Vul | DCEP | 68.044 | -6.18 | 8.96 | 0.727 | 3.626 | 7.067 |
| T Vul | DCEP | 4.435 | -3.01 | 5.75 | 0.064 | 0.514 | 7.760 |
| U Vul | DCEP | 7.991 | -3.69 | 7.13 | 0.603 | 0.595 | 7.584 |
| X Vul | DCEP | 6.320 | -3.42 | 8.85 | 0.742 | 0.943 | 7.532 |
| SV Vul | DCEP | 45.028 | -5.70 | 7.22 | 0.461 | 1.936 | 7.261 |

Notes:

Type: Variability type from the General Catalog of Variable Stars (Samus , N. N. et al. 2007-2009)

P: Period in days from Berdnikov (2006– private communication to S. Andrievsky) or GCVS.

$M_V$: Absolute V magnitude computed assuming listed period is the fundamental except for DCEPS where $P_0$ = P/0.71. Period-Luminosity relation and R (=Av/E(B-V) = 3.23) from Fouqué et al. 2007.

<V>: Intensity mean apparent V magnitude from Fernie 1995. Exceptions are V351 Cep and BB Gem from Schmidt et al. (2005a) and TX Del from Schmidt 2005b.

E(B-V): From Fouqué et al. (2007) or Fernie (1995) as modified by Fouqué et al. Reddening for XY CMa from Yong et al. (2006). BB Her reddening from van Leeuven et al. (2007), BC Pup from Fernie & Hube (1968), and BB Gem from Harris (1985). Reddening for V733 Aql, BD Cas, V379 Cas, V351 Cep, V1334 Cyg, TX Del, and EF Tau calculated using the method of Kovtyukh et al. (2008).

d and Rg: Distance in kiloparsecs from the Sun and the galactic center respectively. The Sun is assumed to be 7.9 kpc from the galactic center (McNamara et al. 2000).

Table 4A
Abundance Data For Cepheids: Carbon through Manganese

| Name | Ref | Sp | [C/H] | [N/H] | [O/H] | [Na/H] | [Mg/H] | [Al/H] | [Si/H] | [S/H] | [Ca/H] | [Sc/H] | [Ti/H] | [V/H] | [Cr/H] | [Mn/H] |
|---|---|---|---|---|---|---|---|---|---|---|---|---|---|---|---|---|
| T Ant | 8 | 1 | -0.36 | 0.13 | -0.43 | 0.15 | -0.25 | -0.09 | -0.15 | -0.18 | -0.22 | -0.19 | -0.17 | -0.29 | -0.27 | -0.24 |
| Eta Aql | 9 | 14 | -0.13 | 0.35 | -0.06 | 0.20 | -0.05 | 0.15 | 0.12 | 0.16 | -0.02 | -0.01 | 0.13 | 0.10 | 0.15 | 0.11 |
| SZ Aql | 10 | 11 | -0.06 | … | -0.03 | 0.25 | -0.08 | 0.31 | 0.16 | 0.25 | 0.10 | 0.07 | 0.07 | 0.04 | 0.21 | 0.20 |
| TT Aql | 10 | 8 | -0.05 | … | 0.02 | 0.27 | -0.11 | 0.25 | 0.08 | 0.26 | 0.06 | 0.10 | 0.06 | -0.03 | 0.14 | 0.14 |
| FF Aql | 11 | 14 | -0.26 | … | -0.09 | 0.23 | -0.26 | 0.12 | 0.05 | 0.02 | -0.01 | -0.07 | 0.05 | 0.03 | 0.09 | 0.07 |
| FM Aql | 1 | 2 | -0.24 | … | -0.19 | 0.32 | 0.00 | 0.33 | 0.17 | 0.24 | 0.17 | -0.17 | 0.15 | 0.10 | 0.19 | 0.11 |
| FN Aql | 1 | 4 | -1.31 | … | -0.08 | 0.19 | -0.21 | 0.10 | 0.00 | -0.02 | -0.07 | -0.07 | 0.04 | 0.01 | -0.04 | -0.12 |
| V496 Aql | 1 | 2 | -0.20 | … | -0.15 | 0.24 | -0.12 | 0.10 | 0.11 | 0.13 | -0.03 | 0.10 | 0.06 | 0.06 | 0.09 | 0.05 |
| V600 Aql | 7 | 1 | -0.25 | … | 0.11 | 0.30 | 0.20 | 0.15 | 0.08 | -0.22 | 0.06 | 0.11 | 0.07 | -0.10 | 0.06 | … |
| V733 Aql | 7 | 1 | -0.07 | … | 0.04 | 0.19 | 0.22 | 0.08 | 0.09 | 0.03 | -0.02 | -0.04 | 0.05 | -0.10 | 0.00 | 0.08 |
| V1162 Aql | 1 | 2 | -0.14 | … | -0.19 | 0.13 | -0.19 | 0.13 | 0.06 | … | -0.03 | -0.21 | -0.03 | -0.02 | 0.02 | -0.01 |
| V1359 Aql | 7 | 1 | 0.16 | … | 0.29 | 0.24 | 0.17 | 0.21 | 0.12 | 0.24 | 0.18 | 0.21 | 0.20 | 0.20 | 0.16 | 0.16 |
| V340 Ara | 2 | 1 | 0.20 | 1.00 | 0.07 | 0.56 | … | 0.35 | 0.35 | 0.51 | 0.21 | 0.33 | 0.30 | 0.21 | 0.09 | 0.25 |
| Y Aur | 7 | 1 | -0.25 | … | -0.42 | -0.06 | … | -0.17 | 0.03 | -0.27 | -0.08 | -0.32 | -0.42 | -0.36 | -0.29 | … |
| RT Aur | 12 | 10 | -0.20 | … | -0.07 | 0.29 | -0.17 | 0.14 | 0.10 | 0.15 | 0.05 | -0.02 | 0.08 | 0.05 | 0.05 | 0.08 |
| RX Aur | 10 | 16 | -0.13 | 0.34 | 0.07 | 0.18 | -0.17 | 0.05 | 0.04 | 0.07 | -0.05 | -0.26 | -0.03 | -0.09 | 0.09 | 0.02 |
| SY Aur | 6 | 1 | -0.41 | 0.34 | -0.11 | 0.23 | -0.09 | -0.05 | 0.05 | -0.02 | -0.05 | -0.10 | 0.00 | 0.01 | -0.06 | -0.13 |
| YZ Aur | 607 | 3 | -0.49 | … | -0.52 | -0.01 | -0.20 | -0.20 | -0.23 | -0.19 | -0.35 | -0.43 | -0.35 | -0.56 | -0.37 | -0.36 |
| AN Aur | 607 | 4 | -0.38 | 0.12 | -0.15 | 0.03 | -0.18 | -0.16 | -0.08 | -0.22 | -0.14 | -0.21 | -0.10 | -0.22 | -0.18 | -0.25 |
| AO Aur | 5 | 1 | … | … | … | -0.03 | … | … | -0.24 | -0.51 | -0.35 | -0.34 | -0.04 | -0.16 | -0.25 | -0.28 |
| BK Aur | 7 | 1 | -0.26 | … | -0.04 | 0.27 | … | 0.20 | 0.14 | -0.04 | 0.25 | 0.01 | 0.32 | -0.04 | 0.19 | … |
| CY Aur | 6 | 1 | -0.59 | 0.15 | -0.41 | 0.01 | -0.38 | -0.31 | -0.27 | -0.39 | -0.37 | -0.34 | -0.29 | -0.22 | -0.36 | -0.36 |
| ER Aur | 6 | 2 | -0.53 | -0.12 | -0.40 | -0.06 | -0.54 | -0.20 | -0.22 | -0.24 | -0.25 | -0.28 | -0.32 | -0.20 | -0.36 | -0.32 |
| V335 Aur | 5 | 1 | … | … | … | -0.15 | … | … | -0.27 | … | … | … | … | … | … | -0.34 |
| RW Cam | 610 | 17 | -0.11 | 0.00 | 0.02 | 0.23 | -0.06 | 0.16 | 0.06 | 0.18 | 0.01 | -0.01 | 0.03 | -0.08 | 0.10 | 0.10 |
| RX Cam | 9 | 9 | -0.15 | … | -0.10 | 0.18 | -0.15 | 0.09 | 0.05 | 0.08 | -0.04 | -0.04 | 0.08 | 0.02 | 0.08 | 0.05 |
| TV Cam | 6 | 1 | -0.39 | 0.40 | … | 0.00 | -0.08 | -0.16 | -0.01 | -0.11 | -0.08 | -0.14 | 0.15 | -0.04 | 0.18 | -0.17 |
| AB Cam | 6 | 1 | -0.40 | 0.36 | -0.29 | 0.10 | -0.18 | -0.04 | -0.02 | 0.00 | -0.07 | -0.13 | -0.16 | -0.07 | -0.11 | -0.19 |
| AD Cam | 6 | 1 | -0.80 | 0.24 | 0.00 | 0.05 | -0.09 | 0.05 | -0.12 | -0.15 | -0.25 | -0.19 | -0.23 | -0.34 | -0.25 | -0.33 |
| L Car | 8 | 1 | 0.06 | -0.02 | 0.25 | 0.05 | 0.28 | 0.14 | 0.06 | 0.28 | -0.07 | -0.07 | 0.02 | 0.01 | -0.04 | -0.05 |
| U Car | 8 | 1 | … | … | … | 0.27 | 0.22 | 0.29 | 0.07 | 0.22 | 0.01 | … | -0.04 | -0.07 | 0.03 | -0.07 |

Table 4A
Abundance Data For Cepheids: Carbon through Manganese

| Name | Ref | Sp | [C/H] | [N/H] | [O/H] | [Na/H] | [Mg/H] | [Al/H] | [Si/H] | [S/H] | [Ca/H] | [Sc/H] | [Ti/H] | [V/H] | [Cr/H] | [Mn/H] |
|---|---|---|---|---|---|---|---|---|---|---|---|---|---|---|---|---|
| V Car | 8 | 1 | -0.23 | 0.48 | -0.27 | 0.09 | -0.07 | 0.09 | 0.06 | 0.04 | -0.05 | -0.10 | 0.03 | 0.03 | -0.05 | -0.14 |
| SX Car | 8 | 1 | -0.30 | 0.25 | -0.30 | 0.20 | -0.15 | -0.02 | 0.05 | 0.06 | -0.06 | -0.11 | -0.07 | -0.23 | -0.13 | -0.17 |
| UW Car | 8 | 1 | -0.23 | 0.39 | -0.28 | 0.15 | -0.12 | 0.04 | 0.07 | 0.11 | -0.03 | -0.07 | 0.03 | -0.13 | -0.12 | -0.11 |
| UX Car | 8 | 1 | -0.15 | 0.44 | -0.04 | 0.20 | -0.08 | 0.03 | 0.09 | 0.12 | 0.00 | 0.04 | 0.01 | 0.03 | 0.00 | -0.12 |
| UY Car | 8 | 1 | -0.10 | 0.37 | -0.15 | 0.14 | 0.03 | 0.12 | 0.08 | 0.18 | -0.08 | 0.05 | 0.02 | -0.05 | 0.01 | -0.04 |
| UZ Car | 8 | 1 | -0.17 | 0.40 | -0.10 | 0.24 | 0.05 | 0.14 | 0.12 | 0.21 | 0.03 | 0.05 | 0.11 | 0.01 | 0.04 | -0.03 |
| VY Car | 8 | 1 | -0.21 | ... | ... | 0.28 | 0.16 | 0.28 | 0.19 | 0.57 | 0.15 | ... | 0.02 | 0.03 | 0.04 | 0.09 |
| WW Car | 8 | 1 | -0.22 | 0.30 | -0.55 | 0.07 | -0.12 | 0.01 | -0.01 | 0.07 | -0.15 | -0.13 | -0.10 | -0.14 | -0.14 | -0.22 |
| WZ Car | 8 | 1 | -0.10 | 0.41 | ... | 0.35 | -0.11 | 0.15 | 0.00 | 0.21 | -0.18 | 0.10 | -0.11 | 0.05 | 0.02 | -0.16 |
| XX Car | 8 | 1 | -0.31 | 0.63 | -0.06 | 0.40 | 0.05 | 0.19 | 0.14 | 0.22 | 0.02 | 0.19 | 0.10 | 0.11 | 0.10 | -0.04 |
| XY Car | 8 | 1 | -0.15 | 0.39 | -0.29 | 0.18 | 0.03 | 0.09 | 0.07 | 0.19 | -0.08 | 0.12 | 0.05 | -0.03 | -0.01 | -0.07 |
| XZ Car | 8 | 1 | -0.07 | 0.57 | ... | 0.36 | 0.08 | 0.24 | 0.21 | 0.24 | 0.06 | 0.05 | 0.04 | 0.11 | 0.14 | 0.06 |
| YZ Car | 8 | 1 | -0.26 | 0.28 | -0.15 | 0.16 | 0.06 | 0.13 | 0.05 | 0.07 | -0.04 | 0.04 | 0.07 | 0.03 | -0.08 | -0.09 |
| AQ Car | 8 | 1 | -0.13 | 0.34 | -0.10 | 0.06 | -0.10 | 0.04 | 0.05 | 0.14 | -0.10 | -0.02 | 0.03 | 0.05 | -0.06 | -0.13 |
| CN Car | 8 | 1 | -0.08 | 0.47 | -0.11 | 0.18 | 0.00 | 0.14 | 0.12 | 0.21 | 0.05 | 0.03 | 0.13 | 0.10 | 0.01 | 0.04 |
| CY Car | 8 | 1 | -0.07 | 0.57 | -0.08 | 0.23 | 0.01 | 0.16 | 0.13 | 0.27 | 0.04 | 0.10 | 0.11 | 0.06 | 0.06 | -0.04 |
| DY Car | 8 | 1 | -0.22 | 0.34 | -0.28 | 0.09 | -0.12 | -0.01 | 0.04 | 0.05 | -0.14 | -0.07 | -0.07 | -0.18 | -0.09 | -0.12 |
| ER Car | 8 | 1 | -0.16 | 0.34 | -0.09 | 0.10 | -0.15 | 0.09 | 0.02 | 0.18 | -0.09 | -0.07 | 0.00 | 0.05 | -0.05 | -0.13 |
| FI Car | 8 | 1 | -0.22 | 0.06 | -0.25 | 0.24 | 0.03 | 0.21 | 0.10 | 0.28 | 0.12 | 0.04 | 0.06 | -0.06 | 0.04 | 0.11 |
| FR Car | 8 | 1 | -0.51 | 0.48 | -0.18 | 0.14 | 0.16 | 0.10 | 0.05 | 0.09 | -0.10 | 0.02 | 0.06 | 0.02 | -0.04 | -0.17 |
| GH Car | 8 | 1 | -0.26 | 0.31 | -0.17 | 0.24 | 0.00 | 0.06 | 0.07 | 0.20 | -0.05 | -0.02 | 0.17 | 0.11 | -0.03 | -0.08 |
| GX Car | 8 | 1 | -0.26 | 0.39 | -0.07 | 0.09 | -0.10 | 0.03 | 0.06 | 0.12 | -0.08 | -0.02 | 0.02 | -0.03 | -0.02 | -0.14 |
| HW Car | 8 | 1 | -0.15 | 0.35 | 0.02 | 0.21 | -0.12 | 0.04 | -0.02 | 0.13 | -0.14 | 0.00 | -0.01 | -0.01 | -0.05 | -0.19 |
| IO Car | 8 | 1 | -0.24 | 0.27 | -0.35 | 0.14 | -0.18 | 0.11 | 0.00 | 0.10 | -0.16 | -0.04 | -0.02 | -0.01 | -0.08 | -0.10 |
| IT Car | 8 | 1 | -0.15 | 0.45 | -0.15 | 0.31 | 0.11 | 0.17 | 0.10 | 0.16 | -0.02 | 0.02 | 0.08 | 0.02 | 0.04 | -0.02 |
| V397 Car | 8 | 1 | -0.15 | 0.42 | -0.14 | 0.17 | -0.03 | 0.10 | 0.05 | 0.16 | 0.01 | 0.01 | 0.01 | 0.05 | -0.07 | -0.09 |
| RW Cas | 7 | 1 | -0.14 | ... | -0.01 | ... | ... | 0.05 | 0.13 | ... | 0.28 | 0.30 | 0.10 | -0.04 | 0.15 | ... |
| RY Cas | 7 | 1 | 0.25 | ... | 0.31 | 0.26 | ... | 0.15 | 0.22 | 0.03 | -0.07 | 0.22 | 0.30 | 0.19 | 0.09 | ... |
| SU Cas | 11 | 13 | -0.13 | ... | -0.02 | 0.26 | -0.22 | 0.10 | 0.09 | 0.13 | 0.07 | -0.10 | 0.09 | 0.10 | 0.10 | 0.10 |
| SW Cas | 7 | 1 | 0.13 | ... | 0.27 | 0.26 | ... | 0.22 | 0.12 | 0.00 | 0.04 | 0.15 | 0.20 | -0.01 | 0.18 | 0.41 |
| SY Cas | 7 | 1 | 0.21 | ... | 0.31 | 0.16 | 0.32 | 0.15 | 0.07 | 0.01 | 0.10 | 0.10 | 0.22 | -0.06 | 0.19 | ... |

Table 4A
Abundance Data For Cepheids: Carbon through Manganese

| Name | Ref | Sp | [C/H] | [N/H] | [O/H] | [Na/H] | [Mg/H] | [Al/H] | [Si/H] | [S/H] | [Ca/H] | [Sc/H] | [Ti/H] | [V/H] | [Cr/H] | [Mn/H] |
|---|---|---|---|---|---|---|---|---|---|---|---|---|---|---|---|---|
| SZ Cas | 7 | 1 | -0.13 | ... | 0.06 | 0.51 | ... | 0.17 | 0.28 | -0.09 | 0.14 | 0.29 | 0.21 | -0.21 | 0.14 | 0.62 |
| TU Cas | 1 | 12 | -0.19 | ... | -0.03 | 0.15 | -0.19 | 0.14 | 0.10 | -0.03 | -0.02 | -0.19 | 0.05 | 0.02 | 0.02 | 0.06 |
| XY Cas | 7 | 1 | -0.24 | ... | -0.09 | 0.45 | ... | 0.10 | 0.13 | -0.47 | 0.14 | -0.10 | 0.14 | -0.23 | 0.00 | ... |
| BD Cas | 1 | 3 | -0.14 | ... | -0.09 | -0.03 | -0.26 | -0.09 | 0.03 | 0.26 | -0.19 | -0.23 | -0.06 | -0.06 | -0.14 | -0.13 |
| CE Cas A | 7 | 1 | -0.16 | ... | -0.04 | 0.56 | ... | 0.01 | 0.33 | -0.18 | 0.29 | 0.35 | 0.31 | -0.16 | 0.29 | ... |
| CE Cas B | 7 | 1 | 0.05 | ... | -0.04 | 0.23 | ... | 0.20 | 0.25 | -0.07 | 0.29 | 0.35 | 0.31 | 0.05 | ... | ... |
| CF Cas | 1 | 5 | -0.19 | ... | 0.06 | 0.09 | -0.21 | 0.10 | 0.01 | 0.10 | -0.01 | -0.04 | 0.00 | -0.06 | 0.06 | -0.01 |
| CH Cas | 7 | 1 | 0.00 | ... | 0.16 | 0.28 | ... | ... | 0.14 | -0.03 | -0.04 | 0.12 | 0.15 | 0.01 | 0.02 | ... |
| CY Cas | 7 | 1 | -0.42 | ... | -0.04 | ... | ... | 0.08 | 0.11 | 0.03 | -0.01 | 0.08 | -0.04 | -0.27 | 0.01 | 0.15 |
| DD Cas | 7 | 1 | 0.02 | ... | 0.07 | 0.30 | 0.20 | 0.10 | 0.10 | -0.05 | 0.00 | 0.10 | 0.23 | -0.10 | 0.01 | ... |
| DF Cas | 1 | 1 | -0.30 | ... | ... | 0.16 | -0.33 | ... | 0.03 | 0.39 | -0.18 | -0.09 | 0.00 | -0.01 | 0.11 | -0.12 |
| DL Cas | 1 | 3 | -0.31 | ... | -0.01 | 0.11 | 0.11 | 0.14 | 0.02 | 0.21 | 0.01 | -0.16 | 0.03 | -0.04 | 0.10 | 0.07 |
| FM Cas | 7 | 1 | -0.45 | ... | -0.21 | 0.17 | 0.20 | 0.30 | 0.09 | -0.31 | 0.14 | -0.08 | 0.11 | -0.42 | -0.08 | ... |
| V379 Cas | 7 | 2 | -0.10 | ... | 0.07 | 0.21 | ... | 0.23 | 0.19 | -0.11 | 0.18 | 0.12 | 0.23 | -0.16 | -0.02 | ... |
| V636 Cas | 11 | 8 | -0.06 | ... | -0.18 | 0.28 | -0.14 | 0.12 | 0.06 | 0.13 | 0.06 | -0.16 | 0.05 | -0.04 | 0.16 | 0.09 |
| V Cen | 108 | 3 | -0.20 | 0.34 | -0.16 | 0.07 | -0.15 | 0.08 | 0.02 | 0.09 | -0.10 | -0.06 | 0.00 | -0.04 | -0.06 | -0.11 |
| QY Cen | 8 | 1 | -0.07 | 0.61 | -0.09 | 0.43 | 0.15 | 0.22 | 0.22 | 0.29 | 0.05 | ... | 0.16 | 0.10 | 0.12 | 0.08 |
| XX Cen | 8 | 1 | -0.07 | 0.56 | -0.03 | 0.30 | 0.02 | 0.22 | 0.13 | 0.28 | 0.05 | 0.14 | 0.14 | 0.16 | 0.08 | 0.05 |
| AY Cen | 8 | 1 | -0.18 | 0.35 | -0.15 | 0.20 | -0.07 | 0.05 | 0.04 | 0.07 | -0.06 | -0.01 | 0.02 | -0.03 | -0.04 | -0.11 |
| AZ Cen | 8 | 1 | -0.21 | 0.35 | -0.10 | 0.27 | -0.07 | 0.02 | 0.06 | 0.14 | -0.13 | -0.04 | -0.04 | 0.00 | -0.07 | -0.17 |
| BB Cen | 8 | 1 | -0.03 | 0.64 | 0.06 | 0.30 | -0.13 | 0.19 | 0.17 | 0.31 | -0.03 | 0.14 | 0.10 | 0.12 | 0.04 | 0.01 |
| KK Cen | 8 | 1 | -0.09 | 0.47 | 0.01 | 0.31 | 0.09 | 0.18 | 0.12 | 0.22 | 0.02 | 0.05 | 0.17 | 0.11 | 0.07 | -0.01 |
| KN Cen | 8 | 1 | 0.45 | ... | 0.62 | ... | 0.59 | 0.53 | 0.43 | 1.02 | 0.25 | 0.43 | 0.22 | 0.24 | 0.29 | 0.38 |
| MZ Cen | 8 | 1 | 0.02 | 0.65 | -0.13 | 0.35 | 0.10 | 0.24 | 0.15 | 0.34 | -0.05 | 0.23 | 0.12 | 0.15 | 0.09 | 0.14 |
| V339 Cen | 8 | 1 | -0.13 | 0.36 | -0.20 | 0.19 | -0.11 | 0.06 | 0.06 | 0.17 | -0.11 | -0.04 | 0.04 | 0.08 | -0.02 | -0.10 |
| V378 Cen | 8 | 1 | -0.28 | 0.35 | -0.09 | 0.19 | -0.24 | 0.06 | 0.07 | 0.15 | -0.17 | -0.07 | 0.05 | 0.02 | -0.06 | -0.13 |
| V381 Cen | 8 | 1 | -0.20 | 0.35 | -0.09 | 0.15 | -0.04 | 0.02 | 0.08 | 0.12 | -0.03 | 0.00 | 0.04 | -0.04 | 0.00 | -0.12 |
| V419 Cen | 8 | 1 | -0.07 | 0.39 | -0.12 | 0.17 | -0.10 | 0.20 | 0.11 | 0.22 | 0.06 | 0.01 | 0.16 | 0.07 | -0.01 | 0.01 |
| V496 Cen | 8 | 1 | -0.18 | 0.39 | -0.16 | 0.18 | -0.02 | 0.12 | 0.07 | 0.13 | -0.07 | 0.01 | -0.01 | 0.02 | -0.02 | -0.08 |
| V659 Cen | 8 | 1 | -0.10 | 0.41 | 0.00 | 0.18 | -0.08 | 0.13 | 0.07 | 0.17 | -0.04 | 0.01 | 0.02 | -0.02 | -0.07 | -0.05 |
| V737 Cen | 8 | 1 | -0.05 | 0.47 | -0.09 | 0.24 | 0.01 | 0.18 | 0.13 | 0.26 | 0.02 | -0.01 | 0.17 | 0.06 | 0.05 | 0.05 |

Table 4A
Abundance Data For Cepheids: Carbon through Manganese

| Name | Ref | Sp | [C/H] | [N/H] | [O/H] | [Na/H] | [Mg/H] | [Al/H] | [Si/H] | [S/H] | [Ca/H] | [Sc/H] | [Ti/H] | [V/H] | [Cr/H] | [Mn/H] |
|---|---|---|---|---|---|---|---|---|---|---|---|---|---|---|---|---|
| Del Cep | 12 | 18 | -0.16 | 0.34 | 0.01 | 0.20 | 0.03 | 0.16 | 0.09 | 0.16 | 0.01 | -0.07 | 0.03 | 0.04 | 0.08 | 0.19 |
| CP Cep | 7 | 1 | -0.39 | … | -0.16 | 0.13 | 0.10 | 0.07 | 0.10 | -0.17 | 0.14 | 0.04 | 0.04 | -0.12 | -0.07 | … |
| CR Cep | 7 | 1 | -0.31 | … | -0.09 | 0.09 | 0.39 | 0.19 | 0.02 | -0.23 | 0.02 | -0.14 | 0.02 | -0.36 | -0.09 | … |
| IR Cep | 107 | 2 | -0.12 | … | 0.04 | 0.36 | -0.08 | 0.24 | 0.19 | 0.12 | 0.08 | -0.01 | 0.19 | 0.09 | -0.01 | -0.02 |
| V351 Cep | 107 | 3 | -0.15 | … | 0.02 | 0.20 | -0.21 | 0.04 | 0.09 | 0.11 | 0.00 | 0.02 | 0.12 | 0.01 | 0.04 | -0.03 |
| AV Cir | 8 | 1 | -0.06 | 0.51 | -0.08 | 0.24 | 0.03 | 0.16 | 0.17 | 0.26 | 0.04 | 0.04 | 0.14 | 0.01 | 0.05 | -0.01 |
| AX Cir | 8 | 2 | -0.25 | 0.32 | -0.07 | 0.09 | -0.05 | 0.01 | -0.02 | 0.06 | -0.13 | -0.14 | -0.05 | -0.11 | -0.12 | -0.25 |
| BP Cir | 8 | 1 | -0.24 | 0.34 | -0.18 | 0.14 | -0.07 | 0.01 | 0.05 | 0.11 | -0.14 | -0.11 | -0.13 | -0.11 | -0.11 | -0.20 |
| RY CMa | 5 | 1 | -0.27 | … | -0.30 | 0.16 | … | … | 0.02 | -0.23 | -0.02 | -0.14 | 0.16 | 0.15 | 0.08 | 0.04 |
| RZ CMa | 7 | 2 | -0.13 | … | 0.03 | 0.19 | 0.32 | -0.12 | 0.06 | -0.19 | 0.01 | 0.04 | 0.02 | -0.17 | -0.04 | … |
| TW CMa | 5 | 1 | -0.68 | … | -0.16 | -0.03 | … | … | -0.10 | -0.38 | -0.19 | -0.25 | 0.14 | 0.10 | -0.11 | -0.23 |
| VZ CMa | 5 | 1 | -0.34 | … | -0.39 | 0.16 | … | … | 0.00 | -0.27 | -0.06 | -0.14 | 0.03 | 0.22 | 0.01 | -0.06 |
| R Cru | 8 | 1 | -0.16 | 0.34 | -0.11 | 0.24 | -0.10 | 0.13 | 0.13 | 0.16 | -0.07 | -0.01 | 0.11 | 0.04 | 0.03 | 0.01 |
| S Cru | 8 | 1 | -0.31 | 0.26 | -0.06 | 0.08 | -0.12 | -0.04 | 0.01 | 0.02 | -0.16 | -0.12 | -0.14 | -0.27 | -0.15 | -0.15 |
| T Cru | 8 | 1 | 0.00 | 0.49 | -0.03 | 0.16 | -0.06 | 0.15 | 0.10 | 0.31 | -0.03 | 0.03 | 0.14 | 0.10 | 0.10 | -0.02 |
| X Cru | 8 | 1 | -0.06 | 0.60 | 0.08 | 0.31 | 0.07 | 0.22 | 0.18 | 0.22 | 0.08 | 0.04 | 0.21 | 0.07 | 0.10 | 0.07 |
| AD Cru | 8 | 1 | -0.13 | 0.34 | -0.06 | 0.24 | 0.04 | 0.06 | 0.09 | 0.15 | -0.06 | -0.01 | -0.01 | 0.11 | 0.02 | -0.06 |
| AG Cru | 8 | 1 | -0.29 | 0.36 | -0.16 | 0.05 | -0.16 | -0.17 | -0.04 | 0.05 | -0.21 | -0.16 | -0.16 | -0.28 | -0.15 | -0.21 |
| BG Cru | 108 | 2 | -0.12 | 0.46 | 0.02 | 0.22 | 0.17 | 0.23 | 0.09 | 0.19 | -0.02 | -0.14 | 0.10 | 0.14 | -0.03 | 0.00 |
| VW Cru | 8 | 1 | -0.15 | 0.35 | -0.07 | 0.15 | 0.01 | 0.10 | 0.08 | 0.17 | -0.04 | 0.04 | 0.07 | 0.04 | 0.04 | -0.10 |
| X Cyg | 10 | 26 | -0.09 | 0.55 | 0.07 | 0.25 | 0.04 | 0.25 | 0.09 | 0.16 | 0.09 | 0.09 | 0.06 | -0.06 | 0.16 | 0.16 |
| SU Cyg | 12 | 12 | -0.29 | … | -0.29 | 0.21 | -0.16 | 0.14 | 0.04 | 0.01 | 0.03 | -0.07 | 0.03 | 0.01 | -0.02 | -0.01 |
| SZ Cyg | 7 | 1 | -0.07 | … | 0.10 | 0.44 | 0.03 | 0.22 | 0.25 | 0.11 | 0.19 | 0.09 | 0.13 | 0.05 | 0.10 | … |
| TX Cyg | 7 | 1 | 0.19 | … | 0.17 | 0.32 | … | 0.21 | 0.22 | 0.03 | 0.15 | 0.09 | 0.45 | 0.02 | 0.20 | … |
| VX Cyg | 7 | 1 | -0.30 | … | 0.17 | 0.41 | -0.16 | 0.13 | 0.18 | -0.09 | 0.20 | 0.14 | 0.11 | -0.02 | 0.10 | … |
| VY Cyg | 7 | 1 | -0.26 | … | 0.06 | 0.23 | 0.19 | 0.15 | 0.06 | -0.29 | 0.06 | 0.02 | 0.13 | -0.17 | 0.00 | … |
| VZ Cyg | 7 | 1 | -0.09 | … | 0.18 | 0.26 | 0.21 | 0.39 | 0.10 | -0.11 | 0.03 | 0.14 | 0.14 | -0.11 | 0.09 | … |
| BZ Cyg | 7 | 1 | 0.10 | … | 0.26 | 0.38 | … | 0.24 | 0.29 | -0.02 | 0.17 | -0.05 | 0.14 | -0.03 | 0.18 | … |
| CD Cyg | 10 | 16 | -0.10 | … | -0.03 | 0.27 | -0.23 | 0.18 | 0.09 | 0.23 | 0.04 | 0.05 | 0.06 | -0.01 | 0.10 | 0.20 |
| DT Cyg | 11 | 14 | -0.05 | … | 0.01 | 0.28 | -0.05 | 0.13 | 0.12 | 0.21 | 0.09 | -0.01 | 0.15 | 0.09 | 0.18 | 0.19 |
| MW Cyg | 7 | 1 | -0.09 | … | 0.14 | 0.15 | 0.08 | 0.18 | 0.06 | -0.05 | -0.01 | 0.05 | 0.07 | -0.13 | -0.03 | … |

Table 4A
Abundance Data For Cepheids: Carbon through Manganese

| Name | Ref | Sp | [C/H] | [N/H] | [O/H] | [Na/H] | [Mg/H] | [Al/H] | [Si/H] | [S/H] | [Ca/H] | [Sc/H] | [Ti/H] | [V/H] | [Cr/H] | [Mn/H] |
|---|---|---|---|---|---|---|---|---|---|---|---|---|---|---|---|---|
| V386 Cyg | 7 | 1 | 0.12 | ... | -0.06 | 0.23 | ... | 0.15 | 0.14 | -0.06 | 0.02 | 0.09 | 0.14 | -0.01 | 0.08 | ... |
| V402 Cyg | 7 | 1 | -0.18 | ... | 0.11 | 0.40 | 0.15 | 0.19 | 0.08 | -0.14 | 0.12 | 0.07 | 0.08 | -0.04 | -0.05 | ... |
| V532 Cyg | 7 | 1 | -0.10 | ... | 0.03 | 0.26 | 0.14 | 0.16 | 0.15 | -0.18 | 0.12 | -0.04 | 0.27 | -0.13 | 0.06 | ... |
| V924 Cyg | 1 | 1 | -0.30 | ... | ... | -0.04 | -0.38 | 0.09 | -0.04 | 0.05 | -0.21 | -0.37 | -0.21 | 0.01 | -0.09 | 0.10 |
| V1154 Cyg | 7 | 1 | -0.19 | ... | 0.00 | 0.11 | 0.53 | 0.09 | 0.04 | -0.34 | 0.13 | -0.02 | 0.08 | -0.26 | -0.05 | ... |
| V1334 Cyg | 11 | 11 | -0.19 | ... | -0.13 | 0.19 | -0.29 | 0.17 | 0.06 | -0.03 | -0.02 | -0.05 | 0.02 | 0.06 | 0.05 | 0.09 |
| V1726 Cyg | 1 | 1 | -0.22 | ... | ... | 0.29 | -0.12 | 0.07 | 0.11 | 0.17 | -0.16 | -0.05 | 0.15 | ... | -0.07 | 0.00 |
| TX Del | 1 | 1 | 0.06 | ... | 0.16 | 0.48 | -0.22 | ... | ... | ... | 0.16 | ... | 0.17 | 0.15 | ... | 0.38 |
| Beta Dor | 1 | 1 | -0.31 | ... | -0.08 | 0.07 | -0.30 | 0.07 | 0.00 | -0.04 | -0.18 | -0.11 | 0.01 | -0.05 | -0.04 | 0.03 |
| Zeta Gem | 11 | 11 | -0.22 | ... | -0.05 | 0.24 | -0.09 | 0.12 | 0.02 | 0.07 | -0.07 | 0.05 | -0.01 | -0.08 | 0.06 | 0.03 |
| W Gem | 9 | 8 | -0.24 | ... | -0.15 | 0.17 | -0.22 | 0.05 | 0.03 | 0.03 | -0.07 | 0.05 | 0.10 | 0.04 | 0.04 | -0.01 |
| RZ Gem | 5 | 1 | -0.32 | ... | -0.23 | -0.06 | ... | ... | -0.12 | -0.33 | -0.27 | -0.13 | -0.03 | -0.12 | 0.00 | -0.22 |
| AA Gem | 5 | 1 | 0.17 | ... | 0.07 | 0.10 | ... | ... | -0.21 | -0.28 | -0.21 | -0.15 | -0.08 | -0.31 | -0.44 | -0.38 |
| AD Gem | 5 | 1 | ... | ... | -0.36 | 0.07 | ... | ... | -0.24 | -0.42 | -0.17 | -0.06 | -0.08 | 0.11 | -0.09 | -0.25 |
| BB Gem | 5 | 1 | -0.31 | ... | -0.45 | 0.13 | ... | ... | -0.07 | -0.37 | -0.31 | -0.10 | 0.24 | 0.12 | 0.06 | -0.20 |
| DX Gem | 5 | 1 | -0.42 | ... | -0.28 | 0.17 | ... | ... | 0.01 | -0.19 | -0.05 | -0.03 | 0.02 | -0.03 | 0.10 | -0.11 |
| BB Her | 1 | 4 | -0.10 | ... | 0.04 | 0.40 | -0.01 | 0.23 | 0.15 | ... | -0.01 | 0.07 | 0.09 | 0.07 | 0.10 | 0.30 |
| V Lac | 7 | 1 | -0.06 | ... | 0.17 | 0.19 | -0.16 | 0.16 | 0.15 | -0.13 | 0.10 | -0.03 | -0.02 | -0.07 | 0.04 | ... |
| X Lac | 7 | 1 | -0.02 | ... | 0.10 | 0.21 | 0.07 | 0.09 | 0.10 | -0.15 | 0.06 | -0.03 | 0.18 | -0.22 | 0.02 | ... |
| Y Lac | 12 | 9 | -0.23 | ... | -0.26 | 0.13 | -0.15 | 0.17 | 0.05 | 0.00 | -0.04 | -0.25 | 0.01 | 0.04 | 0.00 | -0.04 |
| Z Lac | 10 | 9 | -0.22 | ... | -0.11 | 0.23 | -0.21 | 0.11 | 0.04 | 0.12 | -0.08 | -0.10 | -0.01 | -0.10 | 0.08 | 0.06 |
| RR Lac | 7 | 1 | -0.04 | ... | 0.26 | 0.16 | ... | 0.24 | 0.12 | -0.14 | 0.19 | 0.09 | 0.29 | -0.10 | 0.18 | ... |
| BG Lac | 1 | 3 | -0.17 | ... | 0.10 | 0.17 | -0.25 | 0.08 | 0.04 | 0.09 | -0.01 | -0.11 | 0.03 | -0.05 | 0.06 | 0.04 |
| GH Lup | 8 | 1 | -0.02 | 0.32 | -0.04 | 0.23 | -0.07 | 0.19 | 0.08 | 0.30 | -0.03 | 0.00 | 0.04 | -0.05 | 0.06 | 0.00 |
| V473 Lyr | 1 | 2 | -0.34 | ... | -0.24 | 0.01 | -0.14 | -0.06 | -0.03 | 0.09 | -0.10 | -0.05 | -0.06 | -0.04 | -0.08 | -0.22 |
| T Mon | 10 | 20 | -0.23 | 0.59 | 0.04 | 0.41 | 0.17 | 0.22 | 0.15 | 0.39 | 0.08 | 0.01 | 0.14 | 0.02 | 0.23 | 0.16 |
| SV Mon | 10 | 9 | -0.90 | 0.22 | -0.28 | 0.28 | -0.05 | 0.07 | 0.00 | -0.02 | -0.12 | -0.01 | -0.06 | -0.13 | 0.02 | -0.04 |
| TW Mon | 4 | 1 | -0.50 | 0.00 | -0.35 | 0.01 | -0.41 | -0.10 | -0.17 | -0.15 | -0.16 | -0.29 | -0.21 | -0.27 | -0.16 | -0.40 |
| TX Mon | 3 | 1 | -0.41 | 0.29 | -0.16 | -0.08 | -0.26 | -0.07 | -0.09 | -0.17 | -0.24 | -0.16 | -0.17 | -0.28 | -0.16 | -0.25 |
| TZ Mon | 304 | 2 | -0.44 | 0.31 | 0.16 | 0.10 | -0.16 | 0.05 | -0.03 | 0.06 | -0.03 | -0.18 | -0.08 | -0.15 | -0.08 | 0.09 |
| UY Mon | 5 | 1 | -0.28 | ... | -0.24 | 0.13 | ... | ... | -0.08 | -0.22 | -0.07 | -0.04 | 0.19 | 0.09 | 0.01 | -0.19 |

Table 4A
Abundance Data For Cepheids: Carbon through Manganese

| Name | Ref | Sp | [C/H] | [N/H] | [O/H] | [Na/H] | [Mg/H] | [Al/H] | [Si/H] | [S/H] | [Ca/H] | [Sc/H] | [Ti/H] | [V/H] | [Cr/H] | [Mn/H] |
|---|---|---|---|---|---|---|---|---|---|---|---|---|---|---|---|---|
| WW Mon | 4 | 1 | -0.84 | -0.38 | ... | -0.04 | -0.29 | -0.27 | -0.26 | -0.39 | -0.26 | -0.36 | -0.10 | -0.40 | ... | -0.46 |
| XX Mon | 304 | 2 | -0.50 | 0.19 | -0.02 | 0.02 | -0.20 | 0.14 | -0.11 | -0.18 | -0.13 | -0.09 | -0.19 | -0.10 | -0.13 | -0.24 |
| AA Mon | 3 | 1 | -0.49 | 0.35 | -0.19 | 0.04 | -0.22 | -0.14 | -0.12 | -0.12 | -0.26 | -0.21 | -0.23 | 0.02 | -0.26 | -0.29 |
| AC Mon | 103 | 2 | -0.47 | 0.23 | -0.25 | 0.14 | -0.40 | -0.36 | -0.05 | -0.18 | -0.11 | -0.14 | -0.04 | -0.23 | -0.16 | -0.22 |
| CU Mon | 4 | 1 | -0.52 | 0.29 | -0.06 | 0.23 | -0.36 | -0.05 | -0.18 | -0.36 | -0.24 | 0.02 | -0.06 | -0.26 | -0.06 | -0.31 |
| CV Mon | 1 | 1 | -0.25 | ... | 0.02 | 0.03 | -0.32 | -0.05 | 0.01 | 0.08 | -0.19 | -0.14 | 0.10 | 0.30 | 0.04 | -0.05 |
| EE Mon | 3 | 1 | -0.48 | 0.14 | ... | -0.20 | -0.60 | ... | -0.35 | -0.56 | -0.50 | -0.42 | -0.26 | -0.41 | -0.35 | -0.66 |
| EK Mon | 3 | 1 | -0.42 | 0.26 | -0.28 | 0.04 | -0.18 | -0.04 | -0.05 | -0.15 | -0.10 | -0.05 | -0.18 | -0.10 | -0.15 | -0.19 |
| FG Mon | 4 | 1 | -0.48 | 0.26 | -0.46 | 0.09 | -0.30 | -0.01 | -0.10 | -0.13 | -0.03 | -0.23 | -0.19 | -0.22 | -0.11 | -0.39 |
| FI Mon | 4 | 1 | -0.56 | ... | -0.41 | 0.08 | -0.03 | -0.03 | -0.07 | -0.14 | -0.04 | -0.11 | -0.06 | -0.12 | -0.13 | -0.24 |
| V465 Mon | 7 | 1 | 0.10 | ... | 0.22 | 0.11 | ... | ... | 0.10 | -0.21 | -0.10 | 0.00 | 0.26 | -0.15 | -0.11 | ... |
| V495 Mon | 3 | 1 | -0.77 | -0.12 | -0.09 | 0.00 | -0.28 | -0.25 | -0.18 | -0.07 | -0.29 | -0.34 | -0.29 | -0.28 | -0.32 | -0.39 |
| V504 Mon | 3 | 1 | -0.53 | 0.14 | -0.35 | -0.08 | -0.45 | -0.13 | -0.17 | -0.23 | -0.21 | -0.33 | -0.28 | -0.11 | -0.30 | -0.38 |
| V508 Mon | 3 | 1 | -0.77 | 0.07 | -0.54 | -0.01 | -0.28 | -0.15 | -0.18 | -0.19 | -0.22 | -0.21 | -0.13 | -0.22 | -0.22 | -0.41 |
| V510 Mon | 4 | 1 | -0.54 | 0.01 | -0.14 | -0.04 | -0.27 | 0.11 | -0.10 | -0.29 | -0.04 | -0.34 | -0.07 | -0.21 | -0.24 | -0.33 |
| V526 Mon | 1 | 1 | -0.28 | 0.44 | -0.52 | 0.11 | -0.09 | ... | -0.01 | 0.05 | -0.08 | -0.20 | -0.04 | -0.18 | -0.02 | -0.14 |
| R Mus | 8 | 1 | -0.10 | 0.53 | -0.02 | 0.26 | -0.01 | 0.20 | 0.13 | 0.25 | 0.05 | 0.07 | 0.04 | 0.07 | 0.05 | 0.01 |
| S Mus | 8 | 1 | -0.19 | 0.29 | -0.19 | 0.19 | -0.07 | 0.11 | 0.02 | 0.14 | -0.10 | -0.04 | 0.00 | -0.02 | -0.06 | -0.15 |
| RT Mus | 8 | 1 | -0.22 | 0.38 | 0.02 | 0.18 | -0.19 | 0.03 | 0.05 | 0.08 | -0.11 | -0.01 | -0.01 | 0.02 | -0.03 | -0.15 |
| TZ Mus | 8 | 1 | -0.14 | 0.46 | -0.16 | 0.14 | -0.09 | 0.08 | 0.09 | 0.18 | 0.01 | -0.04 | 0.02 | -0.01 | -0.02 | -0.06 |
| UU Mus | 8 | 1 | -0.15 | 0.43 | -0.06 | 0.24 | -0.07 | 0.15 | 0.13 | 0.16 | 0.03 | -0.02 | 0.03 | 0.15 | 0.00 | -0.03 |
| S Nor | 108 | 3 | -0.18 | 0.60 | -0.15 | 0.26 | -0.18 | 0.14 | 0.06 | 0.14 | -0.06 | 0.03 | 0.05 | 0.03 | -0.06 | -0.09 |
| U Nor | 8 | 1 | 0.10 | 0.72 | 0.04 | 0.36 | 0.04 | 0.18 | 0.15 | 0.39 | 0.00 | 0.02 | 0.05 | 0.05 | 0.04 | 0.18 |
| SY Nor | 8 | 1 | 0.24 | 0.69 | 0.21 | 0.41 | 0.16 | 0.29 | 0.31 | 0.50 | 0.13 | 0.37 | 0.19 | 0.24 | 0.20 | 0.22 |
| TW Nor | 8 | 1 | 0.06 | 0.62 | 0.28 | 0.46 | 0.20 | 0.35 | 0.31 | 0.46 | 0.15 | ... | 0.33 | 0.27 | 0.21 | 0.26 |
| GU Nor | 8 | 1 | -0.03 | 0.57 | 0.15 | 0.40 | 0.14 | 0.19 | 0.18 | 0.37 | 0.13 | 0.13 | 0.14 | 0.12 | 0.14 | 0.11 |
| V340 Nor | 108 | 2 | -0.03 | 0.54 | 0.07 | 0.31 | -0.15 | 0.08 | 0.06 | 0.23 | -0.10 | -0.03 | 0.06 | 0.04 | 0.00 | 0.01 |
| Y Oph | 11 | 14 | -0.10 | ... | 0.00 | 0.07 | -0.27 | 0.14 | 0.02 | 0.09 | -0.07 | -0.08 | 0.05 | 0.02 | 0.08 | 0.04 |
| BF Oph | 708 | 2 | -0.05 | 0.40 | -0.05 | 0.20 | 0.03 | 0.06 | 0.11 | 0.02 | -0.01 | 0.01 | 0.06 | -0.07 | 0.02 | -0.07 |
| RS Ori | 104 | 5 | -0.45 | 0.29 | -0.18 | 0.06 | -0.29 | 0.03 | 0.02 | 0.00 | -0.06 | -0.18 | 0.07 | -0.12 | -0.08 | -0.11 |
| CS Ori | 3 | 1 | -0.58 | -0.12 | -0.61 | 0.05 | -0.40 | -0.31 | -0.16 | -0.32 | -0.20 | -0.11 | -0.23 | ... | -0.24 | -0.43 |

Table 4A
Abundance Data For Cepheids: Carbon through Manganese

| Name | Ref | Sp | [C/H] | [N/H] | [O/H] | [Na/H] | [Mg/H] | [Al/H] | [Si/H] | [S/H] | [Ca/H] | [Sc/H] | [Ti/H] | [V/H] | [Cr/H] | [Mn/H] |
|---|---|---|---|---|---|---|---|---|---|---|---|---|---|---|---|---|
| GQ Ori | 107 | 2 | -0.20 | 0.29 | -0.03 | 0.20 | 0.21 | 0.17 | 0.05 | -0.07 | -0.14 | 0.14 | 0.13 | -0.02 | 0.00 | -0.14 |
| SV Per | 7 | 1 | -0.04 | ... | 0.15 | 0.25 | 0.16 | 0.09 | 0.04 | -0.19 | -0.05 | -0.01 | 0.13 | -0.13 | -0.13 | ... |
| UX Per | 6 | 1 | -0.42 | 0.24 | -0.38 | -0.05 | -0.40 | -0.45 | -0.14 | -0.07 | -0.15 | -0.47 | -0.22 | -0.33 | -0.25 | -0.14 |
| VX Per | 10 | 9 | -0.22 | ... | -0.18 | 0.15 | -0.28 | 0.07 | -0.01 | 0.03 | -0.15 | -0.17 | -0.08 | -0.15 | 0.01 | -0.07 |
| AS Per | 7 | 1 | -0.04 | ... | 0.06 | 0.31 | 0.10 | 0.03 | 0.14 | -0.05 | 0.06 | 0.06 | 0.14 | -0.09 | 0.10 | ... |
| AW Per | 1 | 4 | -0.23 | ... | -0.03 | 0.24 | -0.27 | 0.07 | 0.06 | 0.18 | -0.03 | -0.15 | 0.01 | 0.15 | 0.27 | 0.10 |
| BM Per | 607 | 5 | -0.21 | 0.43 | -0.04 | 0.20 | 0.04 | 0.06 | 0.10 | 0.07 | 0.06 | 0.11 | 0.11 | -0.08 | 0.02 | -0.02 |
| HQ Per | 6 | 2 | -0.68 | -0.02 | 0.13 | 0.05 | -0.30 | -0.24 | -0.21 | -0.39 | -0.35 | -0.30 | -0.33 | -0.29 | -0.30 | -0.48 |
| MM Per | 5 | 1 | -0.26 | ... | -0.08 | 0.07 | ... | ... | -0.07 | -0.04 | -0.05 | -0.09 | 0.00 | -0.17 | -0.08 | -0.10 |
| V440 Per | 11 | 10 | -0.27 | ... | -0.12 | 0.07 | -0.39 | 0.02 | -0.02 | -0.05 | -0.18 | -0.08 | -0.02 | -0.04 | 0.00 | -0.12 |
| X Pup | 10 | 7 | -0.21 | ... | -0.22 | 0.17 | -0.25 | 0.10 | -0.03 | 0.02 | -0.09 | -0.12 | -0.05 | -0.17 | 0.03 | -0.08 |
| RS Pup | 4 | 2 | 0.03 | 0.73 | 0.10 | 0.42 | ... | 0.17 | 0.11 | 0.31 | 0.21 | 0.18 | 0.03 | 0.00 | -0.02 | 0.13 |
| VW Pup | 3 | 1 | -0.48 | -0.12 | 0.19 | 0.02 | -0.31 | -0.21 | -0.16 | -0.29 | -0.23 | -0.19 | -0.28 | -0.31 | -0.07 | -0.37 |
| VX Pup | 1 | 1 | ... | ... | ... | 0.08 | ... | ... | -0.06 | ... | -0.33 | -0.13 | -0.01 | -0.02 | -0.16 | -0.10 |
| VZ Pup | 3 | 1 | -0.83 | 0.02 | -0.10 | 0.00 | -0.15 | -0.15 | -0.09 | -0.22 | -0.12 | -0.16 | -0.26 | -0.01 | -0.18 | -0.28 |
| WW Pup | 3 | 1 | -0.64 | -0.16 | ... | 0.01 | -0.33 | -0.12 | -0.14 | -0.20 | -0.22 | -0.31 | -0.25 | -0.19 | -0.20 | -0.31 |
| AD Pup | 3 | 1 | -0.77 | 0.49 | ... | 0.06 | ... | 0.00 | -0.19 | -0.13 | -0.17 | -0.04 | -0.22 | -0.32 | -0.38 | -0.33 |
| AP Pup | 8 | 1 | -0.13 | 0.43 | -0.25 | 0.18 | -0.02 | 0.10 | 0.09 | 0.21 | 0.03 | 0.04 | 0.07 | 0.03 | 0.02 | -0.02 |
| AQ Pup | 4 | 2 | -0.51 | -0.07 | ... | 0.09 | ... | 0.01 | -0.10 | -0.20 | -0.07 | -0.05 | -0.26 | -0.21 | -0.17 | -0.43 |
| AT Pup | 8 | 1 | -0.28 | 0.26 | -0.31 | 0.02 | -0.20 | -0.10 | -0.01 | -0.05 | -0.22 | -0.07 | -0.11 | -0.27 | -0.15 | -0.26 |
| BC Pup | 4 | 2 | -0.65 | -0.04 | -0.55 | 0.01 | -0.30 | -0.05 | -0.15 | -0.27 | -0.15 | -0.15 | -0.10 | -0.14 | -0.19 | -0.19 |
| BN Pup | 3 | 1 | -0.45 | 0.55 | -0.06 | 0.09 | 0.21 | 0.04 | -0.01 | 0.19 | 0.04 | -0.02 | -0.05 | -0.15 | 0.07 | -0.02 |
| CE Pup | 8 | 1 | -0.81 | 0.57 | -0.32 | 0.24 | -0.07 | 0.09 | 0.02 | 0.04 | -0.11 | 0.03 | -0.05 | -0.09 | -0.11 | -0.15 |
| HW Pup | 304 | 2 | -0.51 | 0.08 | -0.24 | -0.10 | -0.49 | -0.10 | -0.16 | -0.24 | -0.25 | -0.25 | -0.18 | -0.23 | -0.27 | -0.31 |
| MY Pup | 108 | 2 | -0.32 | 0.23 | -0.16 | 0.11 | -0.29 | 0.00 | -0.05 | -0.06 | -0.17 | -0.12 | -0.09 | -0.06 | -0.19 | -0.24 |
| NT Pup | 8 | 1 | -1.53 | 0.62 | -0.39 | 0.41 | -0.39 | -0.06 | -0.11 | -0.07 | -0.20 | -0.17 | -0.10 | -0.21 | -0.21 | -0.31 |
| V335 Pup | 7 | 1 | 0.06 | ... | 0.13 | 0.13 | 0.11 | 0.11 | 0.07 | -0.22 | -0.14 | 0.00 | 0.13 | -0.06 | -0.04 | ... |
| RV Sco | 708 | 2 | -0.19 | 0.43 | -0.02 | 0.16 | 0.02 | 0.16 | 0.12 | 0.03 | 0.05 | 0.00 | 0.06 | -0.04 | 0.10 | -0.01 |
| RY Sco | 7 | 1 | 0.04 | ... | 0.06 | 0.44 | 0.60 | 0.28 | 0.13 | 0.07 | 0.03 | 0.28 | 0.09 | -0.01 | 0.14 | ... |
| KQ Sco | 2 | 1 | -0.02 | ... | 0.21 | ... | ... | 0.42 | 0.21 | 0.33 | 0.39 | ... | 0.20 | 0.19 | 0.29 | -0.09 |
| V482 Sco | 8 | 1 | -0.10 | 0.51 | -0.05 | 0.24 | 0.03 | 0.14 | 0.11 | 0.24 | -0.06 | 0.12 | 0.04 | 0.10 | 0.05 | -0.04 |

Table 4A
Abundance Data For Cepheids: Carbon through Manganese

| Name | Ref | Sp | [C/H] | [N/H] | [O/H] | [Na/H] | [Mg/H] | [Al/H] | [Si/H] | [S/H] | [Ca/H] | [Sc/H] | [Ti/H] | [V/H] | [Cr/H] | [Mn/H] |
|---|---|---|---|---|---|---|---|---|---|---|---|---|---|---|---|---|
| V500 Sco | 9 | 5 | -0.15 | ... | -0.12 | 0.14 | -0.17 | 0.08 | 0.04 | 0.02 | -0.04 | -0.05 | 0.03 | 0.05 | 0.02 | 0.00 |
| V636 Sco | 8 | 1 | -0.05 | 0.37 | -0.08 | 0.20 | -0.20 | 0.15 | 0.09 | 0.14 | -0.04 | -0.09 | 0.00 | -0.05 | 0.00 | 0.00 |
| V950 Sco | 8 | 1 | -0.12 | 0.54 | -0.05 | 0.25 | -0.05 | 0.14 | 0.14 | 0.28 | -0.04 | 0.07 | 0.12 | 0.04 | 0.03 | -0.02 |
| Z Sct | 7 | 1 | 0.12 | ... | 0.16 | 0.58 | ... | 0.52 | 0.31 | 0.23 | 0.16 | 0.57 | 0.36 | 0.26 | 0.29 | ... |
| SS Sct | 7 | 1 | -0.32 | ... | -0.04 | 0.22 | ... | 0.24 | 0.21 | -0.12 | 0.18 | 0.02 | 0.05 | -0.23 | 0.07 | ... |
| UZ Sct | 2 | 1 | 0.04 | 0.70 | 0.49 | 0.76 | ... | 0.44 | 0.33 | 0.63 | 0.35 | 0.36 | 0.25 | 0.23 | 0.25 | 0.11 |
| EW Sct | 1 | 3 | -0.07 | ... | -0.04 | 0.07 | -0.10 | 0.15 | 0.08 | 0.15 | -0.01 | -0.09 | 0.09 | 0.08 | 0.05 | 0.08 |
| V367 Sct | 1 | 1 | -0.38 | ... | ... | 0.21 | -0.48 | 0.21 | 0.05 | 0.15 | -0.15 | 0.07 | 0.24 | 0.06 | -0.05 | -0.13 |
| BQ Ser | 1 | 3 | -0.15 | ... | -0.13 | 0.12 | -0.14 | 0.14 | 0.07 | 0.13 | -0.05 | -0.17 | 0.03 | -0.01 | 0.04 | -0.04 |
| S Sge | 9 | 9 | -0.11 | ... | -0.06 | 0.20 | -0.17 | 0.11 | 0.10 | 0.15 | -0.03 | -0.04 | 0.11 | 0.11 | 0.11 | 0.14 |
| U Sgr | 9 | 21 | -0.11 | ... | 0.03 | 0.21 | -0.17 | 0.21 | 0.08 | 0.11 | -0.04 | -0.16 | 0.06 | 0.06 | 0.08 | 0.12 |
| W Sgr | 9 | 8 | -0.18 | ... | -0.14 | 0.18 | -0.25 | 0.07 | 0.05 | 0.08 | -0.07 | -0.12 | 0.08 | 0.06 | 0.03 | 0.04 |
| Y Sgr | 12 | 12 | -0.16 | ... | -0.18 | 0.27 | -0.08 | 0.23 | 0.09 | 0.11 | 0.05 | -0.16 | 0.12 | 0.05 | 0.12 | 0.10 |
| VY Sgr | 2 | 1 | 0.01 | 0.50 | 0.18 | 0.66 | ... | 0.42 | 0.30 | 0.56 | 0.23 | 0.31 | 0.24 | 0.20 | 0.22 | 0.22 |
| WZ Sgr | 10 | 12 | 0.01 | 0.42 | 0.00 | 0.39 | -0.06 | 0.33 | 0.16 | 0.49 | 0.03 | -0.06 | 0.18 | 0.08 | 0.12 | 0.20 |
| XX Sgr | 7 | 1 | -0.21 | ... | 0.07 | 0.27 | 0.34 | 0.34 | 0.11 | -0.15 | 0.04 | 0.04 | 0.11 | -0.02 | 0.10 | ... |
| YZ Sgr | 9 | 8 | -0.10 | ... | -0.12 | 0.25 | -0.17 | 0.17 | 0.11 | 0.14 | -0.02 | 0.06 | 0.11 | 0.08 | 0.10 | 0.11 |
| AP Sgr | 7 | 1 | -0.05 | ... | -0.23 | 0.47 | 0.25 | 0.08 | 0.27 | -0.04 | 0.24 | 0.04 | 0.16 | -0.04 | 0.19 | ... |
| AV Sgr | 2 | 1 | 0.21 | 0.82 | 0.36 | 0.58 | -0.12 | 0.68 | 0.35 | 0.39 | 0.13 | 0.32 | 0.38 | 0.37 | 0.29 | 0.12 |
| BB Sgr | 7 | 1 | -0.06 | ... | -0.13 | 0.36 | ... | 0.15 | 0.14 | -0.17 | 0.12 | 0.08 | 0.22 | -0.10 | 0.09 | ... |
| V350 Sgr | 7 | 1 | 0.12 | ... | 0.23 | 0.23 | 0.19 | 0.32 | 0.13 | -0.04 | 0.08 | 0.17 | 0.20 | 0.01 | 0.02 | ... |
| ST Tau | 1 | 3 | -0.27 | ... | -0.29 | 0.23 | -0.18 | 0.03 | 0.03 | 0.04 | -0.04 | -0.11 | 0.10 | -0.07 | -0.04 | -0.02 |
| SZ Tau | 11 | 16 | -0.17 | ... | -0.03 | 0.27 | -0.18 | 0.10 | 0.07 | 0.03 | -0.02 | -0.07 | 0.02 | 0.04 | 0.14 | 0.12 |
| AE Tau | 3 | 1 | -0.48 | 0.07 | -0.17 | -0.10 | -0.36 | -0.30 | -0.18 | -0.33 | -0.25 | -0.08 | -0.06 | -0.21 | -0.12 | -0.41 |
| EF Tau | 6 | 1 | -0.78 | 0.36 | -0.21 | -0.48 | -0.74 | -0.46 | -0.65 | -0.45 | -0.62 | -0.74 | -0.61 | -0.51 | -0.72 | -1.05 |
| EU Tau | 1 | 2 | -0.24 | ... | -0.05 | 0.24 | -0.28 | -0.01 | 0.04 | 0.09 | -0.05 | -0.07 | 0.03 | -0.05 | -0.02 | -0.05 |
| R TrA | 8 | 1 | -0.23 | 0.36 | -0.08 | 0.26 | 0.08 | 0.14 | 0.18 | 0.29 | 0.04 | 0.04 | 0.10 | 0.16 | 0.07 | 0.03 |
| S TrA | 8 | 1 | -0.08 | 0.51 | -0.13 | 0.28 | -0.01 | 0.15 | 0.12 | 0.28 | 0.04 | 0.12 | 0.12 | 0.09 | 0.07 | 0.04 |
| LR TrA | 8 | 1 | 0.07 | 0.41 | 0.03 | 0.28 | 0.27 | 0.22 | 0.19 | 0.34 | 0.17 | 0.11 | 0.24 | 0.18 | 0.21 | 0.11 |
| T Vel | 408 | 3 | -0.26 | 0.21 | -0.01 | 0.06 | -0.19 | 0.07 | 0.03 | 0.00 | -0.03 | -0.09 | 0.00 | -0.11 | 0.01 | -0.16 |
| V Vel | 8 | 1 | -0.42 | -0.04 | -0.35 | -0.03 | -0.28 | -0.14 | -0.15 | -0.09 | -0.28 | -0.28 | -0.26 | -0.29 | -0.30 | -0.27 |

Table 4A
Abundance Data For Cepheids: Carbon through Manganese

| Name | Ref | Sp | [C/H] | [N/H] | [O/H] | [Na/H] | [Mg/H] | [Al/H] | [Si/H] | [S/H] | [Ca/H] | [Sc/H] | [Ti/H] | [V/H] | [Cr/H] | [Mn/H] |
|---|---|---|---|---|---|---|---|---|---|---|---|---|---|---|---|---|
| RY Vel | 408 | 3 | -0.15 | 0.53 | -0.17 | 0.20 | -0.16 | 0.13 | 0.07 | 0.14 | -0.11 | 0.03 | 0.00 | 0.02 | -0.04 | 0.04 |
| RZ Vel | 408 | 3 | -0.25 | 0.52 | -0.15 | 0.14 | -0.10 | 0.09 | 0.07 | -0.01 | -0.06 | -0.13 | 0.05 | -0.02 | -0.08 | -0.09 |
| ST Vel | 8 | 1 | -0.15 | 0.29 | -0.33 | 0.12 | -0.03 | 0.04 | 0.07 | 0.10 | -0.04 | -0.04 | -0.04 | -0.07 | 0.00 | -0.10 |
| SV Vel | 8 | 1 | -0.16 | 0.50 | -0.16 | 0.25 | -0.04 | 0.07 | 0.09 | 0.15 | -0.11 | -0.01 | 0.02 | 0.14 | -0.04 | -0.06 |
| SW Vel | 108 | 4 | -0.38 | 0.43 | -0.01 | 0.12 | 0.05 | 0.10 | -0.04 | 0.05 | -0.08 | 0.02 | -0.08 | -0.15 | -0.07 | -0.16 |
| SX Vel | 408 | 3 | -0.22 | 0.25 | 0.02 | 0.07 | -0.18 | 0.05 | 0.02 | 0.06 | -0.06 | -0.02 | -0.06 | -0.08 | -0.03 | -0.12 |
| XX Vel | 8 | 1 | -0.24 | 0.34 | -0.29 | 0.11 | -0.17 | -0.04 | -0.01 | 0.07 | -0.17 | -0.02 | -0.07 | -0.17 | -0.08 | -0.18 |
| AE Vel | 8 | 1 | -0.14 | 0.48 | -0.03 | 0.13 | -0.06 | 0.12 | 0.05 | 0.15 | -0.09 | -0.03 | -0.03 | -0.01 | -0.05 | -0.10 |
| AH Vel | 8 | 1 | -0.21 | 0.70 | -0.05 | 0.28 | -0.13 | 0.17 | 0.10 | 0.28 | -0.05 | 0.08 | 0.11 | 0.14 | 0.01 | -0.06 |
| BG Vel | 8 | 1 | -0.18 | 0.38 | 0.02 | 0.18 | -0.04 | 0.08 | 0.03 | 0.07 | -0.06 | -0.14 | 0.01 | -0.07 | -0.10 | -0.13 |
| CS Vel | 8 | 1 | -0.02 | 0.23 | -0.01 | 0.13 | 0.01 | 0.32 | 0.11 | 0.23 | 0.00 | -0.11 | 0.08 | -0.04 | 0.03 | 0.02 |
| CX Vel | 8 | 1 | -0.16 | 0.43 | -0.30 | 0.13 | -0.05 | 0.04 | 0.08 | 0.18 | -0.09 | 0.01 | 0.01 | 0.03 | 0.03 | -0.07 |
| DK Vel | 8 | 1 | -0.15 | 0.50 | 0.03 | 0.15 | 0.00 | 0.09 | 0.08 | 0.20 | -0.02 | -0.03 | -0.05 | -0.06 | -0.05 | -0.09 |
| DR Vel | 8 | 1 | -0.09 | 0.62 | -0.09 | 0.15 | 0.08 | 0.18 | 0.09 | 0.24 | 0.01 | 0.04 | 0.09 | 0.03 | 0.03 | 0.02 |
| EX Vel | 8 | 1 | -0.17 | 0.43 | -0.11 | 0.11 | 0.02 | 0.08 | 0.01 | 0.15 | -0.11 | -0.02 | 0.04 | 0.02 | -0.01 | -0.15 |
| FG Vel | 8 | 1 | -0.35 | 0.25 | -0.06 | 0.16 | -0.04 | 0.04 | 0.00 | 0.06 | -0.11 | -0.14 | -0.06 | -0.09 | -0.09 | -0.20 |
| FN Vel | 8 | 1 | -0.14 | 0.47 | -0.17 | 0.14 | 0.02 | 0.10 | 0.09 | 0.23 | -0.04 | -0.01 | 0.00 | 0.03 | 0.05 | -0.04 |
| S Vul | 10 | 4 | -0.31 | … | -0.20 | 0.22 | … | 0.22 | -0.03 | 0.12 | 0.04 | … | -0.14 | -0.12 | 0.05 | 0.06 |
| T Vul | 12 | 12 | -0.25 | … | -0.09 | 0.15 | -0.13 | 0.10 | 0.04 | 0.04 | -0.01 | -0.09 | 0.05 | 0.01 | -0.01 | 0.00 |
| U Vul | 9 | 7 | -0.11 | … | -0.04 | 0.19 | -0.14 | 0.14 | 0.11 | 0.13 | 0.00 | -0.04 | 0.13 | 0.11 | 0.13 | 0.08 |
| X Vul | 9 | 6 | -0.11 | … | -0.03 | 0.17 | -0.20 | 0.14 | 0.09 | 0.16 | -0.02 | 0.02 | 0.11 | 0.10 | 0.10 | 0.08 |
| SV Vul | 10 | 23 | -0.02 | 0.17 | -0.01 | 0.05 | -0.17 | 0.12 | 0.04 | 0.16 | -0.04 | … | -0.02 | -0.13 | 0.09 | 0.10 |

Notes:
Reference Codes:    1 – Andrievsky et al. (2002a (Paper I)), 2 - Andrievsky et al. (2002b (Paper II)), 3 – Andrievsky et al. (2002c (Paper III)), 4 – Luck et al. (2003 (Paper IV)), 5 – Andrievsky et al. (2004 (Paper V)),  6 – Kovtyukh, Wallerstein, & Andrievsky (2005), 7 – Luck, Kovtyukh, & Andrievsky (2006 (Paper VI), 8 – This paper, 9 – Luck & Andrievsky (2004), 10 – Kovtyukh et al. (2005), 11- Luck et al. (2008) and 12 -  Andrievsky, Luck, & Kovtyukh (2005). For references greater than 100 the reference code is (100 x First Reference) + Second Reference (that is, 408 = references 4 and 8).
Sp:    Total number of spectra (phases) contributing to the abundances.

Table 4B
Abundance Data For Cepheids: Iron through Gadolinium

| Name | Ref | Sp | [Fe/H] | [Co/H] | [Ni/H] | [Cu/H] | [Zn/H] | [Y/H] | [Zr/H] | [La/H] | [Ce/H] | [Pr/H] | [Nd/H] | [Sm/H] | [Eu/H] | [Gd/H] |
|---|---|---|---|---|---|---|---|---|---|---|---|---|---|---|---|---|
| T Ant | 8 | 1 | -0.24 | -0.43 | -0.27 | -0.08 | -0.23 | -0.09 | -0.22 | -0.01 | -0.01 | -0.28 | -0.06 | -0.08 | 0.11 | ... |
| Eta Aql | 9 | 14 | 0.08 | -0.07 | 0.09 | 0.20 | 0.34 | 0.27 | 0.00 | 0.24 | -0.07 | ... | 0.16 | ... | 0.09 | 0.06 |
| SZ Aql | 10 | 11 | 0.17 | -0.02 | 0.12 | 0.06 | 0.41 | 0.37 | 0.00 | 0.27 | -0.09 | ... | 0.15 | ... | 0.14 | 0.13 |
| TT Aql | 10 | 8 | 0.10 | -0.07 | 0.07 | 0.02 | 0.36 | 0.31 | -0.06 | 0.18 | -0.10 | ... | 0.13 | ... | 0.05 | 0.05 |
| FF Aql | 11 | 14 | 0.04 | -0.07 | 0.05 | 0.28 | 0.28 | 0.27 | -0.10 | 0.23 | -0.14 | ... | 0.12 | ... | 0.14 | 0.14 |
| FM Aql | 1 | 2 | 0.08 | 0.02 | 0.09 | 0.26 | ... | 0.16 | -0.01 | 0.23 | -0.17 | ... | 0.14 | ... | 0.09 | ... |
| FN Aql | 1 | 4 | -0.02 | -0.14 | -0.05 | 0.09 | 0.15 | 0.16 | -0.07 | 0.23 | -0.09 | ... | 0.12 | ... | 0.08 | 0.12 |
| V496 Aql | 1 | 2 | 0.05 | -0.09 | 0.03 | ... | 0.33 | 0.14 | -0.10 | 0.09 | -0.23 | ... | 0.02 | ... | -0.01 | -0.09 |
| V600 Aql | 7 | 1 | 0.03 | -0.08 | -0.09 | 0.20 | ... | 0.16 | ... | 0.15 | -0.05 | ... | 0.03 | ... | 0.00 | ... |
| V733 Aql | 7 | 1 | 0.08 | -0.03 | -0.06 | -0.08 | ... | 0.16 | ... | 0.21 | -0.13 | ... | 0.02 | ... | -0.08 | ... |
| V1162 Aql | 1 | 2 | 0.01 | -0.15 | -0.02 | 0.28 | 0.14 | 0.21 | -0.21 | 0.09 | -0.24 | ... | 0.02 | ... | -0.06 | -0.23 |
| V1359 Aql | 7 | 1 | 0.09 | 0.18 | 0.08 | ... | ... | 0.17 | ... | 0.21 | 0.14 | ... | 0.16 | ... | 0.28 | ... |
| V340 Ara | 2 | 1 | 0.31 | 0.15 | 0.39 | 0.29 | 0.46 | 0.34 | 0.20 | 0.19 | -0.01 | ... | 0.05 | ... | 0.29 | 0.21 |
| Y Aur | 7 | 1 | -0.23 | ... | -0.34 | ... | ... | -0.39 | ... | -0.22 | -0.27 | ... | -0.07 | ... | -0.15 | ... |
| RT Aur | 12 | 10 | 0.06 | -0.11 | 0.06 | 0.17 | 0.22 | 0.24 | -0.02 | 0.20 | -0.14 | ... | 0.09 | ... | 0.06 | 0.06 |
| RX Aur | 10 | 16 | -0.01 | -0.14 | -0.05 | 0.11 | 0.29 | 0.17 | -0.04 | 0.30 | -0.13 | -0.40 | 0.16 | ... | 0.16 | 0.18 |
| SY Aur | 6 | 1 | -0.02 | -0.11 | 0.00 | 0.30 | 0.24 | 0.18 | -0.15 | 0.30 | 0.08 | 0.01 | 0.04 | ... | 0.02 | 0.08 |
| YZ Aur | 607 | 3 | -0.36 | -0.41 | -0.39 | -0.46 | ... | -0.30 | -0.32 | 0.10 | -0.38 | -0.36 | -0.22 | ... | -0.17 | -0.09 |
| AN Aur | 607 | 4 | -0.15 | -0.02 | -0.16 | 0.14 | 0.08 | 0.06 | -0.18 | 0.20 | -0.05 | -0.34 | 0.10 | ... | -0.01 | 0.14 |
| AO Aur | 5 | 1 | -0.14 | 0.06 | -0.19 | ... | ... | ... | ... | 0.16 | 0.00 | ... | 0.29 | ... | -0.19 | ... |
| BK Aur | 7 | 1 | 0.17 | 0.08 | 0.04 | 0.32 | 0.09 | 0.35 | ... | ... | -0.09 | ... | 0.31 | ... | 0.14 | ... |
| CY Aur | 6 | 1 | -0.40 | -0.02 | -0.38 | ... | ... | -0.20 | ... | -0.08 | -0.23 | ... | -0.06 | ... | -0.07 | ... |
| ER Aur | 6 | 2 | -0.34 | 0.27 | -0.37 | -0.04 | ... | -0.06 | -0.04 | ... | 0.00 | -0.04 | 0.14 | ... | 0.28 | ... |
| V335 Aur | 5 | 1 | -0.27 | 0.30 | 0.05 | ... | ... | ... | ... | ... | ... | ... | ... | ... | ... | ... |
| RW Cam | 610 | 17 | 0.09 | -0.03 | 0.05 | 0.04 | 0.43 | 0.25 | -0.03 | 0.26 | -0.08 | ... | 0.13 | ... | 0.16 | 0.02 |
| RX Cam | 9 | 9 | 0.04 | -0.14 | 0.03 | 0.20 | 0.25 | 0.27 | 0.00 | 0.26 | -0.05 | ... | 0.21 | ... | 0.11 | -0.09 |
| TV Cam | 6 | 1 | -0.08 | ... | -0.18 | 0.48 | ... | 0.02 | 0.01 | 0.19 | -0.43 | ... | 0.02 | ... | 0.25 | 0.35 |
| AB Cam | 6 | 1 | -0.09 | -0.03 | -0.11 | -0.22 | ... | 0.17 | -0.33 | 0.16 | 0.09 | -0.26 | 0.10 | ... | 0.05 | ... |
| AD Cam | 6 | 1 | -0.22 | -0.15 | -0.19 | -0.30 | ... | -0.10 | ... | 0.11 | -0.09 | ... | -0.02 | ... | 0.02 | ... |
| L Car | 8 | 1 | 0.05 | -0.09 | 0.02 | 0.10 | ... | 0.11 | 0.04 | 0.19 | 0.03 | -0.10 | -0.03 | ... | 0.09 | ... |

Table 4B
Abundance Data For Cepheids: Iron through Gadolinium

| Name | Ref | Sp | [Fe/H] | [Co/H] | [Ni/H] | [Cu/H] | [Zn/H] | [Y/H] | [Zr/H] | [La/H] | [Ce/H] | [Pr/H] | [Nd/H] | [Sm/H] | [Eu/H] | [Gd/H] |
|---|---|---|---|---|---|---|---|---|---|---|---|---|---|---|---|---|
| U Car | 8 | 1 | 0.01 | -0.18 | 0.02 | -0.26 | ... | 0.22 | ... | ... | -0.04 | -0.24 | -0.05 | ... | 0.16 | ... |
| V Car | 8 | 1 | 0.01 | -0.10 | -0.03 | -0.04 | -0.15 | 0.10 | 0.13 | -0.02 | 0.12 | -0.29 | 0.09 | 0.01 | 0.12 | ... |
| SX Car | 8 | 1 | -0.09 | -0.05 | -0.09 | 0.10 | -0.17 | 0.01 | ... | 0.19 | -0.02 | -0.42 | -0.10 | ... | 0.01 | ... |
| UW Car | 8 | 1 | -0.06 | -0.02 | -0.02 | 0.20 | -0.10 | 0.03 | ... | 0.00 | -0.13 | ... | 0.02 | 0.04 | -0.03 | ... |
| UX Car | 8 | 1 | 0.02 | -0.03 | -0.04 | 0.12 | ... | 0.15 | 0.25 | ... | 0.15 | -0.42 | 0.06 | ... | 0.13 | ... |
| UY Car | 8 | 1 | 0.03 | -0.09 | -0.03 | 0.16 | -0.15 | 0.20 | 0.31 | 0.36 | 0.17 | ... | 0.01 | ... | 0.13 | ... |
| UZ Car | 8 | 1 | 0.07 | -0.04 | 0.07 | 0.15 | -0.17 | 0.17 | 0.12 | 0.28 | 0.01 | -0.22 | 0.09 | ... | 0.08 | ... |
| VY Car | 8 | 1 | 0.26 | -0.01 | 0.26 | ... | ... | 0.46 | ... | 0.56 | 0.32 | 0.13 | 0.33 | ... | 0.46 | ... |
| WW Car | 8 | 1 | -0.07 | -0.17 | -0.16 | -0.21 | -0.29 | 0.02 | 0.05 | 0.15 | -0.05 | ... | -0.04 | ... | -0.07 | ... |
| WZ Car | 8 | 1 | 0.03 | -0.13 | -0.08 | -0.11 | -0.39 | 0.16 | 0.10 | ... | 0.07 | -0.19 | 0.11 | 0.13 | 0.23 | 0.27 |
| XX Car | 8 | 1 | 0.11 | 0.02 | 0.09 | 0.07 | -0.05 | 0.32 | 0.24 | 0.01 | 0.28 | 0.04 | 0.18 | ... | 0.19 | 0.30 |
| XY Car | 8 | 1 | 0.04 | -0.11 | 0.01 | 0.03 | -0.16 | 0.25 | 0.19 | 0.18 | 0.08 | 0.12 | 0.11 | ... | 0.19 | ... |
| XZ Car | 8 | 1 | 0.14 | 0.06 | 0.07 | 0.20 | -0.02 | 0.23 | 0.07 | 0.22 | 0.24 | -0.17 | 0.11 | 0.11 | 0.23 | ... |
| YZ Car | 8 | 1 | 0.02 | -0.11 | 0.02 | -0.04 | 0.11 | 0.18 | 0.11 | 0.21 | 0.09 | -0.19 | 0.02 | 0.05 | 0.09 | ... |
| AQ Car | 8 | 1 | 0.06 | -0.08 | -0.03 | -0.11 | -0.14 | 0.19 | 0.16 | 0.06 | 0.14 | -0.19 | 0.06 | 0.00 | 0.14 | ... |
| CN Car | 8 | 1 | 0.06 | -0.02 | 0.04 | 0.22 | -0.05 | 0.20 | 0.01 | 0.15 | 0.13 | -0.08 | 0.03 | -0.02 | 0.08 | ... |
| CY Car | 8 | 1 | 0.10 | -0.03 | 0.01 | 0.03 | -0.03 | 0.22 | 0.04 | 0.23 | 0.17 | -0.11 | 0.08 | 0.03 | 0.15 | 0.18 |
| DY Car | 8 | 1 | -0.07 | ... | -0.15 | 0.08 | -0.25 | 0.04 | 0.18 | ... | -0.01 | ... | -0.03 | ... | 0.01 | ... |
| ER Car | 8 | 1 | 0.03 | -0.19 | -0.03 | -0.07 | -0.20 | 0.15 | 0.11 | 0.03 | 0.07 | -0.31 | 0.04 | -0.08 | 0.13 | ... |
| FI Car | 8 | 1 | 0.06 | 0.03 | 0.03 | -0.13 | 0.42 | 0.09 | 0.21 | 0.38 | 0.10 | -0.18 | 0.11 | ... | 0.31 | 0.20 |
| FR Car | 8 | 1 | 0.02 | -0.13 | -0.06 | -0.11 | -0.48 | 0.16 | 0.10 | 0.24 | 0.13 | -0.27 | 0.09 | ... | 0.14 | 0.12 |
| GH Car | 8 | 1 | -0.01 | -0.08 | 0.00 | 0.26 | -0.22 | 0.13 | 0.15 | 0.26 | 0.05 | -0.30 | 0.00 | ... | 0.16 | ... |
| GX Car | 8 | 1 | 0.01 | -0.10 | -0.05 | 0.04 | -0.22 | 0.14 | 0.21 | 0.23 | 0.09 | -0.26 | 0.00 | ... | 0.05 | ... |
| HW Car | 8 | 1 | 0.04 | -0.09 | -0.03 | -0.17 | -0.37 | 0.15 | 0.27 | 0.01 | 0.18 | -0.03 | 0.09 | ... | 0.09 | 0.18 |
| IO Car | 8 | 1 | -0.05 | -0.20 | -0.02 | 0.00 | -0.12 | 0.10 | 0.01 | 0.01 | 0.03 | -0.01 | -0.06 | -0.07 | 0.11 | ... |
| IT Car | 8 | 1 | 0.06 | -0.05 | 0.04 | 0.03 | -0.20 | 0.22 | 0.08 | 0.14 | 0.03 | -0.29 | 0.02 | ... | 0.05 | 0.06 |
| V397 Car | 8 | 1 | 0.03 | -0.22 | -0.06 | 0.08 | -0.22 | 0.13 | 0.23 | 0.06 | 0.09 | ... | 0.05 | -0.05 | 0.09 | ... |
| RW Cas | 7 | 1 | 0.22 | 0.15 | 0.12 | ... | ... | 0.23 | ... | ... | 0.00 | ... | 0.11 | ... | 0.23 | ... |
| RY Cas | 7 | 1 | 0.26 | ... | 0.06 | ... | 0.24 | 0.61 | ... | ... | 0.35 | ... | 0.40 | ... | 0.53 | ... |
| SU Cas | 11 | 13 | 0.06 | -0.04 | 0.04 | 0.30 | 0.17 | 0.26 | 0.02 | 0.22 | -0.05 | ... | 0.14 | ... | 0.09 | 0.03 |
| SW Cas | 7 | 1 | 0.13 | -0.03 | -0.03 | -0.08 | ... | 0.31 | ... | 0.30 | -0.12 | ... | 0.16 | ... | 0.15 | ... |

Table 4B
Abundance Data For Cepheids: Iron through Gadolinium

| Name | Ref | Sp | [Fe/H] | [Co/H] | [Ni/H] | [Cu/H] | [Zn/H] | [Y/H] | [Zr/H] | [La/H] | [Ce/H] | [Pr/H] | [Nd/H] | [Sm/H] | [Eu/H] | [Gd/H] |
|---|---|---|---|---|---|---|---|---|---|---|---|---|---|---|---|---|
| SY Cas | 7 | 1 | 0.04 | 0.31 | -0.03 | 0.49 | 0.44 | 0.11 | ... | 0.42 | 0.01 | ... | 0.14 | ... | 0.10 | ... |
| SZ Cas | 7 | 1 | 0.04 | ... | 0.03 | ... | ... | 0.23 | ... | ... | 0.00 | ... | 0.32 | ... | 0.28 | ... |
| TU Cas | 1 | 12 | 0.03 | -0.08 | -0.04 | 0.15 | 0.46 | 0.17 | 0.01 | 0.24 | -0.05 | ... | 0.08 | ... | 0.11 | 0.17 |
| XY Cas | 7 | 1 | 0.03 | -0.05 | -0.07 | 0.28 | ... | 0.20 | ... | 0.16 | -0.19 | ... | -0.08 | ... | -0.09 | ... |
| BD Cas | 1 | 3 | -0.07 | 0.09 | -0.26 | -0.14 | ... | 0.01 | 0.13 | 0.41 | 0.19 | ... | -0.25 | ... | ... | 0.30 |
| CE Cas A | 7 | 1 | 0.18 | 0.45 | 0.13 | ... | 0.68 | 0.32 | ... | ... | -0.17 | ... | -0.24 | ... | 0.17 | ... |
| CE Cas B | 7 | 1 | 0.22 | 0.40 | 0.11 | ... | 0.25 | 0.56 | ... | ... | 0.30 | ... | 0.30 | ... | 0.42 | ... |
| CF Cas | 1 | 5 | -0.01 | -0.15 | -0.03 | -0.11 | 0.25 | 0.11 | -0.19 | 0.14 | -0.17 | ... | 0.04 | ... | 0.06 | -0.02 |
| CH Cas | 7 | 1 | 0.17 | ... | 0.11 | ... | 0.18 | 0.70 | ... | ... | 0.11 | ... | 0.20 | ... | 0.50 | ... |
| CY Cas | 7 | 1 | 0.06 | 0.05 | -0.06 | ... | 0.16 | 0.21 | ... | 0.23 | 0.00 | ... | 0.03 | ... | 0.22 | ... |
| DD Cas | 7 | 1 | 0.10 | 0.03 | -0.06 | 0.62 | 0.24 | 0.24 | ... | 0.09 | -0.02 | ... | 0.06 | ... | 0.09 | ... |
| DF Cas | 1 | 1 | 0.13 | -0.30 | 0.04 | -0.43 | ... | 0.06 | 0.22 | ... | 0.07 | ... | 0.38 | ... | 0.28 | ... |
| DL Cas | 1 | 3 | -0.01 | -0.04 | 0.00 | -0.19 | 0.47 | 0.21 | 0.16 | 0.12 | 0.09 | ... | 0.12 | ... | 0.11 | -0.05 |
| FM Cas | 7 | 1 | -0.09 | -0.16 | -0.15 | -0.02 | ... | -0.06 | ... | -0.01 | -0.25 | ... | -0.14 | ... | -0.25 | ... |
| V379 Cas | 7 | 2 | 0.06 | 0.07 | 0.00 | 0.24 | 0.51 | 0.16 | ... | 0.30 | -0.06 | ... | 0.12 | ... | 0.12 | ... |
| V636 Cas | 11 | 8 | 0.07 | -0.08 | 0.09 | 0.04 | 0.35 | 0.18 | -0.07 | 0.14 | -0.15 | ... | 0.11 | ... | 0.02 | -0.03 |
| V Cen | 108 | 3 | -0.01 | -0.17 | -0.02 | 0.02 | -0.03 | 0.15 | 0.10 | 0.19 | 0.08 | -0.35 | 0.05 | -0.05 | 0.12 | ... |
| QY Cen | 8 | 1 | 0.16 | 0.08 | 0.13 | 0.22 | 0.14 | 0.27 | 0.15 | 0.18 | 0.14 | -0.27 | 0.10 | ... | 0.28 | ... |
| XX Cen | 8 | 1 | 0.16 | 0.01 | 0.10 | 0.08 | 0.00 | 0.29 | 0.23 | 0.13 | 0.23 | -0.09 | 0.17 | 0.17 | 0.19 | ... |
| AY Cen | 8 | 1 | 0.01 | -0.07 | -0.04 | -0.04 | -0.25 | 0.18 | 0.11 | 0.11 | 0.14 | -0.10 | 0.08 | ... | 0.09 | ... |
| AZ Cen | 8 | 1 | -0.05 | -0.03 | -0.07 | 0.08 | 0.06 | 0.06 | 0.08 | 0.10 | 0.03 | -0.31 | 0.00 | ... | 0.04 | ... |
| BB Cen | 8 | 1 | 0.13 | 0.08 | 0.11 | 0.13 | -0.15 | 0.25 | 0.13 | 0.16 | 0.11 | -0.10 | 0.07 | ... | 0.25 | ... |
| KK Cen | 8 | 1 | 0.12 | 0.00 | 0.07 | 0.05 | -0.16 | 0.25 | 0.12 | 0.26 | 0.04 | -0.27 | 0.02 | ... | 0.20 | ... |
| KN Cen | 8 | 1 | 0.35 | 0.31 | 0.37 | ... | ... | 0.63 | 0.40 | 0.66 | 0.44 | 0.04 | 0.36 | ... | ... | 0.67 |
| MZ Cen | 8 | 1 | 0.20 | 0.01 | 0.18 | 0.01 | -0.16 | 0.20 | 0.18 | 0.23 | -0.07 | -0.29 | 0.02 | ... | 0.16 | ... |
| V339 Cen | 8 | 1 | 0.04 | -0.15 | 0.01 | -0.02 | -0.11 | 0.17 | 0.06 | 0.08 | 0.08 | -0.15 | -0.03 | -0.08 | 0.14 | ... |
| V378 Cen | 8 | 1 | -0.02 | -0.03 | -0.01 | -0.02 | -0.22 | 0.23 | 0.17 | 0.15 | 0.08 | 0.02 | 0.12 | ... | 0.19 | ... |
| V381 Cen | 8 | 1 | 0.02 | -0.04 | -0.02 | 0.06 | -0.16 | 0.11 | 0.22 | 0.17 | -0.02 | -0.33 | 0.01 | ... | 0.08 | ... |
| V419 Cen | 8 | 1 | 0.07 | -0.05 | 0.02 | 0.09 | -0.36 | 0.21 | 0.17 | 0.36 | 0.12 | -0.40 | 0.08 | ... | 0.16 | 0.18 |
| V496 Cen | 8 | 1 | 0.00 | -0.02 | -0.05 | 0.17 | ... | 0.13 | 0.12 | 0.31 | 0.15 | ... | -0.04 | ... | 0.08 | ... |
| V659 Cen | 8 | 1 | 0.07 | -0.09 | 0.00 | 0.02 | -0.10 | 0.12 | 0.03 | 0.20 | 0.14 | -0.25 | 0.03 | -0.03 | 0.23 | ... |

Table 4B
Abundance Data For Cepheids: Iron through Gadolinium

| Name | Ref | Sp | [Fe/H] | [Co/H] | [Ni/H] | [Cu/H] | [Zn/H] | [Y/H] | [Zr/H] | [La/H] | [Ce/H] | [Pr/H] | [Nd/H] | [Sm/H] | [Eu/H] | [Gd/H] |
|---|---|---|---|---|---|---|---|---|---|---|---|---|---|---|---|---|
| V737 Cen | 8 | 1 | 0.13 | -0.04 | 0.05 | 0.11 | -0.01 | 0.22 | 0.05 | 0.06 | 0.18 | -0.21 | 0.08 | -0.02 | 0.18 | … |
| Del Cep | 12 | 18 | 0.09 | -0.06 | 0.06 | 0.12 | 0.22 | 0.24 | -0.01 | 0.21 | -0.10 | … | 0.16 | … | 0.06 | 0.11 |
| CP Cep | 7 | 1 | -0.01 | -0.07 | -0.13 | … | 0.28 | 0.05 | … | 0.12 | -0.12 | … | 0.03 | … | 0.00 | … |
| CR Cep | 7 | 1 | -0.06 | 0.00 | -0.19 | 0.10 | … | 0.05 | … | 0.14 | -0.20 | … | 0.00 | … | -0.11 | … |
| IR Cep | 107 | 2 | 0.05 | 0.01 | -0.03 | -0.03 | 0.33 | 0.13 | 0.16 | 0.16 | 0.02 | … | 0.09 | … | 0.12 | … |
| V351 Cep | 107 | 3 | 0.02 | -0.04 | -0.02 | 0.03 | 0.17 | 0.20 | -0.01 | 0.22 | -0.06 | … | 0.13 | … | 0.08 | 0.16 |
| AV Cir | 8 | 1 | 0.10 | 0.06 | 0.06 | 0.09 | -0.01 | 0.20 | 0.14 | 0.00 | 0.12 | -0.04 | -0.02 | -0.05 | 0.07 | … |
| AX Cir | 8 | 2 | -0.06 | -0.26 | -0.14 | -0.13 | -0.34 | -0.01 | 0.03 | 0.04 | -0.08 | -0.49 | -0.10 | -0.45 | 0.01 | 0.05 |
| BP Cir | 8 | 1 | -0.06 | -0.22 | -0.08 | 0.08 | -0.19 | 0.09 | 0.04 | 0.05 | -0.02 | -0.25 | -0.03 | -0.12 | 0.12 | … |
| RY CMa | 5 | 1 | 0.02 | 0.12 | 0.02 | … | … | … | -0.09 | 0.20 | -0.31 | … | 0.34 | … | -0.08 | … |
| RZ CMa | 7 | 2 | -0.03 | -0.01 | -0.05 | 0.45 | … | 0.23 | … | 0.23 | -0.02 | … | 0.13 | … | 0.12 | … |
| TW CMa | 5 | 1 | -0.19 | … | -0.05 | … | … | … | -0.10 | 0.28 | -0.04 | … | 0.44 | … | -0.15 | … |
| VZ CMa | 5 | 1 | -0.06 | … | -0.02 | … | … | … | -0.18 | 0.33 | -0.02 | … | 0.41 | … | 0.22 | … |
| R Cru | 8 | 1 | 0.08 | 0.10 | 0.02 | 0.08 | -0.03 | 0.19 | 0.00 | 0.13 | 0.10 | -0.21 | 0.01 | -0.09 | 0.12 | … |
| S Cru | 8 | 1 | -0.12 | -0.08 | -0.17 | 0.03 | … | 0.00 | 0.00 | … | 0.00 | … | -0.14 | … | -0.06 | … |
| T Cru | 8 | 1 | 0.09 | -0.07 | 0.02 | 0.02 | -0.08 | 0.23 | 0.11 | 0.17 | 0.19 | -0.20 | 0.17 | 0.17 | 0.16 | … |
| X Cru | 8 | 1 | 0.14 | 0.04 | 0.08 | 0.07 | -0.01 | 0.23 | 0.22 | 0.16 | 0.19 | -0.18 | 0.14 | 0.00 | 0.18 | … |
| AD Cru | 8 | 1 | 0.06 | -0.08 | -0.01 | 0.07 | -0.18 | 0.16 | 0.15 | 0.07 | 0.12 | -0.05 | 0.10 | 0.00 | 0.25 | 0.22 |
| AG Cru | 8 | 1 | -0.13 | … | -0.14 | 0.08 | -0.33 | -0.04 | 0.11 | 0.10 | -0.19 | … | -0.22 | … | -0.06 | … |
| BG Cru | 108 | 2 | 0.02 | -0.01 | -0.06 | -0.19 | -0.40 | 0.08 | 0.08 | … | 0.23 | -0.11 | 0.00 | … | 0.23 | 0.38 |
| VW Cru | 8 | 1 | 0.10 | -0.13 | 0.01 | 0.03 | 0.16 | 0.21 | 0.14 | 0.17 | 0.13 | -0.21 | 0.02 | … | 0.22 | … |
| X Cyg | 10 | 26 | 0.10 | -0.03 | 0.06 | 0.01 | 0.52 | 0.28 | -0.01 | 0.23 | -0.07 | … | 0.12 | … | 0.11 | 0.13 |
| SU Cyg | 12 | 12 | -0.03 | -0.08 | -0.01 | 0.22 | 0.14 | 0.15 | 0.01 | 0.24 | -0.11 | … | 0.15 | … | 0.08 | 0.10 |
| SZ Cyg | 7 | 1 | 0.09 | 0.07 | 0.02 | 0.43 | 0.39 | 0.22 | … | 0.21 | 0.00 | … | 0.03 | … | 0.15 | … |
| TX Cyg | 7 | 1 | 0.20 | … | 0.10 | 0.36 | 0.34 | 0.45 | … | 0.44 | 0.23 | … | 0.30 | … | 0.13 | … |
| VX Cyg | 7 | 1 | 0.09 | 0.11 | 0.04 | … | … | 0.14 | … | 0.16 | -0.01 | … | 0.01 | … | 0.17 | … |
| VY Cyg | 7 | 1 | 0.00 | -0.10 | -0.11 | 0.09 | 0.10 | 0.17 | … | 0.21 | 0.01 | … | 0.09 | … | 0.01 | … |
| VZ Cyg | 7 | 1 | 0.05 | -0.06 | -0.07 | 0.13 | … | 0.17 | … | 0.14 | 0.01 | … | 0.08 | … | 0.11 | … |
| BZ Cyg | 7 | 1 | 0.19 | … | 0.06 | … | … | 0.17 | … | 0.31 | -0.07 | … | 0.15 | … | 0.12 | … |
| CD Cyg | 10 | 16 | 0.11 | -0.02 | 0.07 | 0.05 | 0.35 | 0.25 | -0.04 | 0.24 | -0.07 | … | 0.12 | … | 0.11 | 0.08 |
| DT Cyg | 11 | 14 | 0.10 | 0.07 | 0.09 | 0.23 | 0.20 | 0.31 | -0.03 | 0.22 | -0.03 | … | 0.19 | … | 0.18 | 0.20 |

Table 4B
Abundance Data For Cepheids: Iron through Gadolinium

| Name | Ref | Sp | [Fe/H] | [Co/H] | [Ni/H] | [Cu/H] | [Zn/H] | [Y/H] | [Zr/H] | [La/H] | [Ce/H] | [Pr/H] | [Nd/H] | [Sm/H] | [Eu/H] | [Gd/H] |
|---|---|---|---|---|---|---|---|---|---|---|---|---|---|---|---|---|
| MW Cyg | 7 | 1 | 0.09 | -0.05 | -0.09 | -0.01 | ... | 0.24 | ... | 0.15 | 0.04 | ... | 0.13 | ... | 0.06 | ... |
| V386 Cyg | 7 | 1 | 0.11 | ... | -0.03 | ... | 0.21 | 0.35 | ... | 0.41 | 0.10 | ... | 0.42 | ... | 0.41 | ... |
| V402 Cyg | 7 | 1 | 0.02 | -0.06 | -0.09 | 0.01 | ... | 0.15 | ... | 0.16 | 0.03 | ... | 0.01 | ... | 0.01 | ... |
| V532 Cyg | 7 | 1 | -0.04 | -0.08 | -0.09 | 0.36 | 0.38 | 0.12 | ... | 0.09 | 0.00 | ... | -0.05 | ... | -0.03 | ... |
| V924 Cyg | 1 | 1 | -0.09 | ... | -0.14 | 0.12 | ... | -0.18 | ... | ... | ... | ... | 0.04 | ... | ... | ... |
| V1154Cyg | 7 | 1 | -0.10 | 0.05 | -0.12 | 0.13 | 0.14 | 0.08 | ... | 0.18 | -0.10 | ... | 0.05 | ... | -0.06 | ... |
| V1334Cyg | 11 | 11 | 0.03 | -0.32 | 0.08 | 0.29 | ... | 0.18 | -0.08 | 0.27 | -0.09 | ... | 0.16 | ... | 0.17 | ... |
| V1726Cyg | 1 | 1 | -0.02 | ... | -0.15 | ... | ... | 0.14 | ... | ... | ... | ... | 0.24 | ... | 0.31 | ... |
| TX Del | 1 | 1 | 0.24 | ... | 0.17 | ... | ... | 0.07 | ... | 0.13 | -0.34 | ... | -0.38 | ... | ... | ... |
| Beta Dor | 1 | 1 | -0.01 | -0.19 | -0.04 | -0.45 | 0.14 | 0.02 | 0.03 | 0.18 | 0.05 | ... | -0.05 | ... | 0.04 | 0.20 |
| Zeta Gem | 11 | 11 | 0.00 | -0.16 | -0.01 | -0.02 | 0.20 | 0.21 | -0.09 | 0.14 | -0.11 | ... | 0.04 | ... | 0.05 | 0.00 |
| W Gem | 9 | 8 | -0.01 | -0.14 | -0.05 | 0.10 | 0.18 | 0.23 | -0.03 | 0.29 | 0.01 | ... | 0.19 | ... | 0.12 | 0.06 |
| RZ Gem | 5 | 1 | -0.12 | -0.06 | -0.07 | ... | ... | ... | -0.29 | 0.32 | 0.01 | ... | 0.36 | ... | 0.00 | ... |
| AA Gem | 5 | 1 | -0.24 | -0.35 | -0.11 | ... | ... | ... | 0.03 | 0.37 | -0.05 | ... | 0.45 | ... | 0.16 | ... |
| AD Gem | 5 | 1 | -0.18 | ... | -0.18 | ... | ... | ... | ... | 0.05 | -0.11 | ... | 0.22 | ... | 0.04 | ... |
| BB Gem | 5 | 1 | -0.09 | 0.08 | -0.09 | ... | ... | ... | -0.12 | 0.22 | 0.31 | ... | 0.12 | ... | 0.14 | ... |
| DX Gem | 5 | 1 | -0.02 | ... | -0.11 | ... | ... | ... | -0.12 | 0.41 | 0.01 | ... | 0.42 | ... | 0.04 | ... |
| BB Her | 1 | 4 | 0.15 | -0.04 | 0.15 | 0.19 | 0.39 | 0.32 | 0.00 | 0.11 | -0.17 | ... | 0.02 | ... | 0.08 | ... |
| V Lac | 7 | 1 | 0.00 | ... | -0.09 | 0.63 | 0.15 | 0.26 | ... | 0.19 | 0.14 | ... | 0.33 | ... | 0.10 | ... |
| X Lac | 7 | 1 | -0.02 | 0.00 | -0.10 | 0.24 | 0.12 | 0.18 | ... | 0.19 | 0.00 | ... | 0.14 | ... | -0.02 | ... |
| Y Lac | 12 | 9 | -0.04 | -0.15 | -0.09 | 0.18 | 0.08 | 0.15 | -0.09 | 0.22 | -0.06 | ... | 0.14 | ... | 0.07 | 0.33 |
| Z Lac | 10 | 9 | 0.01 | -0.14 | -0.03 | 0.07 | 0.26 | 0.20 | -0.07 | 0.25 | -0.05 | ... | 0.11 | ... | 0.07 | 0.06 |
| RR Lac | 7 | 1 | 0.13 | 0.02 | 0.00 | 0.18 | ... | 0.28 | ... | 0.44 | 0.00 | ... | 0.29 | ... | 0.08 | ... |
| BG Lac | 1 | 3 | -0.01 | -0.13 | -0.03 | 0.10 | 0.28 | 0.14 | -0.12 | 0.07 | -0.17 | ... | 0.05 | ... | 0.02 | 0.18 |
| GH Lup | 8 | 1 | 0.08 | -0.15 | -0.02 | 0.04 | -0.03 | 0.11 | 0.16 | 0.13 | 0.07 | -0.37 | 0.04 | ... | 0.08 | ... |
| V473 Lyr | 1 | 2 | -0.06 | -0.13 | -0.11 | -0.10 | 0.12 | 0.10 | 0.02 | 0.20 | 0.11 | ... | 0.03 | ... | -0.02 | 0.00 |
| T Mon | 10 | 20 | 0.13 | 0.06 | 0.12 | -0.15 | 0.54 | 0.34 | 0.03 | 0.35 | 0.07 | ... | 0.20 | ... | 0.21 | 0.20 |
| SV Mon | 10 | 9 | -0.03 | -0.23 | -0.08 | -0.10 | 0.31 | 0.21 | -0.08 | 0.24 | -0.10 | ... | 0.14 | ... | 0.04 | 0.07 |
| TW Mon | 4 | 1 | -0.24 | -0.18 | -0.33 | -0.34 | 0.06 | -0.08 | -0.29 | 0.17 | -0.18 | ... | -0.09 | ... | -0.03 | ... |
| TX Mon | 3 | 1 | -0.14 | -0.24 | -0.20 | -0.04 | 0.15 | 0.06 | -0.27 | 0.23 | 0.10 | -0.12 | -0.10 | ... | 0.07 | -0.23 |
| TZ Mon | 304 | 2 | -0.07 | -0.17 | -0.14 | -0.01 | 0.20 | -0.06 | -0.16 | 0.10 | -0.17 | -0.06 | -0.17 | ... | 0.00 | ... |

Table 4B
Abundance Data For Cepheids: Iron through Gadolinium

| Name | Ref | Sp | [Fe/H] | [Co/H] | [Ni/H] | [Cu/H] | [Zn/H] | [Y/H] | [Zr/H] | [La/H] | [Ce/H] | [Pr/H] | [Nd/H] | [Sm/H] | [Eu/H] | [Gd/H] |
|---|---|---|---|---|---|---|---|---|---|---|---|---|---|---|---|---|
| UY Mon | 5 | 1 | -0.08 | 0.29 | -0.03 | … | … | … | -0.05 | 0.32 | -0.27 | … | 0.42 | … | -0.06 | … |
| WW Mon | 4 | 1 | -0.29 | 0.03 | -0.30 | -0.14 | -0.04 | -0.06 | … | 0.04 | -0.11 | … | 0.38 | … | -0.05 | … |
| XX Mon | 304 | 2 | -0.14 | -0.22 | -0.19 | -0.12 | 0.23 | 0.03 | -0.20 | 0.09 | -0.03 | -0.08 | 0.05 | … | 0.05 | 0.04 |
| AA Mon | 3 | 1 | -0.21 | … | -0.30 | 0.29 | … | 0.07 | … | … | 0.03 | … | -0.11 | … | -0.32 | … |
| AC Mon | 103 | 2 | -0.14 | … | -0.23 | -0.71 | … | -0.03 | … | 0.26 | -0.20 | … | 0.11 | … | 0.23 | … |
| CU Mon | 4 | 1 | -0.26 | -0.33 | -0.27 | -0.15 | 0.10 | 0.08 | … | 0.24 | 0.22 | -0.23 | 0.13 | … | 0.31 | … |
| CV Mon | 1 | 1 | -0.03 | … | -0.08 | -0.05 | … | 0.01 | 0.00 | 0.19 | -0.03 | … | 0.29 | … | 0.13 | … |
| EE Mon | 3 | 1 | -0.51 | … | -0.58 | … | … | -0.40 | … | … | -0.27 | … | -0.23 | … | … | … |
| EK Mon | 3 | 1 | -0.10 | -0.06 | -0.20 | -0.96 | … | -0.07 | 0.05 | … | -0.22 | … | 0.06 | … | 0.31 | … |
| FG Mon | 4 | 1 | -0.20 | -0.16 | -0.13 | -0.16 | 0.16 | -0.03 | -0.34 | -0.13 | 0.03 | -0.02 | -0.01 | … | 0.01 | 0.31 |
| FI Mon | 4 | 1 | -0.18 | 0.09 | -0.14 | -0.18 | 0.10 | 0.04 | … | 0.03 | -0.05 | -0.29 | 0.02 | … | 0.11 | … |
| V465 Mon | 7 | 1 | 0.03 | … | -0.05 | 0.32 | 0.08 | 0.29 | … | 0.38 | 0.08 | … | 0.23 | … | 0.22 | … |
| V495 Mon | 3 | 1 | -0.26 | -0.25 | -0.36 | -0.31 | … | -0.10 | -0.44 | 0.05 | 0.04 | -0.20 | -0.14 | … | -0.17 | … |
| V504 Mon | 3 | 1 | -0.31 | … | -0.36 | 0.20 | … | -0.17 | -0.24 | … | -0.26 | -0.39 | -0.30 | … | -0.15 | … |
| V508 Mon | 3 | 1 | -0.25 | -0.35 | -0.35 | -0.30 | 0.11 | -0.17 | -0.16 | 0.06 | -0.17 | -0.54 | -0.24 | … | -0.27 | -0.14 |
| V510 Mon | 4 | 1 | -0.19 | … | -0.22 | -0.08 | 0.07 | -0.09 | -0.29 | 0.00 | -0.16 | … | 0.02 | … | -0.18 | -0.17 |
| V526 Mon | 1 | 1 | -0.13 | 0.04 | -0.07 | … | … | … | -0.11 | 0.41 | … | … | 0.25 | … | 0.16 | … |
| R Mus | 8 | 1 | 0.10 | -0.02 | 0.09 | 0.13 | -0.01 | 0.26 | 0.13 | 0.12 | 0.20 | -0.20 | 0.12 | 0.03 | 0.22 | … |
| S Mus | 8 | 1 | -0.02 | -0.16 | -0.08 | -0.12 | -0.26 | 0.10 | 0.09 | 0.09 | 0.03 | -0.11 | -0.03 | … | 0.06 | … |
| RT Mus | 8 | 1 | 0.02 | -0.06 | -0.05 | 0.12 | -0.24 | 0.15 | 0.27 | 0.25 | 0.13 | -0.17 | 0.06 | … | 0.13 | … |
| TZ Mus | 8 | 1 | -0.01 | -0.07 | 0.00 | -0.33 | 0.01 | 0.10 | … | 0.28 | 0.01 | … | -0.03 | … | 0.11 | … |
| UU Mus | 8 | 1 | 0.05 | -0.05 | 0.02 | 0.12 | 0.00 | 0.20 | 0.06 | 0.20 | 0.14 | -0.26 | 0.01 | 0.07 | 0.20 | … |
| S Nor | 108 | 3 | 0.06 | -0.08 | -0.04 | -0.09 | -0.18 | 0.14 | 0.14 | 0.31 | -0.06 | -0.32 | 0.01 | … | 0.04 | … |
| U Nor | 8 | 1 | 0.15 | -0.04 | 0.09 | 0.25 | 0.17 | 0.26 | 0.16 | 0.03 | 0.10 | -0.04 | 0.07 | -0.08 | 0.15 | 0.07 |
| SY Nor | 8 | 1 | 0.31 | 0.21 | 0.29 | 0.30 | 0.11 | 0.39 | 0.28 | 0.24 | 0.27 | -0.03 | 0.28 | … | 0.35 | … |
| TW Nor | 8 | 1 | 0.28 | 0.18 | 0.29 | 0.24 | 0.14 | 0.31 | … | 0.35 | … | -0.16 | 0.16 | … | 0.43 | … |
| GU Nor | 8 | 1 | 0.15 | 0.02 | 0.12 | 0.27 | -0.06 | 0.25 | 0.08 | 0.09 | 0.08 | -0.26 | -0.05 | … | 0.11 | 0.14 |
| V340 Nor | 108 | 2 | 0.04 | -0.08 | 0.04 | -0.04 | 0.02 | 0.11 | 0.17 | 0.15 | -0.11 | -0.39 | -0.05 | 0.15 | 0.06 | 0.12 |
| Y Oph | 11 | 14 | 0.06 | -0.10 | 0.06 | 0.05 | 0.20 | 0.31 | -0.05 | 0.24 | -0.07 | … | 0.17 | … | 0.13 | -0.02 |
| BF Oph | 708 | 2 | 0.03 | 0.00 | -0.05 | 0.10 | -0.04 | 0.19 | 0.26 | 0.19 | 0.05 | -0.29 | -0.01 | … | 0.09 | … |
| RS Ori | 104 | 5 | -0.10 | -0.01 | -0.12 | 0.11 | 0.21 | 0.15 | -0.12 | 0.18 | -0.19 | -0.31 | 0.05 | … | 0.00 | 0.07 |

Table 4B
Abundance Data For Cepheids: Iron through Gadolinium

| Name | Ref | Sp | [Fe/H] | [Co/H] | [Ni/H] | [Cu/H] | [Zn/H] | [Y/H] | [Zr/H] | [La/H] | [Ce/H] | [Pr/H] | [Nd/H] | [Sm/H] | [Eu/H] | [Gd/H] |
|---|---|---|---|---|---|---|---|---|---|---|---|---|---|---|---|---|
| CS Ori | 3 | 1 | -0.26 | ... | -0.25 | 0.19 | ... | -0.09 | ... | 0.21 | 0.07 | ... | -0.17 | ... | -0.03 | ... |
| GQ Ori | 107 | 2 | 0.01 | 0.05 | -0.12 | 0.20 | 0.17 | 0.32 | ... | 0.30 | 0.04 | ... | 0.05 | ... | 0.16 | ... |
| SV Per | 7 | 1 | 0.01 | -0.12 | -0.15 | 0.02 | 0.04 | 0.26 | ... | 0.28 | 0.11 | ... | 0.16 | ... | 0.17 | ... |
| UX Per | 6 | 1 | -0.21 | 0.27 | -0.21 | 0.43 | 0.27 | -0.18 | ... | 0.05 | -0.25 | 0.33 | -0.03 | ... | -0.17 | ... |
| VX Per | 10 | 9 | -0.04 | -0.15 | -0.09 | 0.00 | 0.19 | 0.13 | -0.14 | 0.21 | -0.11 | ... | 0.09 | ... | 0.05 | -0.20 |
| AS Per | 7 | 1 | 0.10 | 0.05 | -0.08 | 0.19 | 0.24 | 0.16 | ... | 0.10 | 0.02 | ... | 0.07 | ... | 0.10 | ... |
| AW Per | 1 | 4 | 0.01 | -0.01 | 0.04 | 0.57 | 0.51 | 0.11 | -0.02 | 0.25 | -0.11 | ... | 0.10 | ... | 0.11 | -0.12 |
| BM Per | 607 | 5 | 0.02 | -0.03 | -0.05 | -0.09 | 0.27 | 0.19 | 0.11 | 0.31 | 0.02 | -0.33 | 0.09 | ... | 0.20 | 0.11 |
| HQ Per | 6 | 2 | -0.31 | -0.37 | -0.33 | -0.22 | -0.10 | -0.09 | -0.15 | -0.06 | -0.26 | -0.28 | -0.11 | ... | -0.10 | 0.07 |
| MM Per | 5 | 1 | -0.01 | -0.03 | -0.15 | ... | ... | ... | -0.16 | 0.18 | 0.06 | ... | 0.44 | ... | -0.08 | ... |
| V440 Per | 11 | 10 | -0.04 | -0.06 | -0.08 | 0.14 | 0.10 | 0.27 | -0.05 | 0.31 | -0.07 | ... | 0.18 | ... | 0.15 | 0.16 |
| X Pup | 10 | 7 | -0.03 | -0.26 | -0.07 | -0.08 | 0.22 | 0.14 | -0.03 | 0.28 | -0.07 | ... | 0.21 | ... | 0.11 | -0.04 |
| RS Pup | 4 | 2 | 0.17 | -0.01 | 0.06 | -0.09 | ... | 0.18 | 0.09 | 0.33 | 0.09 | -0.09 | 0.13 | ... | 0.24 | 0.13 |
| VW Pup | 3 | 1 | -0.19 | -0.26 | -0.17 | -0.64 | ... | -0.09 | ... | 0.19 | 0.08 | -0.38 | -0.06 | ... | -0.08 | ... |
| VX Pup | 1 | 1 | -0.13 | ... | -0.18 | 0.25 | ... | 0.10 | ... | ... | ... | ... | 0.35 | ... | -0.15 | ... |
| VZ Pup | 3 | 1 | -0.16 | ... | -0.28 | 0.10 | ... | -0.06 | ... | ... | -0.11 | -0.21 | 0.02 | ... | 0.02 | 0.32 |
| WW Pup | 3 | 1 | -0.18 | ... | -0.21 | -0.34 | -0.01 | 0.04 | ... | 0.15 | 0.00 | -0.35 | -0.04 | ... | -0.09 | ... |
| AD Pup | 3 | 1 | -0.24 | -0.48 | -0.37 | -0.47 | ... | -0.11 | -0.20 | 0.27 | -0.14 | -0.29 | -0.12 | ... | -0.06 | -0.03 |
| AP Pup | 8 | 1 | 0.06 | -0.15 | 0.00 | 0.14 | -0.10 | 0.17 | 0.16 | 0.13 | 0.15 | -0.24 | 0.08 | 0.05 | 0.19 | ... |
| AQ Pup | 4 | 2 | -0.14 | -0.34 | -0.20 | -0.46 | ... | 0.05 | -0.33 | 0.13 | -0.17 | -0.27 | 0.00 | ... | 0.03 | 0.12 |
| AT Pup | 8 | 1 | -0.14 | ... | -0.15 | -0.38 | -0.30 | 0.01 | 0.13 | 0.27 | 0.11 | ... | 0.04 | ... | 0.07 | ... |
| BC Pup | 4 | 2 | -0.23 | -0.03 | -0.19 | -0.14 | -0.20 | -0.14 | -0.50 | 0.08 | -0.11 | 0.01 | 0.19 | ... | 0.21 | ... |
| BN Pup | 3 | 1 | 0.01 | -0.23 | -0.03 | -0.24 | ... | 0.24 | -0.17 | 0.30 | 0.00 | -0.15 | 0.03 | ... | 0.07 | ... |
| CE Pup | 8 | 1 | -0.04 | -0.20 | -0.01 | 0.04 | -0.19 | 0.27 | 0.16 | 0.42 | 0.15 | -0.21 | 0.13 | ... | 0.30 | ... |
| HW Pup | 304 | 2 | -0.25 | -0.15 | -0.25 | -0.25 | 0.17 | -0.07 | 0.00 | -0.08 | -0.05 | -0.22 | -0.08 | ... | -0.02 | ... |
| MY Pup | 108 | 2 | -0.13 | -0.16 | -0.08 | -0.16 | -0.09 | 0.04 | 0.04 | 0.12 | -0.04 | 0.18 | 0.01 | ... | 0.04 | ... |
| NT Pup | 8 | 1 | -0.15 | -0.45 | -0.20 | -0.27 | -0.41 | 0.10 | -0.08 | 0.11 | 0.07 | -0.06 | 0.04 | -0.36 | 0.04 | -0.06 |
| V335 Pup | 7 | 1 | -0.01 | ... | -0.12 | 0.22 | -0.04 | 0.34 | ... | 0.43 | 0.20 | ... | 0.34 | ... | 0.32 | ... |
| RV Sco | 708 | 2 | 0.06 | 0.04 | -0.01 | 0.09 | -0.20 | 0.17 | 0.25 | 0.25 | 0.02 | -0.32 | 0.10 | ... | 0.03 | ... |
| RY Sco | 7 | 1 | 0.09 | 0.13 | 0.01 | 0.19 | ... | 0.39 | ... | 0.45 | 0.05 | ... | 0.24 | ... | 0.26 | ... |
| KQ Sco | 2 | 1 | 0.16 | 0.12 | 0.13 | -0.06 | ... | 0.40 | ... | 0.05 | -0.31 | ... | 0.13 | ... | 0.02 | ... |

Table 4B
Abundance Data For Cepheids: Iron through Gadolinium

| Name | Ref | Sp | [Fe/H] | [Co/H] | [Ni/H] | [Cu/H] | [Zn/H] | [Y/H] | [Zr/H] | [La/H] | [Ce/H] | [Pr/H] | [Nd/H] | [Sm/H] | [Eu/H] | [Gd/H] |
|---|---|---|---|---|---|---|---|---|---|---|---|---|---|---|---|---|
| V482 Sco | 8 | 1 | 0.07 | -0.06 | 0.01 | 0.10 | ... | 0.23 | 0.20 | 0.21 | 0.14 | -0.27 | 0.02 | ... | 0.18 | ... |
| V500 Sco | 9 | 5 | 0.01 | -0.08 | 0.00 | 0.09 | 0.21 | 0.23 | -0.12 | 0.20 | -0.03 | ... | 0.11 | ... | 0.07 | -0.24 |
| V636 Sco | 8 | 1 | 0.07 | -0.16 | 0.02 | -0.06 | -0.01 | 0.15 | 0.11 | 0.19 | -0.08 | -0.32 | -0.09 | ... | 0.06 | ... |
| V950 Sco | 8 | 1 | 0.11 | -0.03 | 0.05 | 0.13 | -0.06 | 0.18 | 0.08 | 0.03 | 0.10 | -0.18 | 0.07 | 0.00 | 0.14 | ... |
| Z Sct | 7 | 1 | 0.29 | 0.28 | 0.24 | 0.38 | ... | 0.36 | ... | 0.24 | 0.03 | ... | 0.14 | ... | 0.29 | ... |
| SS Sct | 7 | 1 | 0.06 | ... | -0.01 | 0.48 | 0.29 | 0.19 | ... | 0.40 | ... | ... | 0.16 | ... | 0.05 | ... |
| UZ Sct | 2 | 1 | 0.33 | 0.19 | 0.31 | 0.43 | 0.78 | 0.43 | 0.05 | 0.29 | 0.09 | ... | -0.01 | ... | 0.25 | 0.38 |
| EW Sct | 1 | 3 | 0.04 | -0.10 | -0.01 | 0.11 | 0.34 | 0.22 | -0.12 | 0.29 | -0.07 | ... | 0.17 | ... | 0.06 | ... |
| V367 Sct | 1 | 1 | -0.01 | 0.03 | -0.01 | -0.02 | ... | -0.06 | -0.15 | 0.45 | ... | ... | 0.18 | ... | 0.36 | ... |
| BQ Ser | 1 | 3 | -0.04 | -0.17 | -0.07 | 0.08 | 0.35 | 0.13 | -0.10 | 0.13 | -0.09 | ... | 0.22 | ... | 0.07 | 0.20 |
| S Sge | 9 | 9 | 0.08 | -0.05 | 0.07 | 0.18 | 0.39 | 0.26 | -0.01 | 0.23 | -0.08 | ... | 0.15 | ... | 0.07 | 0.04 |
| U Sgr | 9 | 21 | 0.08 | -0.08 | 0.04 | 0.00 | 0.27 | 0.19 | -0.01 | 0.18 | -0.03 | ... | 0.09 | ... | 0.07 | 0.08 |
| W Sgr | 9 | 8 | 0.02 | -0.06 | -0.01 | 0.17 | 0.23 | 0.20 | -0.07 | 0.23 | -0.04 | ... | 0.10 | ... | 0.03 | 0.10 |
| Y Sgr | 12 | 12 | 0.05 | -0.09 | 0.07 | 0.24 | 0.23 | 0.23 | -0.04 | 0.15 | -0.18 | ... | 0.08 | ... | 0.01 | 0.01 |
| VY Sgr | 2 | 1 | 0.26 | 0.24 | 0.33 | 0.24 | 0.78 | ... | -0.04 | 0.25 | -0.01 | ... | -0.07 | ... | 0.20 | 0.18 |
| WZ Sgr | 10 | 12 | 0.19 | 0.10 | 0.18 | 0.10 | 0.48 | 0.25 | -0.01 | 0.28 | 0.04 | ... | 0.12 | ... | 0.24 | 0.19 |
| XX Sgr | 7 | 1 | 0.10 | 0.10 | -0.02 | 0.21 | ... | 0.26 | ... | 0.43 | 0.10 | ... | 0.27 | ... | 0.09 | ... |
| YZ Sgr | 9 | 8 | 0.06 | -0.01 | 0.06 | 0.11 | 0.25 | 0.30 | -0.03 | 0.19 | -0.10 | ... | 0.09 | ... | 0.07 | 0.08 |
| AP Sgr | 7 | 1 | 0.10 | ... | 0.06 | ... | 0.17 | 0.28 | ... | 0.02 | 0.05 | ... | ... | ... | 0.14 | ... |
| AV Sgr | 2 | 1 | 0.34 | 0.11 | 0.38 | 0.41 | 0.61 | 0.13 | 0.14 | 0.21 | 0.00 | ... | 0.10 | ... | 0.39 | 0.35 |
| BB Sgr | 7 | 1 | 0.08 | 0.15 | 0.00 | 0.21 | 0.38 | 0.18 | ... | 0.22 | -0.09 | ... | -0.03 | ... | 0.04 | ... |
| V350 Sgr | 7 | 1 | 0.18 | 0.15 | 0.03 | 0.13 | ... | 0.33 | ... | 0.26 | 0.05 | ... | 0.16 | ... | 0.15 | ... |
| ST Tau | 1 | 3 | -0.05 | -0.33 | -0.04 | 0.19 | 0.02 | 0.18 | -0.13 | 0.17 | -0.10 | ... | 0.14 | ... | 0.04 | 0.13 |
| SZ Tau | 11 | 16 | 0.07 | -0.14 | 0.05 | 0.20 | 0.26 | 0.21 | 0.00 | 0.25 | -0.03 | ... | 0.11 | ... | 0.14 | 0.05 |
| AE Tau | 3 | 1 | -0.19 | -0.45 | -0.22 | -0.14 | ... | 0.08 | -0.18 | 0.32 | -0.27 | ... | -0.08 | ... | 0.11 | ... |
| EF Tau | 6 | 1 | -0.74 | ... | -0.82 | -0.36 | ... | -0.67 | ... | -0.80 | -0.68 | ... | -0.33 | ... | -0.55 | ... |
| EU Tau | 1 | 2 | -0.06 | -0.02 | -0.08 | 0.21 | ... | 0.07 | -0.08 | 0.16 | -0.18 | ... | -0.02 | ... | 0.01 | 0.18 |
| R TrA | 8 | 1 | 0.06 | -0.07 | 0.00 | 0.18 | 0.03 | 0.18 | -0.16 | 0.27 | 0.11 | -0.33 | 0.09 | 0.15 | 0.14 | ... |
| S TrA | 8 | 1 | 0.12 | -0.07 | 0.06 | 0.07 | ... | 0.30 | 0.20 | 0.31 | 0.16 | -0.17 | 0.07 | ... | 0.17 | ... |
| LR TrA | 8 | 1 | 0.25 | 0.12 | 0.15 | 0.13 | 0.01 | 0.26 | 0.24 | 0.33 | 0.09 | -0.17 | 0.04 | ... | 0.22 | 0.26 |
| T Vel | 408 | 3 | -0.02 | -0.18 | -0.08 | -0.07 | 0.13 | 0.19 | -0.13 | 0.15 | 0.04 | -0.12 | 0.02 | ... | -0.02 | 0.03 |

**Table 4B**
Abundance Data For Cepheids: Iron through Gadolinium

| Name | Ref | Sp | [Fe/H] | [Co/H] | [Ni/H] | [Cu/H] | [Zn/H] | [Y/H] | [Zr/H] | [La/H] | [Ce/H] | [Pr/H] | [Nd/H] | [Sm/H] | [Eu/H] | [Gd/H] |
|---|---|---|---|---|---|---|---|---|---|---|---|---|---|---|---|---|
| V Vel  | 8   | 1  | -0.23 | -0.29 | -0.27 | -0.09 | -0.29 | -0.13 | -0.03 | …    | -0.01 | …     | -0.20 | …     | -0.04 | …    |
| RY Vel | 408 | 3  | 0.01  | 0.04  | 0.00  | 0.18  | 0.29  | 0.20  | 0.00  | 0.26 | 0.09  | 0.02  | 0.21  | 0.06  | 0.25  | 0.34 |
| RZ Vel | 408 | 3  | -0.02 | -0.13 | -0.04 | 0.38  | 0.41  | 0.19  | -0.23 | 0.32 | -0.06 | 0.10  | 0.10  | …     | 0.12  | …    |
| ST Vel | 8   | 1  | 0.00  | -0.04 | -0.03 | 0.13  | -0.19 | 0.17  | 0.12  | 0.25 | 0.11  | …     | -0.05 | …     | 0.04  | …    |
| SV Vel | 8   | 1  | 0.08  | -0.10 | -0.02 | 0.03  | -0.20 | 0.22  | 0.12  | 0.08 | 0.12  | -0.21 | 0.15  | 0.04  | 0.18  | …    |
| SW Vel | 108 | 4  | -0.07 | -0.22 | -0.12 | -0.35 | 0.11  | 0.14  | -0.03 | 0.23 | -0.09 | …     | 0.14  | …     | 0.09  | 0.11 |
| SX Vel | 408 | 3  | -0.02 | -0.19 | -0.06 | -0.07 | 0.16  | 0.17  | -0.10 | 0.13 | -0.03 | -0.23 | -0.02 | -0.08 | 0.05  | 0.01 |
| XX Vel | 8   | 1  | -0.05 | -0.27 | -0.10 | 0.10  | -0.25 | 0.10  | 0.19  | …    | 0.01  | …     | 0.06  | …     | 0.03  | …    |
| AE Vel | 8   | 1  | 0.05  | -0.15 | 0.01  | -0.13 | -0.07 | 0.14  | 0.06  | 0.22 | 0.04  | -0.06 | -0.02 | …     | 0.09  | …    |
| AH Vel | 8   | 1  | 0.10  | -0.03 | 0.06  | -0.16 | -0.07 | 0.26  | 0.14  | 0.17 | 0.24  | -0.06 | 0.14  | 0.08  | 0.21  | …    |
| BG Vel | 8   | 1  | -0.01 | -0.21 | -0.06 | -0.15 | -0.12 | 0.08  | -0.15 | 0.06 | -0.01 | -0.36 | -0.05 | 0.05  | 0.01  | …    |
| CS Vel | 8   | 1  | 0.08  | -0.11 | 0.05  | -0.07 | 0.25  | 0.13  | 0.04  | 0.20 | 0.03  | …     | 0.06  | …     | 0.05  | …    |
| CX Vel | 8   | 1  | 0.06  | -0.07 | 0.00  | 0.11  | -0.33 | 0.17  | 0.14  | 0.23 | 0.05  | …     | -0.02 | …     | 0.05  | …    |
| DK Vel | 8   | 1  | -0.02 | 0.00  | -0.04 | 0.11  | -0.08 | 0.09  | 0.10  | 0.08 | 0.01  | -0.23 | -0.04 | -0.39 | 0.14  | …    |
| DR Vel | 8   | 1  | 0.08  | -0.07 | 0.03  | 0.12  | 0.22  | 0.16  | 0.16  | 0.16 | 0.07  | -0.21 | 0.05  | …     | 0.14  | 0.10 |
| EX Vel | 8   | 1  | 0.05  | 0.01  | -0.03 | -0.10 | -0.28 | 0.15  | 0.11  | -0.02| 0.14  | -0.13 | 0.07  | 0.34  | 0.20  | 0.12 |
| FG Vel | 8   | 1  | -0.05 | -0.18 | -0.07 | -0.15 | 0.26  | 0.01  | -0.20 | 0.08 | -0.02 | -0.33 | 0.04  | …     | -0.15 | …    |
| FN Vel | 8   | 1  | 0.06  | -0.07 | -0.03 | 0.00  | 0.25  | 0.15  | 0.20  | 0.27 | 0.14  | -0.24 | 0.00  | …     | 0.06  | …    |
| S Vul  | 10  | 4  | -0.01 | -0.23 | -0.02 | -0.04 | …     | 0.21  | -0.09 | 0.18 | -0.25 | …     | 0.12  | …     | 0.02  | -0.01|
| T Vul  | 12  | 12 | 0.01  | -0.12 | -0.01 | 0.14  | 0.13  | 0.15  | -0.01 | 0.24 | -0.08 | …     | 0.14  | …     | 0.08  | 0.07 |
| U Vul  | 9   | 7  | 0.09  | -0.04 | 0.11  | 0.22  | 0.31  | 0.27  | -0.04 | 0.13 | -0.06 | …     | 0.19  | …     | 0.07  | 0.02 |
| X Vul  | 9   | 6  | 0.07  | -0.05 | 0.08  | 0.09  | 0.35  | 0.20  | -0.01 | 0.15 | -0.06 | …     | 0.13  | …     | 0.07  | 0.06 |
| SV Vul | 10  | 23 | 0.05  | -0.15 | 0.03  | -0.11 | 0.26  | 0.24  | -0.08 | 0.20 | -0.13 | …     | 0.06  | …     | 0.04  | -0.04|

Notes:     Same as for Table 4A.

**Table 5**
[Fe/H] Comparisons

| Star | Papers I-VI Source | [Fe/H] | R | Current | D1 | D2 |
|---|---|---|---|---|---|---|
| l Car | … | … | 0.00 | 0.05 | … | 0.05 |
| U Car | … | … | 0.15 | 0.01 | … | -0.14 |
| V Car | … | … | -0.04 | 0.01 | … | 0.05 |
| WZ Car | … | … | 0.13 | 0.03 | … | -0.10 |
| V Cen | 1 | 0.04 | 0.00 | -0.03 | -0.07 | -0.03 |
| KN Cen | … | … | 0.13 | 0.35 | … | 0.22 |
| XX Cen | … | … | 0.04 | 0.16 | … | 0.12 |
| BG Cru | 1 | -0.02 | … | 0.06 | 0.08 | … |
| GH Lup | … | … | 0.03 | 0.08 | … | 0.05 |
| S Mus | … | … | 0.16 | -0.02 | … | -0.18 |
| UU Mus | … | … | 0.08 | 0.05 | … | -0.03 |
| S Nor | 1 | 0.05 | 0.03 | 0.07 | 0.02 | 0.04 |
| U Nor | … | … | 0.10 | 0.15 | … | 0.05 |
| V340 Nor | 1 | 0.00 | … | 0.08 | 0.08 | … |
| BF Oph | 7 | 0.00 | … | 0.05 | 0.05 | … |
| AP Pup | … | … | -0.07 | 0.06 | … | 0.13 |
| CE Pup | … | … | … | -0.15 | … | … |
| MY Pup | 1 | -0.12 | … | -0.14 | -0.02 | … |
| NT Pup | … | … | … | -0.15 | … | … |
| RV Sco | 7 | 0.10 | … | 0.03 | -0.07 | … |
| T Vel | 4 | -0.02 | 0.06 | -0.02 | 0.00 | -0.08 |
| RY Vel | 4 | -0.03 | -0.07 | 0.10 | 0.13 | 0.17 |
| RZ Vel | 4 | -0.07 | -0.16 | 0.07 | 0.14 | 0.23 |
| SW Vel | 1 | 0.01 | -0.20 | -0.16 | -0.17 | 0.04 |
| SX Vel | 4 | -0.03 | 0.00 | 0.00 | 0.03 | 0.00 |
| | | | | Mean | 0.02 | 0.03 |
| | | | | N | 12 | 18 |
| | | | | Max | 0.14 | 0.23 |
| | | | | Min | -0.17 | -0.18 |
| | | | | Sigma | 0.09 | 0.12 |

Notes

Source: 1 – Andrievsky et al. (2002a (Paper I)), 4 – Luck et al. (2003 (Paper IV)), 7 – Luck, Kovtyukh, & Andrievsky (2006 (Paper VI)).
R: Romaniello et al. (2008).
Current: This Paper.
D1 & D2: This Paper – Papers I-VI and This Paper – Romaniello et al. (2008).

**Table 6**
Abundance Gradients: Species = a* Rg + b

| Species | Gradient | | Uncertainty | | Sigma | N |
|---|---|---|---|---|---|---|
| | a | b | a | b | | |
| [C/H] | -0.077 | 0.442 | 0.006 | 0.052 | 0.183 | 265 |
| [N/H] | -0.053 | 0.811 | 0.006 | 0.058 | 0.176 | 150 |
| [O/H] | -0.038 | 0.240 | 0.006 | 0.050 | 0.169 | 253 |
| [Na/H] | -0.044 | 0.564 | 0.004 | 0.033 | 0.118 | 266 |
| [Mg/H] | -0.048 | 0.326 | 0.006 | 0.052 | 0.172 | 229 |
| [Al/H] | -0.053 | 0.546 | 0.004 | 0.033 | 0.114 | 251 |
| [Si/H] | -0.049 | 0.468 | 0.003 | 0.024 | 0.086 | 269 |
| [S/H] | -0.076 | 0.699 | 0.005 | 0.045 | 0.159 | 264 |
| [Ca/H] | -0.040 | 0.300 | 0.004 | 0.032 | 0.115 | 269 |
| [Sc/H] | -0.045 | 0.356 | 0.004 | 0.037 | 0.130 | 261 |
| [Ti/H] | -0.043 | 0.398 | 0.004 | 0.033 | 0.119 | 269 |
| [V/H] | -0.047 | 0.365 | 0.004 | 0.032 | 0.113 | 267 |
| [Cr/H] | -0.046 | 0.385 | 0.004 | 0.031 | 0.110 | 266 |
| [Mn/H] | -0.062 | 0.466 | 0.005 | 0.040 | 0.139 | 229 |
| [Fe/H] | -0.055 | 0.475 | 0.003 | 0.027 | 0.095 | 270 |
| [Co/H] | -0.018 | 0.093 | 0.005 | 0.043 | 0.141 | 236 |
| [Ni/H] | -0.054 | 0.419 | 0.003 | 0.028 | 0.100 | 270 |
| [Cu/H] | -0.041 | 0.391 | 0.007 | 0.061 | 0.205 | 236 |
| [Zn/H] | -0.003 | 0.112 | 0.010 | 0.088 | 0.244 | 193 |
| [Y/H] | -0.044 | 0.523 | 0.004 | 0.038 | 0.132 | 256 |
| [Zr/H] | -0.047 | 0.404 | 0.005 | 0.044 | 0.127 | 195 |
| [La/H] | -0.019 | 0.347 | 0.005 | 0.039 | 0.129 | 244 |
| [Ce/H] | -0.021 | 0.173 | 0.004 | 0.039 | 0.136 | 262 |
| [Pr/H] | 0.002 | -0.211 | 0.007 | 0.060 | 0.146 | 112 |
| [Nd/H] | -0.006 | 0.130 | 0.004 | 0.039 | 0.137 | 268 |
| [Eu/H] | -0.021 | 0.278 | 0.004 | 0.037 | 0.127 | 264 |
| [Gd/H] | -0.021 | 0.278 | 0.008 | 0.072 | 0.143 | 96 |
| [Sm/H] | -0.055 | 0.412 | 0.025 | 0.194 | 0.140 | 39 |

Note: Sigma is the standard deviation of the fit using N stars.

**Figures**

Figure 1: The gradient in [Fe/H] for the 270 Cepheid variables from this and preceding studies. The red line is a simple least squares fit to the entire dataset which yields a gradient d[Fe/H]/dR$_G$ = -0.055 dex kpc$^{-1}$. The blue line is a LOWESS fit which uses localized weighting. The LOWESS fit shows a flattening of the gradient in the 8 – 10 kpc region. However, the flattening is controlled by a small number of stars at the 9.5 kpc distance mark. CE Pup and TW Nor at R$_G$ = 14.7 and 5.8 kpc have typical error bars for distance and abundance indicated. Note that the distance uncertainties vary with R$_G$ with the error minimizing at R$_G$ = 7.9 kpc – the solar galactocentric radius. The stars in green comes from prior work in this series while the yellow symbols are for stars in the Carina region. This color scheme is used in all figures which follow.

Figure 2: The azimuthal variation in [Fe/H] for two annuli near the solar galactocentric radius. There is no discernible dependence on galactocentric angle. The formal slope is −4.56E-4 dex/degree with an uncertainty of ±5.6E-4. The confidence interval indicates the data is consistent with no gradient. The solar region is at an angle of 0°.

Figure 3: The distribution of program Cepheids in the galactic plane.

Figure 4: Distance above or below the plane versus galactocentric radius. The stars at large distances from the plane do not perturb the derived gradients.

Figure 5: A contour map of the [Fe/H] abundances versus spatial position. The salient features of this plot are the smooth variation of the [Fe/H] ratio with position, and the presence of the "metallicity island" at (1,3) kpc. The island is dependent on the abundances of less than 10 stars and hence, is rather uncertain.

Figure 6: The [Fe/H], [C/Fe], [O/Fe], and [Na/Fe] ratios versus log(Period). There is no discernible dependence of the ratios on log(P) which can be interpreted as the degree of processing and mixing is comparable in all of these objects.

Figure 7: The gradient in [C/H] and [C/Fe]. Note that the gradient in [C/H] is steeper than that in [Fe/H] which manifests in the non-zero slope in [C/Fe]. For the [C/Fe] data we also show the 95% confidence interval.

Figure 8: [O/H] and [O/Fe] versus galactocentric distance. The scatter in oxygen is significant but the gradient is comparable to that found in B stars. For the [O/Fe] data we also show the 95% confidence interval.

Figure 9: The gradient data for [Si/H], [Y/H], [Nd/H], and [Eu/H]. Nd is the one element which shows no gradient but one must wonder if the gradient is hidden in the scatter.

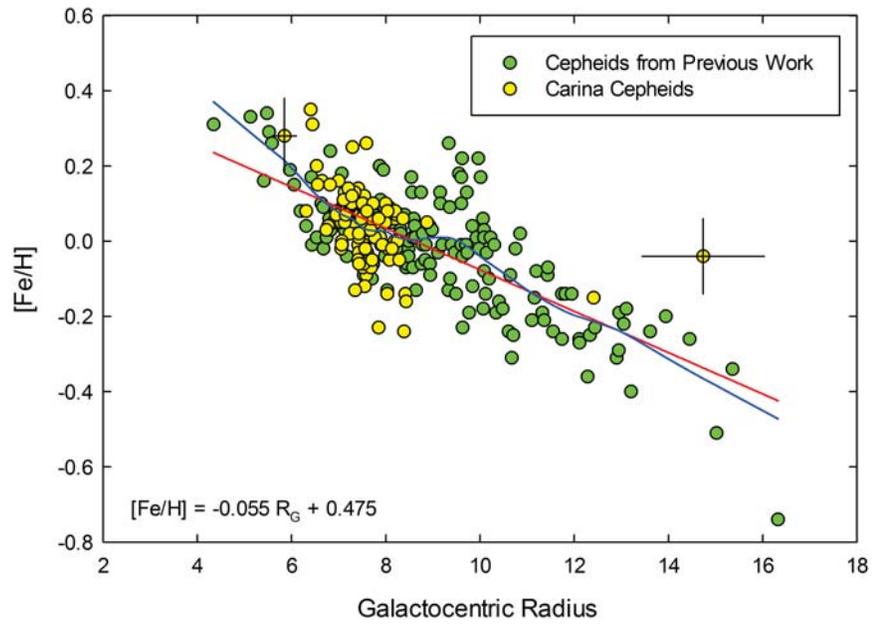

**Figure 1**

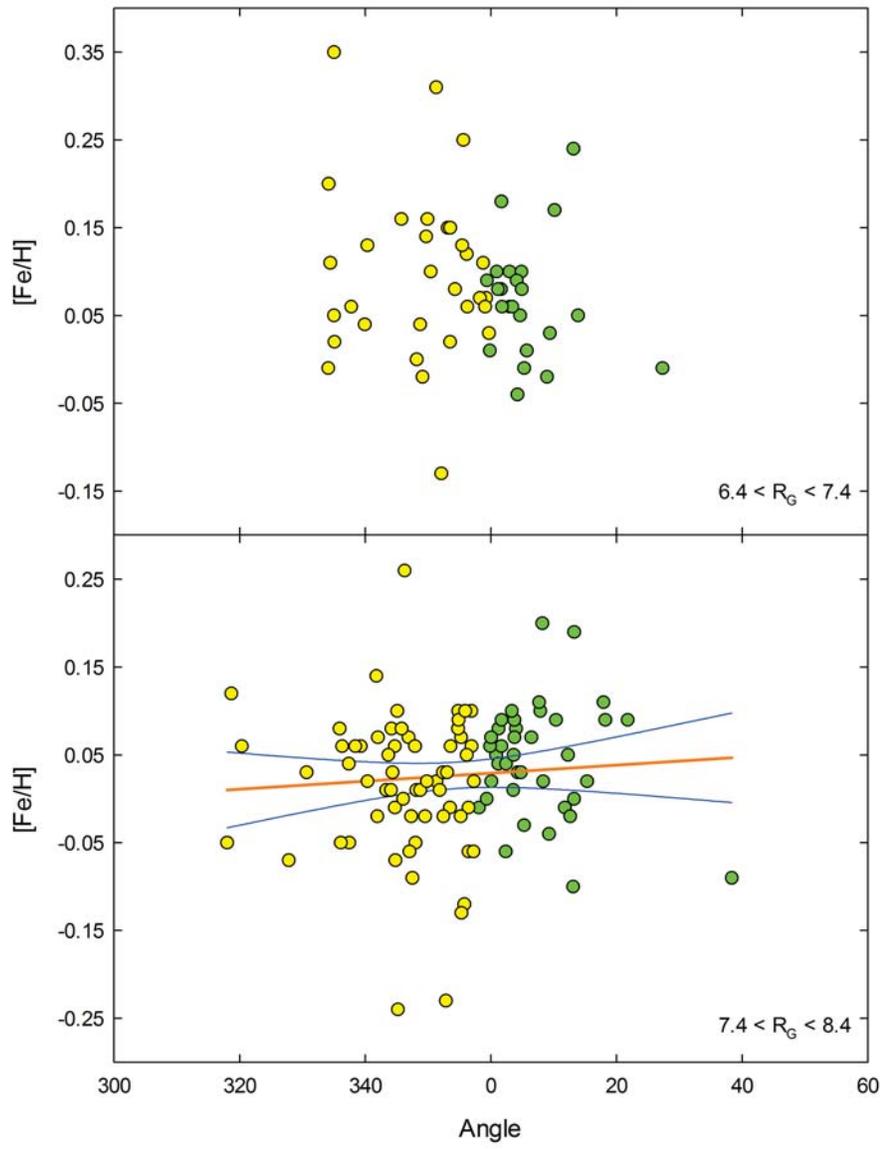

**Figure 2**

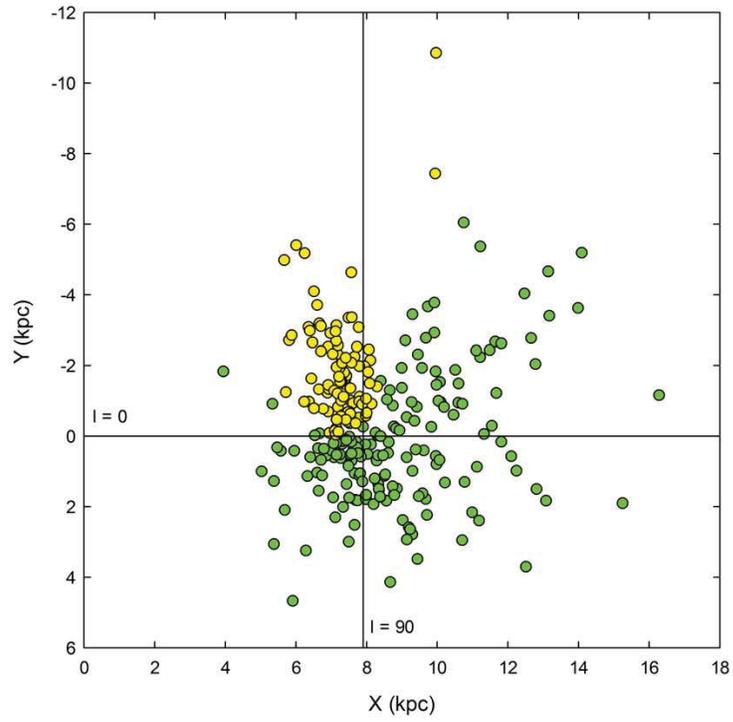

**Figure 3**

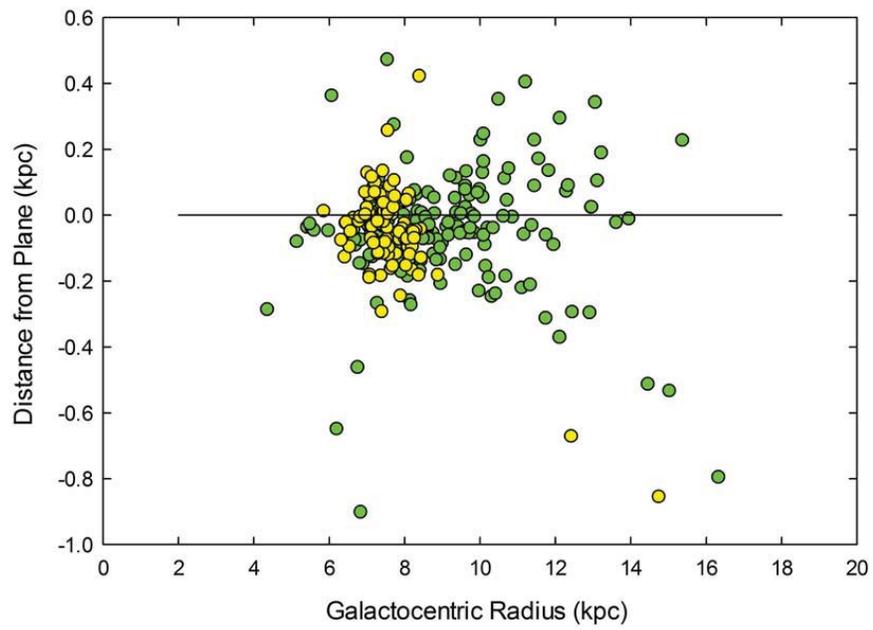

**Figure 4**

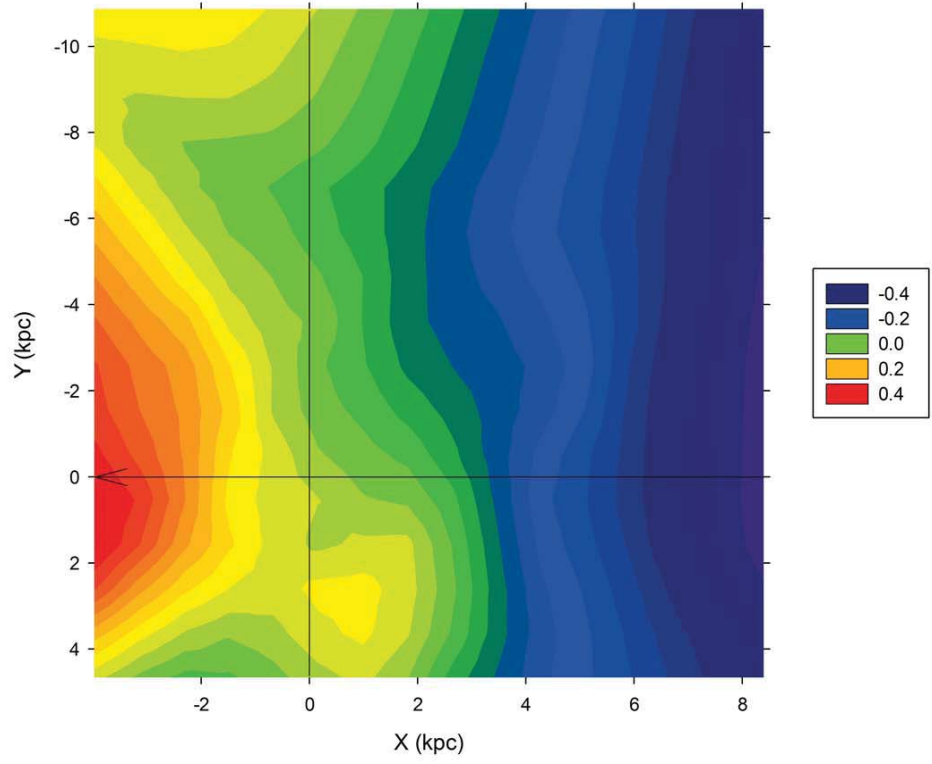

**Figure 5**

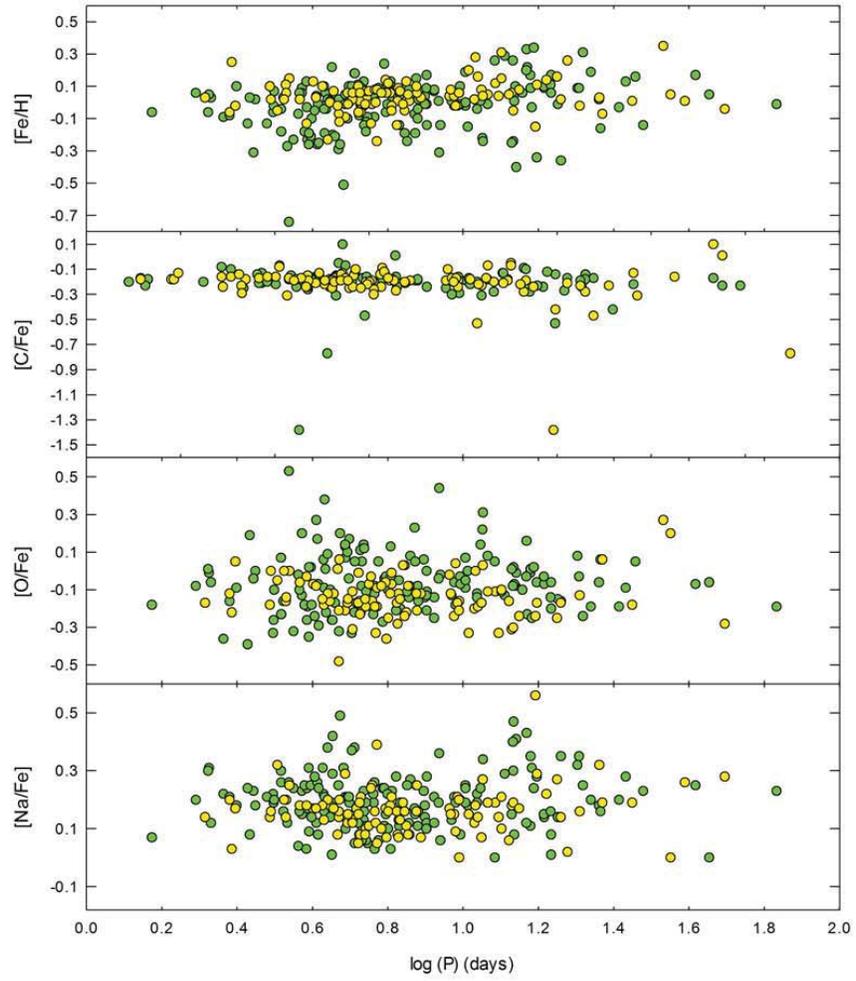

**Figure 6**

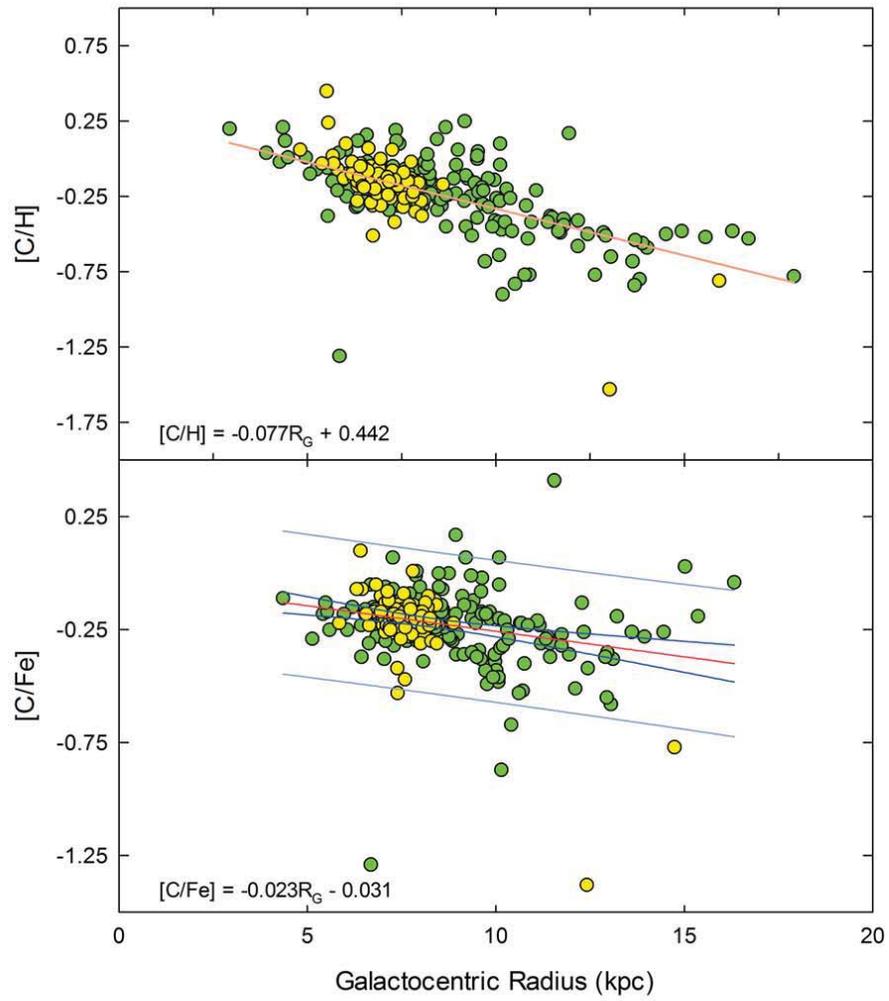

**Figure 7**

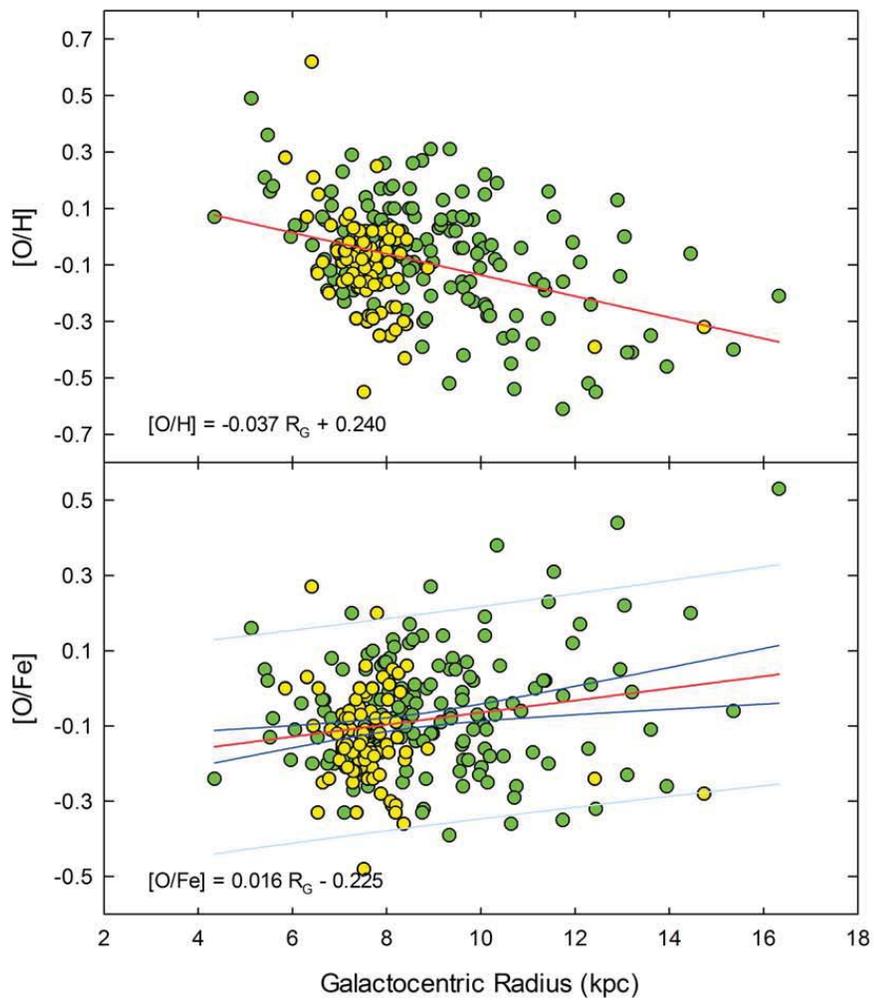

**Figure 8**

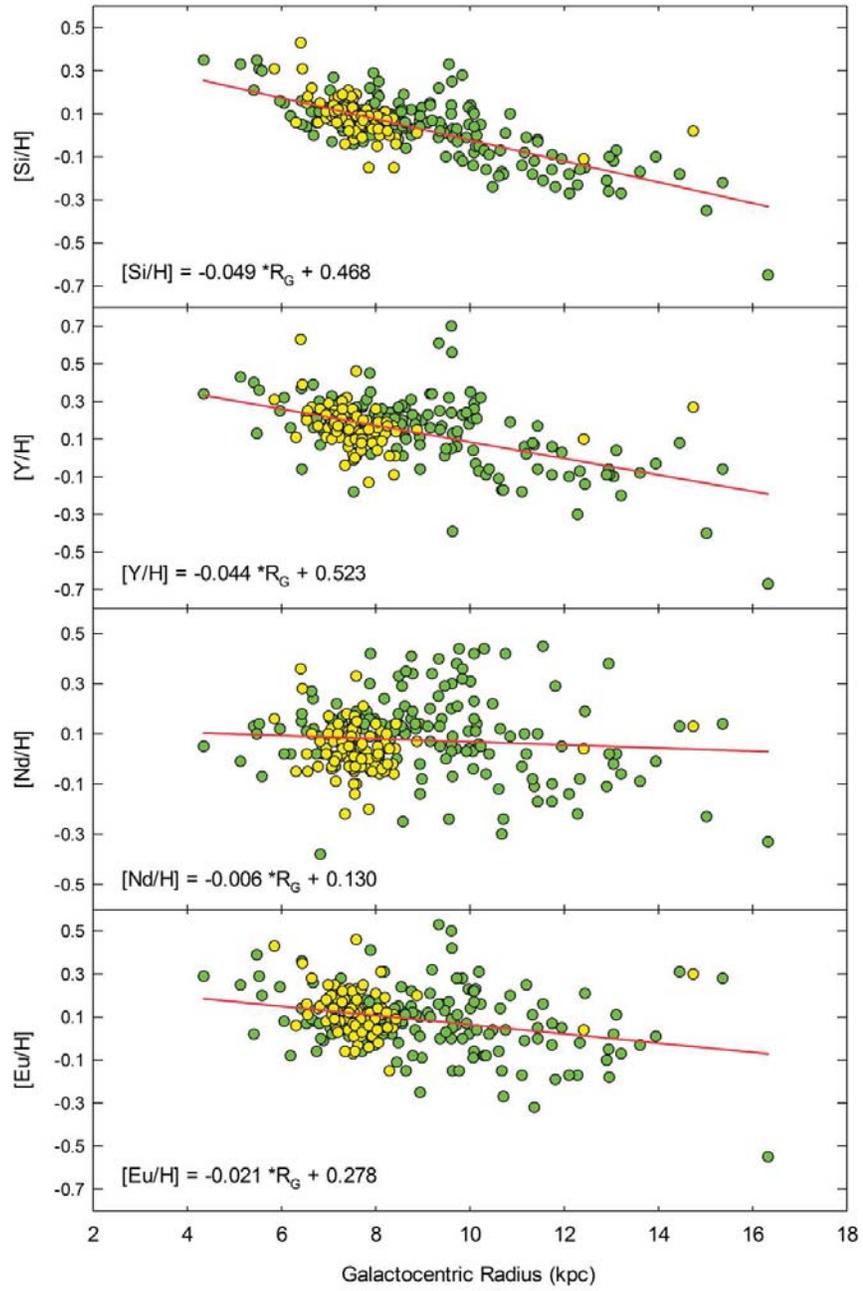

Figure 9